\documentclass[floatfix,twocolumn,showpacs,aps,tightenlines]{revtex4}
\usepackage{graphicx}
\usepackage{color}
\usepackage{psfrag}
\usepackage{dcolumn}
\usepackage{bm}

\usepackage{amssymb}
\usepackage{amsmath}
\usepackage{amsfonts}
\usepackage{longtable}

\newcommand{\Spar}{\tilde S_\|}
\newcommand{\Dpar}{\tilde \Delta_\|}
\newcommand{\dmt}{\delta{m}}

\newcommand{\beq}{\begin{equation}}
\newcommand{\eeq}{\end{equation}}
\newcommand{\KMS}{\rm km\,s^{-1}}

\newcommand{\be}{\begin{equation}}
\newcommand{\ee}{\end{equation}}
\newcommand{\bes}{\begin{subequations}}
\newcommand{\ees}{\end{subequations}}

\newcommand{\hn}{\hat{{n}}}
\newcommand{\hL}{\hat{{L}}}

\newcommand{\hl}{\hat{{\lambda}}}

\usepackage{bm}

\begin{document}

\title{Remnant mass, spin, and recoil
from spin aligned black-hole binaries}

\author{James Healy} 
\author{Carlos O. Lousto}
\author{Yosef Zlochower} 
\affiliation{Center for Computational Relativity and Gravitation,
School of Mathematical Sciences,
Rochester Institute of Technology, 78 Lomb Memorial Drive, Rochester,
 New York 14623}

\date{\today}

\begin{abstract}

We perform a set of 36 nonprecessing black-hole binary simulations
with spins either aligned or counteraligned with the orbital angular
momentum in order to model the final mass, spin, and recoil of the
merged black hole as a function of the individual black hole spin
magnitudes and the mass ratio of the progenitors.  We find that the
maximum recoil for these configurations is $V_{max}=526\pm23\,\KMS$,
which occurs when the progenitor spins are maximal, the mass ratio is
$q_{max}=m_1/m_2=0.623\pm0.038$, the smaller black-hole spin is
aligned with the orbital angular momentum, and the larger black-hole
spin is counteraligned ($\alpha_1=-\alpha_2=1$).  This maximum recoil
is  about $80\,\KMS$ larger than previous estimates, but most
importantly, because the maximum occurs for smaller mass ratios, the
probability for a merging binary to recoil faster than $400\,\KMS$ can
be as large as $17\%$, while the probability for recoils faster than
$250\, \KMS$ can be as large as  45\%.
We provide explicit phenomenological formulas
for the final mass, spin, and recoil as a function of the individual BH
spins and the mass difference between the two black holes. Here we
include terms up through fourth-order in the initial spins and mass
difference, and find excellent agreement (within a few percent) with
independent results available in the literature.
The maximum radiated energy is $E_{\rm rad}/m\approx11.3\%$ and final
spin $\alpha_{\rm rem}^{\rm max}\approx0.952$ for equal mass,
aligned maximally spinning binaries.
\end{abstract}

 \maketitle

\section{Introduction}\label{sec:intro}

The fully nonlinear simulations of merging black-hole binaries (BHBs)
that were enabled by the 2005 breakthroughs in numerical
relativity~\cite{Pretorius:2005gq, Campanelli:2005dd, Baker:2005vv}
revealed some unexpectedly large effects.  Perhaps one of the most
striking is that the merger remnant can recoil away from the center of
mass by thousands of km/s for BHBs with spins at least partially in
the orbital plane~\cite{Campanelli:2007ew, Gonzalez:2007hi,
Lousto:2011kp, Lousto:2012su, Lousto:2012gt}.  Such recoils, if
common, would have major implications for structure formation and the
evolution of galaxies, as well as the retention of BHs in globular
clusters and the formation of intermediate mass BHs.
The probability of these large recoils depends on the
distribution of mass ratios and spins of the progenitor binaries. While the detailed modeling of
those recoil velocities from merging BHBs as a function of the
individual spins (magnitudes and directions) of the BHs and the mass
ratio is well underway~\cite{Lousto:2012gt, Lousto:2010xk,
Zlochower:2010sn, Lousto:2009mf, Lousto:2008dn}, the major effort
required to simulate BHBs in a realistic astrophysical environment
started more recently~\cite{Noble:2012xz, Bode:2011tq,
Bogdanovic:2010he, Bode:2009mt, Gold:2013zma, Shapiro:2013qsa,
Farris:2012ux, Farris:2011vx, Palenzuela:2010nf, Palenzuela:2009hx,
Megevand:2009yx, Palenzuela:2009yr}.  Analyses of Newtonian and
post-Newtonian simulations appear to indicate that accretion dynamics
will skew the spin distributions away from configurations that favor
very large recoils~\cite{Bogdanovic:2007hp, Dotti:2009vz, Miller:2013gya}
because these effects tend to align (or counter-align) the spins with
the orbital angular momentum. 
In addition, during the late stages
of the BHB evolution,
post-Newtonian resonance effects \cite{Schnittman:2004vq, Gerosa:2013laa}
tend to further align the BH spins with each other and the 
orbital angular momentum (or counteralign them azimuthally).
On the other hand, recent studies of  chaotic \cite{Dotti:2012qw}
and partially chaotic \cite{Sesana:2014bea}
accretion suggest misalignment of spins can also be a common evolutionary
scenario for BHBs, possibly allowing for the merger remnant
to escape from large galaxies \cite{Komossa:2012cy, Gerosa:2014gja}.

In this paper we simulate the late-inspiral and merger
stages of BHBs in configurations where the spins are exactly aligned
or counter-aligned with the orbital angular momentum.  By doing so,
we are able to 
quantify how large the recoil can be when coherent accretion effects dominate
the distribution of spin directions, thus providing a lower bound
to the recoil of the BH remnant. The aligned-spin case also provides
the optimal configuration for the radiation of gravitational energy
and angular momentum.
Here we provide a unified, higher-order phenomenological model of
the remnant mass, spin, and recoil from the merger
of two BHs with different masses and different spin magnitudes (either 
aligned or antialigned spins).


This paper is organized as follows, in Sec.
\ref{sec:recoils} we review the current status of 
the modeling of the remnant recoil. In Sec.~\ref{sec:evolution}
we
review the numerical techniques used for our evolutions of
the BHBs and the subsequent analyses of the progenitor and remnant
properties.
In Sec.~\ref{sec:fits}
we present the explicit form of the new phenomenological
formulas for the final mass, spin and recoil of the  merger remnant.
We apply these formulas to
astrophysically motivated distributions of the mass ratios and
spins of the progenitor binaries to obtain probabilities for
 a given recoil, final remnant mass and spin.
In Sec.~\ref{sec:discussion}
we discuss the consequences of our results.

\section{Model of recoils on the orbital plane}\label{sec:recoils}

Beginning in Ref.~\cite{Campanelli:2007ew}, we developed a 
 heuristic model for the gravitational recoil of a merging binary.
The model for the in-plane recoil 
was based on PN-inspired fitting formulas  combined with the 
results of~\cite{Gonzalez:2006md,
Herrmann:2007ac, Koppitz:2007ev}
(a similar model was developed independently
in \cite{Baker:2008md}). Here we use the PN-inspired variables
\begin{eqnarray*}
  m = m_1 + m_2,\\
\delta m = \frac{m_1 - m_2}{m},\\
 \vec S = \vec S_1 + \vec S_2,\\
  \vec \Delta  = m (\vec S_2/m_2 - \vec S_1/m_1),
\end{eqnarray*}
where $m_i$ is the mass of BH $i=1,2$ and $\vec S_i$ is the spin of BH
$i$.
We also use the auxiliary variables
\begin{eqnarray*}
 \eta = \frac{m_1 m_2}{m^2},\\
 q=\frac{m_1}{m_2},\\
 \vec \alpha_i = \vec S_i/m_i^2,
\end{eqnarray*}
where $|\vec \alpha_i| \leq 1$ is the dimensionless spin of BH $i$,
and we use the convention that $m_1 \leq m_2$ and hence $q\leq 1$.

The in-plane recoil can be split (at least approximately) into two
components: a part  due solely to unequal masses and a part due to the
out-of-plane components of the spins of the two BHs. To lowest order
in the spin, the formula is given by,
\begin{equation}\label{eq:empirical}
\vec{V}_{\rm recoil}(q,\vec\alpha_i)=v_m\,\hat{e}_1+
v_\perp(\cos(\xi)\,\hat{e}_1+\sin(\xi)\,\hat{e}_2),
\end{equation}
where
\begin{subequations}
\begin{equation}\label{eq:vm}
v_m=-A\eta^2 \delta m\left(1+B\,\eta\right),
\end{equation}
\begin{equation}\label{eq:vperp}
v_\perp=H\eta^2\left[\frac{\Delta_\|}{m^2} - H_S \delta m
\frac{S_\|}{m^2}\right].
\end{equation}
\end{subequations}
Here the  index $\perp$ and $\|$
refer to components perpendicular to and parallel to
 the orbital angular momentum
during the short period around merger when most of the recoil is
generated,
while $\hat{e}_1,\hat{e}_2$ are orthogonal unit vectors in the
orbital plane, and $\xi$ measures the angle between the ``unequal mass''
and ``spin'' contributions to the recoil velocity in the orbital plane
(See Fig.~\ref{fig:VmxiVs}).
This formula can be extended by
adding additional nonlinear terms (as we will show in this paper). 
The coefficients are given by
$A = 1.2\times 10^{4}\,\KMS$~\cite{Gonzalez:2006md},
$B = -0.93$~\cite{Gonzalez:2006md},
 and $H = (6.9\pm0.5)\times 10^{3}\,\KMS$~\cite{Lousto:2007db}.
We will study in detail how $\xi$ depends on the configurations here ($\xi$
was initially studied in Ref.~\cite{Lousto:2007db}, where it was found
that  $\xi \sim 145^\circ$ for a range of quasicircular
configurations).

\subsection{Post-Newtonian analysis}\label{subsec:PN}

Here we use the leading-order post-Newtonian expressions for the
radiated linear momentum to get a qualitative understanding of
 the full numerical results.
As seen in Eq.\ (3.31) of Ref.~\cite{Kidder:1995zr},
the instantaneous radiated
linear momentum due to the asymmetry in the masses of the
binary is given by
\begin{eqnarray}
\dot{\vec{P}}_N=&&-\frac{8\eta^2\delta{m}}{105}
\left(\frac{m}{r}\right)^4
\left[(5\,V_T^2-2\,V_r^2+4m/r)\,V_r\,\hn\right.\nonumber\\
&&\left.-(12\,V_r^2+50\,V_T^2+8m/r)\,V_T\,\hl\right],
\end{eqnarray}
and the radiated linear momentum due to the leading-order spin-orbit coupling
is 
\beq
\dot{\vec{P}}_{SO}=-\frac{8\eta^2\,m}{15\,r^5}
\left[4\,V_T\,V_r\,\hn+2(V_T^2-V_r^2)\hl\right]\Delta_\|, 
\eeq
where
$V_T$ and $V_r$ are the tangential and radial velocities,
respectively.
The velocity is given by,  $\vec{V}=V_T\hl+V_r\hn$, where
$\hn=(\vec{r}_1-\vec{r}_2)/|\vec{r}_1-\vec{r}_2|$ and
$\hn\times\hl=\hL$.

For a quasicircular orbit, the angle between these two 
components of the instantaneous radiated
linear momentum is given by
\beq\label{eq:signDelta}
\cos\xi^\Delta=\left[-1+\frac{15625}{6728}\frac{V_r^2}{V_T^2}+...\right]
\text{sign}(\delta{m}\vec{\Delta}\cdot\hL),
\eeq
and hence, for circular orbits, i.e. $V_r=0$,
\beq
\cos\xi_c^\Delta=-\text{sign}(\delta{m}\vec{\Delta}\cdot\hL).
\eeq
Thus the two components are opposite of each other when  $\vec \Delta$ is aligned
with the orbital angular momentum (corotating orbits for our
configurations). Similarly, for our counter-rotating configurations,
the two components add constructively.

On the other hand, for orbits dominated by the radial motion instead
(i.e., $V_T \ll V_r$) the angle $\xi$ has the form,
\beq
\cos\xi^\Delta=\left[\frac{4\,V_T}{V_r}\frac{(r\,V_r^2+2M)}{(-r\,V_r^2+2M)}+...\right]
\text{sign}(\delta{m}\vec{\Delta}\cdot\hL).
\eeq
Hence, in the near-headon case (i.e. $V_T\approx 0$) ,
we have
\beq
\cos\xi_h^\Delta\approx 0,
\eeq
and the two components of the recoil are perpendicular to each other.

The next leading term in the spin orbit contribution to the recoil
[see Eqs.~(4.7)-(4.9) of
Ref.~\cite{Racine:2008kj}] is proportional to $\delta m S_\|$.
For circular orbits the angle $\cos\xi_c^S$ between the unequal mass
recoil and the terms in the recoil proportional to $\delta m \vec S$
is given by
\beq\label{eq:signS}
\cos\xi_c^S=-\text{sign}(\vec{S}\cdot\hL),
\eeq
while for headon collisions [see  Eqs.~(3.17) of
Ref.~\cite{Racine:2008kj}]
the two components are perpendicular and
\beq\label{eq:hS}
\cos\xi_h^S\approx 0.
\eeq

\begin{figure} \includegraphics[width=0.8\columnwidth]{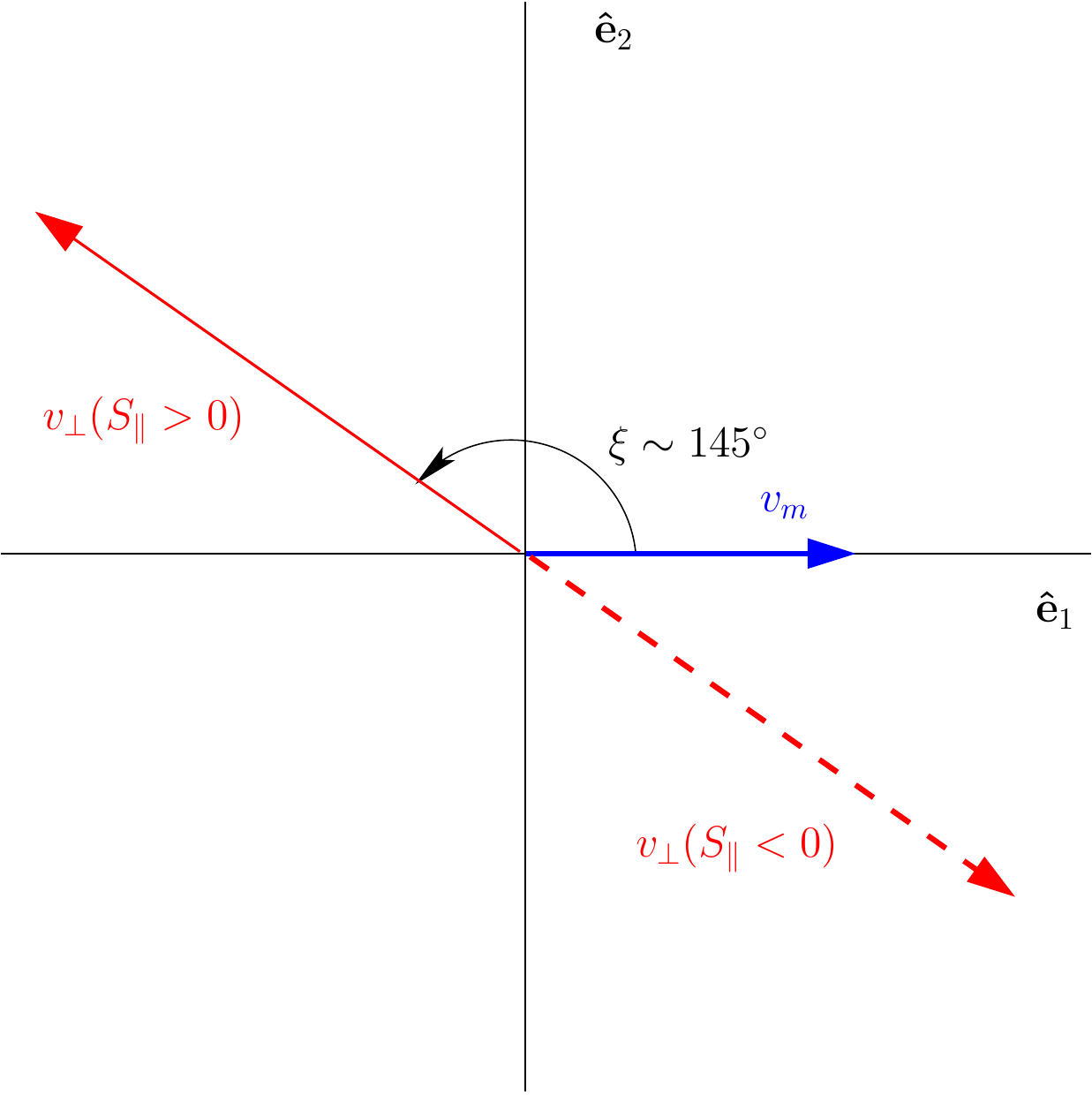}
\caption{
  A sketch showing how the angle $\xi$ between the unequal mass
  contribution to the recoil and the spin dependent contribution to the
  recoil depends on the sign of $S_\|$ (with similar behavior for the
  $\delta m \Delta_\|$ dependent term).  The two components essentially
  add if the net spin is counteraligned with the orbital angular
  momentum and subtract if the spin is aligned with the orbital angular
  momentum.}
 \label{fig:VmxiVs}
\end{figure}

In a full numerical simulation, the inspiral of two BHs
is neither circular nor headon, and hence we  expect a value
of $\xi$ that lies between $90^\circ$ and either $180^\circ$
(corotating) or $0^\circ$ (counterrotating) (see
Fig.~\ref{fig:VmxiVs}).
In particular, because of the hangup effect \cite{Campanelli:2006uy}
we expect
that aligned-spin configurations, which have tighter (i.e., more
circular) orbits,
 have $\xi\approx180^\circ$, while counter-aligned configurations, which
inspiral much more quickly, should have $\xi \lesssim 90^\circ$.

Based on the above analysis, we expect $\cos \xi$ to be a discontinuous
function with a finite jump 
 when $\vec S\cdot\vec L$ and $\delta m
\Delta\cdot \vec L$ change sign. While we can model $\xi$ as a discontinuous
function, there is a way around this. Note that the magnitude
of the in-plane recoil is given by
\begin{equation}
  V^2 = (v_m + |v_\perp|\cos\xi )^2 + v_\perp^2\sin^2\xi,
\end{equation}
where $|v_\perp|$ is the magnitude of spin contribution to the in-plane
recoil. The important thing to note is that while
$\cos \xi$ is discontinuous, the recoil itself should be continuous.
For this to be true, the sign change in $\cos \xi$ can only 
occur when $v_\perp=0$, i.e., we expect that $|v_\perp|\cos\xi$ is
continuous. We can therefore express the product  $|v_\perp|\cos\xi$ as
a product of two continuous functions $v_\perp$, which we will allow to
be positive or negative, and $\cos \tilde \xi$, where $\cos \tilde
\xi$ has a fixed sign (for historical reasons, we chose $\cos \tilde
\xi$ to be negative) and $|\cos\tilde \xi| = |\cos \xi|$.
Finally, the magnitude of the recoil is given by,
\begin{equation}
  V^2 = (v_m + v_\perp\cos\tilde\xi )^2 + v_\perp^2\sin^2\tilde \xi,
\label{eq:vplane}
\end{equation}
where $v_\perp$ can be negative and ${\rm sign}(v_\perp) \cos \tilde \xi$ is the
cosine of the angle between unequal-mass and spin components of the
recoil.

\section{Numerical Simulations}\label{sec:evolution}

We use the TwoPunctures thorn~\cite{Ansorg:2004ds} to generate initial
puncture data~\cite{Brandt97b} for the BHB simulations. These data are
characterized by mass parameters $m_p$ (which
are not the horizon masses), as well as the momentum and spin,  of
each BH.  We evolve these BHB data sets using the {\sc
LazEv}~\cite{Zlochower:2005bj} implementation of the moving puncture
approach~\cite{Campanelli:2005dd, Baker:2005vv} with the conformal
function $W=\sqrt{\chi}=\exp(-2\phi)$ suggested by
Ref.~\cite{Marronetti:2007wz}.  For the runs presented here, we use
centered, eighth-order finite differencing in
space~\cite{Lousto:2007rj} and a fourth-order Runge Kutta time
integrator. (Note that we do not upwind the advection terms.) Our code
uses the {\sc Cactus}/{\sc EinsteinToolkit}~\cite{cactus_web,
einsteintoolkit} infrastructure.  We use the {\sc
Carpet}~\cite{Schnetter-etal-03b} mesh refinement driver to provide a
``moving boxes'' style of mesh refinement.

We locate the apparent horizons using the {\sc AHFinderDirect}
code~\cite{Thornburg2003:AH-finding} and measure the horizon spin
using the isolated horizon (IH) algorithm detailed
in~\cite{Dreyer02a}.

For the computation of the radiated energy and linear momentum
we use the formulas in \cite{Campanelli:1998jv} which are expressed
directly in terms of the Weyl scalar $\psi_4$. To extract the radiation of
angular momentum components, we
use formulas based on ``flux-linkages''~\cite{Winicour_AMGR} and
explicitly written in terms of $\psi_4$  in
\cite{Campanelli:1998jv, Lousto:2007mh}.

To generate the initial data parameters, we use 3PN quasicircular
orbital parameters with a given initial orbital
frequency $\omega_i$.
In practice this leads
to an initial eccentricity of the order of $e_{i}\sim10^{-2}$
that radiates after a few orbits to about
$e_f\sim 10^{-3}$,  which is small enough  for modeling the remnant in
astrophysical applications. Table~\ref{tab:ecc} provides 
explicit values for all the initial data parameters used in each of
the  runs presented in this paper. We also provide the initial and
final eccentricity and total number of orbits in the table.

We evolve these data sets using the grid refinement structure and
global resolution
discussed in the Appendix. In the Appendix, we also describe in detail
the errors in our results due to finite extraction radii and finite truncation
errors, as well as how we extrapolate from finite radii to null
infinity.  

\begin{figure}
  \includegraphics[width=\columnwidth]{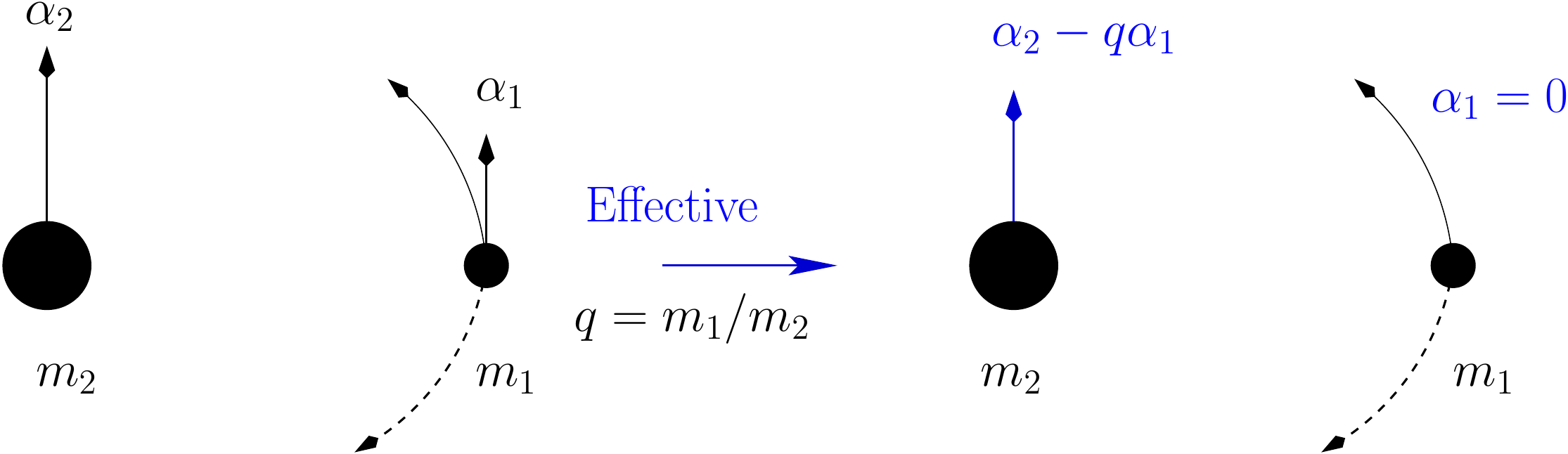}
  \caption{The (counter)aligned spin configuration UU (DD) and
   its effective counterpart 0U (0D) (dashed counterorbiting).}
  \label{fig:effective}
\end{figure}

\begin{figure}
  \includegraphics[width=2.0in]{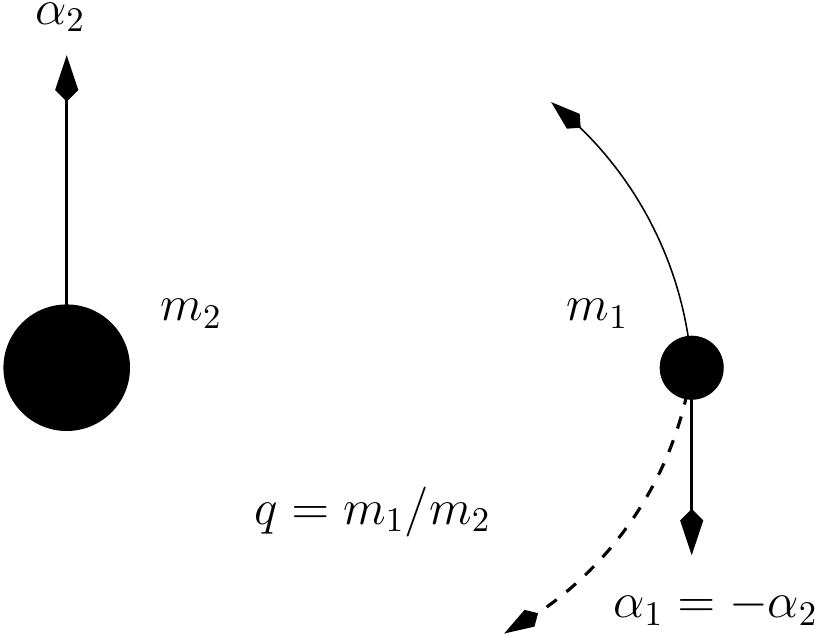}
  \caption{The DU and UD (dashed counterorbiting) configurations.}
  \label{fig:DUUD}
\end{figure}

\begin{figure}
  \includegraphics[angle=270,width=0.49\columnwidth]{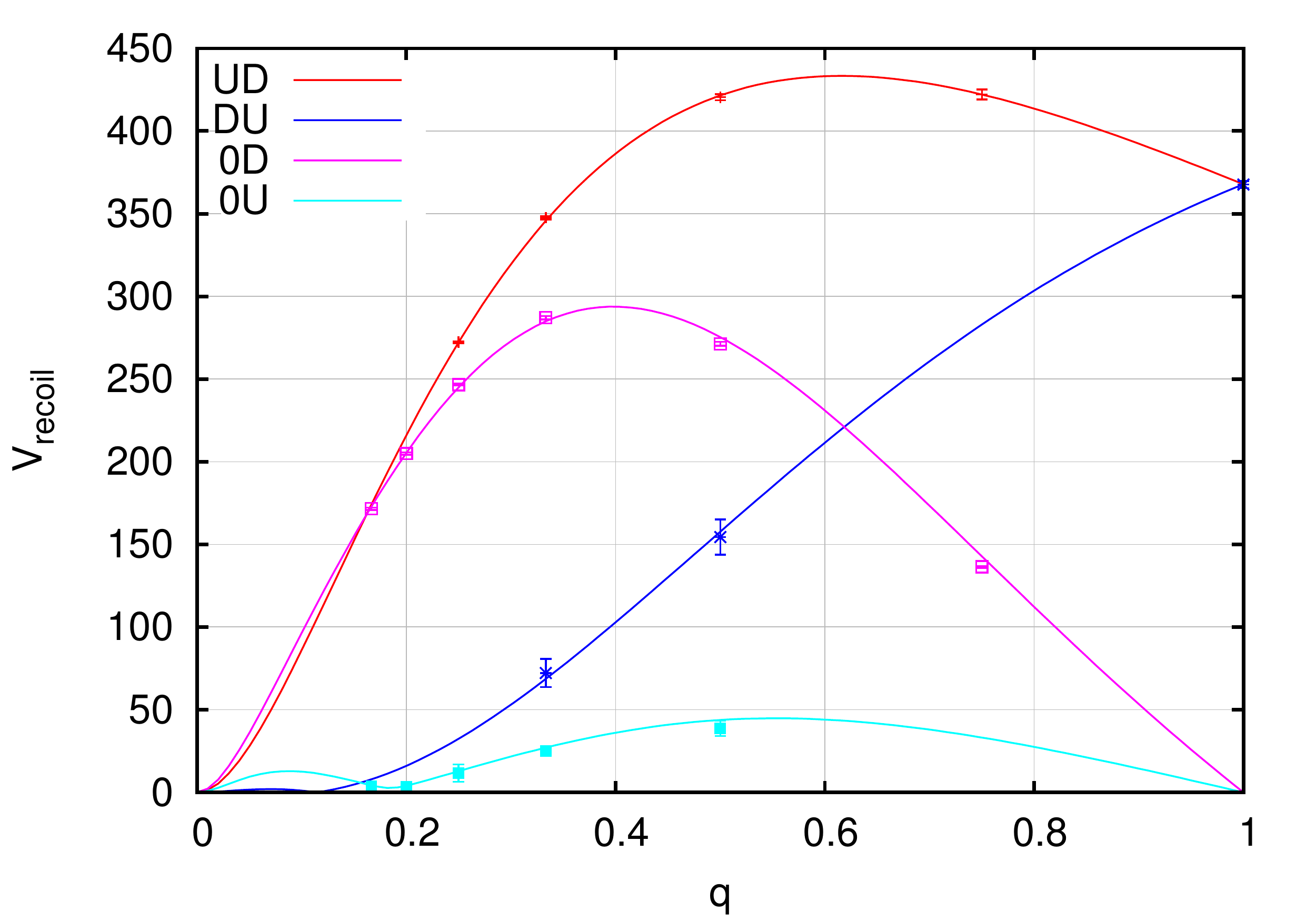}
  \includegraphics[angle=270,width=0.49\columnwidth]{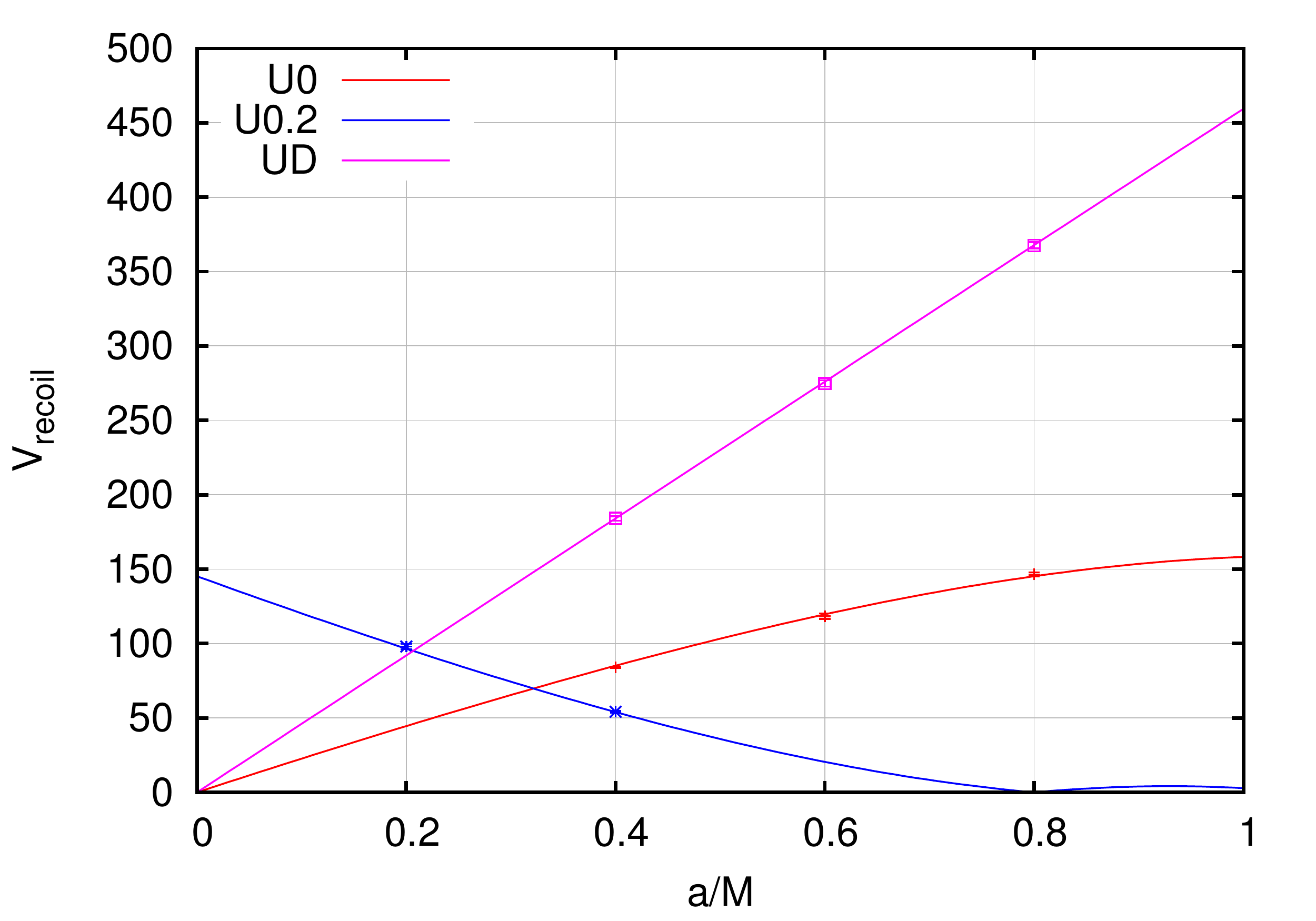}
  \caption{The recoils for the families UD/DU/0U/0D and for equal masses
cases as given in Table~\ref{tab:ID}}
  \label{fig:fig1ab}
\end{figure}

In order to cover the three-dimensional parameter space of the
aligned-spin BHBs, we consider several
families of physically
motivated configurations.
We denote our configurations by
XY, where X=U,D, or 0 denotes the spin of the smaller BH (i.e.,
aligned, counteraligned, or zero) and Y denotes the spin of the larger
BH.
If accretion tends to align the spins, then the UD, DU, UU, and DD
configurations should be among the most probable.
The 0U and 0D configurations, depicted in
 Fig.~\ref{fig:effective}, are interesting
in that if
the recoil as given in Eq.~(\ref{eq:vperp}) is dominated by the leading
$\Delta_\|$ dependence, then a 0U or 0D configuration is an 
effective counterpart to a UU or DD configuration where both BHs are spinning
with the same dimensionless spin $\alpha$ and
$\alpha_\text{Effective}=\pm(1-q)\alpha'$
 (i.e., a 0U with spin
$\alpha_\text{Effective}$ should give the same recoil as a UU/DD
configuration with $\alpha_2 = \alpha_1 = \alpha'$).
Since $\alpha_\text{Effective}$ is smaller than $\alpha'$ we can
apparently simulate maximal UU and DD configurations with non-maximal
0U and 0D configurations.
We thus study BHBs with the smaller BH nonspinning and 
the larger BH spinning with spin
$\alpha_\text{Effective}=\pm(1-q)$ for
$q=1/2,1/3,1/4,1/5,1/6$. A first analysis of those simulations 
suggested that although leading, the $\Delta_\|$ dependence
in Eq.~(\ref{eq:vperp}) is not sufficient to model the recoils with
high accuracy.  
We thus consider additional  families (see Fig.~\ref{fig:DUUD})
of BHBs, with specific spins $\alpha\leq 0.8$
in a UD or DU configuration.
In addition to showing the importance of the total spin $S_\|$ to the
recoil, we also found from these  configurations  that the maximum recoil
occurs for $q_{max}\sim0.62$, as shown in Fig.~\ref{fig:fig1ab}.

We chose other configurations to selectively activate or deactivate
blocks of terms in the expansion formulas for the recoil and radiated
energy-momentum. 
Thus some simulations have
only one of the variables $\delta{m}$, $S_\|$, and $\Delta_\|$
nonvanishing and others have all of them nonvanishing. This provides a
means of fitting all terms and then verifying the fit for more general
cases.
The complete set of initial data 
parameters are given in Table~\ref{tab:ID}. In the table, the runs
are labeled by the  mass ratio, spin magnitude of the black hole 1
(the smaller BH)
and spin magnitude of the spin 2 (the larger BH).

\begin{widetext}

\begin{table}
\caption{Initial data parameters for the quasi-circular
configurations with a non-spinning smaller mass black hole (labeled 1),
and a larger mass spinning black hole (labeled 2). The punctures are located
at $\vec r_1 = (x_1,0,0)$ and $\vec r_2 = (x_2,0,0)$, with momenta
$P=\pm (0, P,0)$, spins $\vec S_i = (0, 0, S_i)$, mass parameters
$m^p/m$, horizon (Christodoulou) masses $m^H/m$, total ADM mass
$M_{\rm ADM}$, and dimensionless spins $a/m_H = S/m_H^2$. 
The configuration are denoted by QX\_Y\_Z, where X gives the mass
ratio $m^H_1 / m^H_2$, Y gives the  spin of the smaller BH
($a_1/m_H^2$), and Z
gives the spin of the larger BH $(a_2/m_H^2)$. (*) Note that the $q=1/10$ binary
also had an initial radial momentum of $P_r/m = -0.0001685$.
}\label{tab:ID}
\begin{ruledtabular}
\begin{tabular}{lcccccccccccc}
Config.   & $x_1/m$ & $x_2/m$  & $P/m$    & $m^p_1/m$ & $m^p_2/m$ & $S_1/m^2$ & $S_2/m^2$ & $m^H_1/m$ & $m^H_2/m$ & $M_{\rm ADM}/m$ & $a_1/m_1^H$ & $a_2/m_2^H$\\
\hline
Q1.000\_0.00\_0.00   & -4.7666 & 4.7666 & 0.099322 & 0.48523 & 0.48523 & 0 & 0 & 0.5 & 0.5 & 0.98931 & 0 & 0 \\
Q1.000\_0.00\_0.40   & -4.6378 & 4.523 & 0.1004 & 0.48472 & 0.45144 & 0 & 0.1 & 0.5 & 0.5 & 0.9888 & 0 & 0.4 \\
Q1.000\_0.00\_0.60   & -4.5759 & 4.4035 & 0.10101 & 0.48445 & 0.40145 & 0 & 0.15 & 0.5 & 0.5 & 0.9886 & 0 & 0.6 \\
Q1.000\_0.00\_0.80   & -4.5152 & 4.2852 & 0.10159 & 0.48418 & 0.30103 & 0 & 0.2 & 0.5 & 0.5 & 0.98842 & 0 & 0.8 \\
Q1.000\_0.20\_0.80   & -4.4307 & 4.3878 & 0.10071 & 0.47635 & 0.30108 & 0.05 & 0.2 & 0.5 & 0.5 & 0.98838 & 0.2 & 0.8 \\
Q1.000\_0.40\_-0.40  & -5.0346 & 4.9785 & 0.095751 & 0.45266 & 0.45259 & -0.1 & 0.1 & 0.5 & 0.5 & 0.98971 & -0.4 & 0.4 \\
Q1.000\_0.40\_0.80   & -4.405 & 4.3766 & 0.10025 & 0.45098 & 0.30107 & 0.1 & 0.2 & 0.5 & 0.5 & 0.98829 & 0.4 & 0.8 \\
Q1.000\_-0.60\_0.60  & -4.8029 & 4.7172 & 0.09907 & 0.40219 & 0.4021 & -0.15 & 0.15 & 0.5 & 0.5 & 0.98937 & -0.6 & 0.6 \\
Q1.000\_-0.80\_0.80  & -4.9832 & 4.5267 & 0.09905 & 0.30178 & 0.30168 & -0.2 & 0.2 & 0.5 & 0.5 & 0.98951 & -0.8 & 0.8 \\
Q0.750\_0.00\_-0.25  & -6.0062 & 4.5158 & 0.091564 & 0.41524 & 0.54417 & 0 & -0.081633 & 0.42857 & 0.57143 & 0.99034 & 0 & -0.25 \\
Q0.750\_-0.80\_0.45  & -5.7814 & 4.2576 & 0.094019 & 0.25784 & 0.50845 & -0.14694 & 0.14694 & 0.42857 & 0.57143 & 0.99011 & -0.8 & 0.45 \\
Q0.750\_0.80\_-0.45  & -5.6572 & 4.308 & 0.093655 & 0.25774 & 0.50848 & 0.14694 & -0.14694 & 0.42857 & 0.57143 & 0.98998 & 0.8 & -0.45 \\
Q0.750\_-0.80\_-0.60 & -6.2721 & 4.6997 & 0.091817 & 0.25847 & 0.46326 & -0.14694 & -0.19592 & 0.42857 & 0.57143 & 0.99111 & -0.8 & -0.6 \\
Q0.750\_0.80\_0.60   & -5.05 & 3.7787 & 0.097289 & 0.25676 & 0.46102 & 0.14694 & 0.19592 & 0.42857 & 0.57143 & 0.98858 & 0.8 & 0.6 \\
Q0.750\_0.80\_-0.80  & -6.1633 & 4.7098 & 0.089327 & 0.25845 & 0.34767 & 0.14694 & -0.26122 & 0.42857 & 0.57143 & 0.99083 & 0.8 & -0.8 \\
Q0.500\_0.00\_-0.50  & -6.9641 & 3.5416 & 0.084316 & 0.32093 & 0.58184 & 0 & -0.22222 & 0.33333 & 0.66667 & 0.99136 & 0 & -0.5 \\
Q0.500\_0.00\_0.50   & -6.2598 & 3.1299 & 0.087209 & 0.31969 & 0.58068 & 0 & 0.22222 & 0.33333 & 0.66667 & 0.99027 & 0 & 0.5 \\
Q0.500\_-0.80\_0.20  & -6.6141 & 3.2581 & 0.086691 & 0.19907 & 0.64372 & -0.088889 & 0.088889 & 0.33333 & 0.66667 & 0.99096 & -0.8 & 0.2 \\
Q0.500\_0.80\_-0.20  & -6.487 & 3.2681 & 0.086176 & 0.19898 & 0.64368 & 0.088889 & -0.088889 & 0.33333 & 0.66667 & 0.99076 & 0.8 & -0.2 \\
Q0.500\_-0.80\_-0.40 & -6.682 & 3.3317 & 0.088305 & 0.19909 & 0.60963 & -0.088889 & -0.17778 & 0.33333 & 0.66667 & 0.9913 & -0.8 & -0.4 \\
Q0.500\_0.80\_0.40   & -5.9213 & 2.9452 & 0.089199 & 0.19825 & 0.60838 & 0.088889 & 0.17778 & 0.33333 & 0.66667 & 0.9898 & 0.8 & 0.4 \\
Q0.500\_-0.80\_-0.80 & -7.3137 & 3.674 & 0.083939 & 0.19978 & 0.4078 & -0.088889 & -0.35556 & 0.33333 & 0.66667 & 0.99207 & -0.8 & -0.8 \\
Q0.500\_-0.80\_0.80  & -6.0831 & 2.6801 & 0.091117 & 0.19815 & 0.4061 & -0.088889 & 0.35556 & 0.33333 & 0.66667 & 0.98988 & -0.8 & 0.8 \\
Q0.500\_0.80\_-0.80  & -7.0541 & 3.8437 & 0.081882 & 0.19978 & 0.40782 & 0.088889 & -0.35556 & 0.33333 & 0.66667 & 0.99179 & 0.8 & -0.8 \\
Q0.500\_0.80\_0.80   & -5.7338 & 2.8246 & 0.089267 & 0.19801 & 0.40601 & 0.088889 & 0.35556 & 0.33333 & 0.66667 & 0.98938 & 0.8 & 0.8 \\
Q0.333\_0.00\_-0.67  & -7.8557 & 2.6696 & 0.07199 & 0.23933 & 0.57549 & 0 & -0.375 & 0.25 & 0.75 & 0.99283 & 0 & -0.66667 \\
Q0.333\_0.00\_0.67   & -6.8651 & 2.2087 & 0.074136 & 0.23799 & 0.57436 & 0 & 0.375 & 0.25 & 0.75 & 0.99145 & 0 & 0.66667 \\
Q0.333\_-0.80\_0.80  & -6.545 & 1.8547 & 0.078325 & 0.14731 & 0.45916 & -0.05 & 0.45 & 0.25 & 0.75 & 0.99112 & -0.8 & 0.8 \\
Q0.333\_0.80\_-0.80  & -7.7455 & 2.8644 & 0.071192 & 0.1488 & 0.46071 & 0.05 & -0.45 & 0.25 & 0.75 & 0.99301 & 0.8 & -0.8 \\
Q0.250\_0.00\_-0.75  & -8.7925 & 2.2393 & 0.059859 & 0.19121 & 0.54795 & 0 & -0.48 & 0.2 & 0.8 & 0.99415 & 0 & -0.75 \\
Q0.250\_0.00\_0.75   & -7.0934 & 1.7028 & 0.064078 & 0.18938 & 0.54664 & 0 & 0.48 & 0.2 & 0.8 & 0.9925 & 0 & 0.75 \\
Q0.250\_0.80\_-0.80  & -8.4636 & 2.1489 & 0.061171 & 0.11861 & 0.49276 & 0.032 & -0.512 & 0.2 & 0.8 & 0.99409 & 0.8 & -0.8 \\
Q0.200\_0.00\_-0.80  & -9.4341 & 1.9209 & 0.051156 & 0.15919 & 0.51448 & 0 & -0.55556 & 0.16667 & 0.83333 & 0.99506 & 0 & -0.8 \\
Q0.200\_0.00\_0.80   & -7.2578 & 1.3894 & 0.055956 & 0.15726 & 0.51325 & 0 & 0.55556 & 0.16667 & 0.83333 & 0.99338 & 0 & 0.8 \\
Q0.167\_0.00\_-0.83  & -9.0003 & 1.5299 & 0.04788 & 0.1357 & 0.47918 & 0 & -0.61224 & 0.14286 & 0.85714 & 0.99542 & 0 & -0.83333 \\
Q0.167\_0.00\_0.83   & -7.2953 & 1.2159 & 0.049656 & 0.13442 & 0.4784 & 0 & 0.61224 & 0.14286 & 0.85714 & 0.99408 & 0 & 0.83333 \\
Q0.100\_0.00\_0.00 ${}^*$   &  7.6331 & -0.7532 & 0.036699 & 0.08524 & 0.90740 & 0 & 0 & 0.09129 & 0.91255 & 1.0000 & 0 & 0 \\
\end{tabular}
\end{ruledtabular}
\end{table}

\begin{table}
\caption{The mass and spin of the BHBs in Table~\ref{tab:ID} after the 
BHs had time to equilibrate ($t/M=150$).}\label{tab:IDr}
\begin{ruledtabular}
\begin{tabular}{lccccccc}
Config.  & $m^r_1/m$ & $m^r_2/m$ & $\alpha^r_1$ & $\alpha^r_2$ & $\delta m_r$ & $S_r/m^2_r$ & $\Delta_r/m^2_r$ \\
\hline
Q1.000\_0.00\_0.00 &	$0.500001$ & $0.500001$ & $-0.000002$ & $-0.000002$ & $0.000000$ & $-0.000001$ & $0.000000$\\
Q1.000\_0.00\_0.40 &	$0.499998$ & $0.500005$ & $-0.000002$ & $0.399919$ & $-0.000007$ & $0.099981$ & $0.199962$\\
Q1.000\_0.00\_0.60 &	$0.499998$ & $0.499975$ & $-0.000002$ & $0.600030$ & $0.000023$ & $0.150000$ & $0.300009$\\
Q1.000\_0.00\_0.80 &	$0.499999$ & $0.499804$ & $-0.000001$ & $0.800605$ & $0.000195$ & $0.200073$ & $0.400225$\\
Q1.000\_0.20\_0.80 &	$0.500006$ & $0.499804$ & $0.199991$ & $0.800585$ & $0.000202$ & $0.250083$ & $0.300196$\\
Q1.000\_0.40\_-0.40 &	$0.500006$ & $0.500007$ & $-0.400013$ & $0.399986$ & $-0.000001$ & $-0.000006$ & $0.399999$\\
Q1.000\_0.40\_0.80 &	$0.500009$ & $0.499803$ & $0.399974$ & $0.800587$ & $0.000206$ & $0.300099$ & $0.200183$\\
Q1.000\_-0.60\_0.60 &	$0.499976$ & $0.499980$ & $-0.600096$ & $0.600029$ & $-0.000004$ & $-0.000014$ & $0.600063$\\
Q1.000\_-0.80\_0.80 &	$0.499801$ & $0.499808$ & $-0.800702$ & $0.800576$ & $-0.000007$ & $-0.000026$ & $0.800639$\\
Q0.750\_0.00\_-0.25 &	$0.428573$ & $0.571432$ & $-0.000002$ & $-0.250015$ & $-0.142858$ & $-0.081638$ & $-0.142865$\\
Q0.750\_-0.80\_0.45 &	$0.428415$ & $0.571439$ & $-0.800654$ & $0.449990$ & $-0.143045$ & $-0.000011$ & $0.600241$\\
Q0.750\_0.80\_-0.45 &	$0.428417$ & $0.571428$ & $0.800563$ & $-0.450042$ & $-0.143033$ & $-0.000016$ & $-0.600235$\\
Q0.750\_-0.80\_-0.60 &	$0.428413$ & $0.571397$ & $-0.800656$ & $-0.600070$ & $-0.143011$ & $-0.343001$ & $0.000133$\\
Q0.750\_0.80\_0.60 &	$0.428422$ & $0.571391$ & $0.800523$ & $0.600059$ & $-0.142995$ & $0.342972$ & $-0.000093$\\
Q0.750\_0.80\_-0.80 &	$0.428412$ & $0.571165$ & $0.800579$ & $-0.800777$ & $-0.142813$ & $-0.114398$ & $-0.800692$\\
Q0.500\_0.00\_-0.50 &	$0.333333$ & $0.666646$ & $-0.000003$ & $-0.500057$ & $-0.333320$ & $-0.222243$ & $-0.333367$\\
Q0.500\_0.00\_0.50 &	$0.333333$ & $0.666647$ & $-0.000003$ & $0.500073$ & $-0.333321$ & $0.222250$ & $0.333380$\\
Q0.500\_-0.80\_0.20 &	$0.333239$ & $0.666684$ & $-0.800560$ & $0.199993$ & $-0.333471$ & $-0.000010$ & $0.400141$\\
Q0.500\_0.80\_-0.20 &	$0.333239$ & $0.666670$ & $0.800469$ & $-0.200003$ & $-0.333462$ & $-0.000001$ & $-0.400120$\\
Q0.500\_-0.80\_-0.40 &	$0.333243$ & $0.666681$ & $-0.800542$ & $-0.399969$ & $-0.333463$ & $-0.266713$ & $0.000124$\\
Q0.500\_0.80\_0.40 &	$0.333244$ & $0.666666$ & $0.800412$ & $0.400002$ & $-0.333452$ & $0.266713$ & $-0.000065$\\
Q0.500\_-0.80\_-0.80 &	$0.333233$ & $0.666348$ & $-0.800574$ & $-0.800723$ & $-0.333254$ & $-0.444809$ & $-0.266894$\\
Q0.500\_-0.80\_0.80 &	$0.333246$ & $0.666358$ & $-0.800499$ & $0.800638$ & $-0.333244$ & $0.266823$ & $0.800592$\\
Q0.500\_0.80\_-0.80 &	$0.333232$ & $0.666330$ & $0.800530$ & $-0.800798$ & $-0.333244$ & $-0.266891$ & $-0.800708$\\
Q0.500\_0.80\_0.80 &	$0.333245$ & $0.666344$ & $0.800398$ & $0.800701$ & $-0.333237$ & $0.444774$ & $0.266924$\\
Q0.333\_0.00\_-0.67 &	$0.249994$ & $0.749857$ & $-0.000002$ & $-0.666947$ & $-0.499938$ & $-0.375127$ & $-0.500189$\\
Q0.333\_0.00\_0.67 &	$0.249997$ & $0.749855$ & $-0.000004$ & $0.666920$ & $-0.499931$ & $0.375108$ & $0.500168$\\
Q0.333\_-0.80\_0.80 &	$0.249953$ & $0.749624$ & $-0.800179$ & $0.800758$ & $-0.499883$ & $0.400322$ & $0.800613$\\
Q0.333\_0.80\_-0.80 &	$0.249944$ & $0.749607$ & $0.800395$ & $-0.800865$ & $-0.499888$ & $-0.400372$ & $-0.800748$\\
Q0.250\_0.00\_-0.75 &	$0.200004$ & $0.799697$ & $-0.000002$ & $-0.750583$ & $-0.599873$ & $-0.480297$ & $-0.600419$\\
Q0.250\_0.00\_0.75 &	$0.199998$ & $0.799695$ & $-0.000003$ & $0.750566$ & $-0.599881$ & $0.480291$ & $0.600408$\\
Q0.250\_0.80\_-0.80 &	$0.199967$ & $0.799568$ & $0.800375$ & $-0.800890$ & $-0.599880$ & $-0.480459$ & $-0.800787$\\
Q0.200\_0.00\_-0.80 &	$0.166673$ & $0.832872$ & $0.000001$ & $-0.800897$ & $-0.666502$ & $-0.556069$ & $-0.667348$\\
Q0.200\_0.00\_0.80 &	$0.166665$ & $0.832869$ & $0.000004$ & $0.800884$ & $-0.666515$ & $0.556068$ & $0.667342$\\
Q0.167\_0.00\_-0.83 &	$0.142855$ & $0.856538$ & $-0.000002$ & $-0.834514$ & $-0.714117$ & $-0.612992$ & $-0.715227$\\
Q0.167\_0.00\_0.83 &	$0.142855$ & $0.856533$ & $-0.000001$ & $0.834513$ & $-0.714115$ & $0.612989$ & $0.715226$\\
\end{tabular}
\end{ruledtabular}
\end{table}

\end{widetext}

Since the initial data assumes conformal flatness and pure longitudinal
extrinsic curvature, it contains initial distortions that are either
 radiated 
away or absorbed by the BHs
 during the first orbital period. After this initial transient period,
the BH masses and spins settle to their equilibrium values.
In Table~\ref{tab:IDr}
we give the values for the
horizon mass and spin after this transient period has ended.

\begin{table}
\caption{Table of the initial orbital frequency $m\omega_i$,
number of orbits to merger, $N$, and the initial and final eccentricities, 
$e_i$ and $e_f$.}\label{tab:ecc}
\begin{ruledtabular}
\begin{tabular}{lcccc}
Config.   & $m\omega_i$ & $N$ & $e_i$ & $e_f$ \\
\hline
Q1.000\_0.00\_0.00 &	$0.0300$ & $5.3$ & $0.028$ & $0.005$\\
Q1.000\_0.00\_0.40 &	$0.0313$ & $5.4$ & $0.021$ & $0.006$\\
Q1.000\_0.00\_0.60 &	$0.0320$ & $5.6$ & $0.021$ & $0.005$\\
Q1.000\_0.00\_0.80 &	$0.0327$ & $5.7$ & $0.020$ & $0.004$\\
Q1.000\_0.20\_0.80 &	$0.0324$ & $6.2$ & $0.020$ & $0.002$\\
Q1.000\_0.40\_-0.40 &	$0.0280$ & $5.9$ & $0.023$ & $0.002$\\
Q1.000\_0.40\_0.80 &	$0.0324$ & $6.6$ & $0.020$ & $0.002$\\
Q1.000\_-0.60\_0.60 &	$0.0300$ & $5.1$ & $0.022$ & $0.005$\\
Q1.000\_-0.80\_0.80 &	$0.0300$ & $5.1$ & $0.022$ & $0.005$\\
Q0.750\_0.00\_-0.25 &	$0.0263$ & $6.2$ & $0.024$ & $0.005$\\
Q0.750\_-0.80\_0.45 &	$0.0280$ & $5.7$ & $0.022$ & $0.003$\\
Q0.750\_0.80\_-0.45 &	$0.0280$ & $6.1$ & $0.023$ & $0.006$\\
Q0.750\_-0.80\_-0.60 &	$0.0254$ & $4.6$ & $0.024$ & $0.005$\\
Q0.750\_0.80\_0.60 &	$0.0320$ & $7.1$ & $0.020$ & $0.004$\\
Q0.750\_0.80\_-0.80 &	$0.0250$ & $6.6$ & $0.025$ & $0.005$\\
Q0.500\_0.00\_-0.50 &	$0.0265$ & $5.4$ & $0.020$ & $0.005$\\
Q0.500\_0.00\_0.50 &	$0.0300$ & $7.2$ & $0.024$ & $0.003$\\
Q0.500\_-0.80\_0.20 &	$0.0287$ & $5.6$ & $0.019$ & $0.003$\\
Q0.500\_0.80\_-0.20 &	$0.0287$ & $6.2$ & $0.019$ & $0.005$\\
Q0.500\_-0.80\_-0.40 &	$0.0287$ & $4.0$ & $0.018$ & $0.004$\\
Q0.500\_0.80\_0.40 &	$0.0320$ & $6.8$ & $0.018$ & $0.004$\\
Q0.500\_-0.80\_-0.80 &	$0.0254$ & $4.1$ & $0.019$ & $0.008$\\
Q0.500\_-0.80\_0.80 &	$0.0330$ & $6.1$ & $0.017$ & $0.002$\\
Q0.500\_0.80\_-0.80 &	$0.0250$ & $6.1$ & $0.021$ & $0.005$\\
Q0.500\_0.80\_0.80 &	$0.0330$ & $7.8$ & $0.017$ & $0.002$\\
Q0.333\_0.00\_-0.67 &	$0.0265$ & $5.0$ & $0.013$ & $0.008$\\
Q0.333\_0.00\_0.67 &	$0.0310$ & $8.6$ & $0.014$ & $0.003$\\
Q0.333\_-0.80\_0.80 &	$0.0345$ & $7.1$ & $0.012$ & $0.003$\\
Q0.333\_0.80\_-0.80 &	$0.0260$ & $5.3$ & $0.014$ & $0.006$\\
Q0.250\_0.00\_-0.75 &	$0.0248$ & $5.7$ & $0.014$ & $0.005$\\
Q0.250\_0.00\_0.75 &	$0.0320$ & $10.1$ & $0.011$ & $0.003$\\
Q0.250\_0.80\_-0.80 &	$0.0260$ & $5.2$ & $0.009$ & $0.008$\\
Q0.200\_0.00\_-0.80 &	$0.0238$ & $6.4$ & $0.015$ & $0.006$\\
Q0.200\_0.00\_0.80 &	$0.0325$ & $11.3$ & $0.011$ & $0.002$\\
Q0.167\_0.00\_-0.83 &	$0.0265$ & $4.4$ & $0.014$ & $0.008$\\
Q0.167\_0.00\_0.83 &	$0.0330$ & $12.8$ & $0.010$ & $0.002$\\
\end{tabular}
\end{ruledtabular}
\end{table}

After the BHs in the progenitor BHB merge,
we measure the remnants mass, spin, and recoil velocity.
We measure the recoil velocity
from the radiation of linear momentum at infinity, as this is the most
reliable and gauge invariant way of computing recoils. The resulting
recoil velocities
are given in Table~\ref{tab:kicks}.
In order
to produce accurate results, we extracted the waveform at different finite
radii and extrapolated to infinity. Here we chose observer locations
equidistant in $1/r$, where the largest extraction radius was
$102.6m$.
We fit the finite-radius results for the recoil, energy
radiated, and angular momentum radiated as a linear and quadratic
function in $1/r$, and we use the difference between these two fits
as an estimate for the error.
As we discuss in the Appendix, other sources
of error come from the finite numerical resolution and the maximum $\ell$-mode
used in the extraction. Based on our assessment of those errors (see
Appendix), we compute the recoil using the $\ell=2$ through
$\ell=6$ modes.

\begin{widetext}

\begin{table}
\caption{The recoil velocity as calculated using $\ell_{max}=6$ and $r_{max}=102.6m$.}\label{tab:kicks}
\begin{ruledtabular}
\begin{tabular}{llrrr}
Run & Config. & $V_x$ & $V_y$ & $V$\\
\hline
1 & Q1.000\_0.00\_0.00 &	$0.0 \pm 0.0$ & 	$0.0 \pm 0.0$ & 	$0.0 \pm 0.0$ \\
2 & Q1.000\_0.00\_0.40 &	$-24.36 \pm 1.16$ & 	$-80.37 \pm 0.23$ & 	$83.97 \pm 0.40$ \\
3 & Q1.000\_0.00\_0.60 &	$54.72 \pm 1.65$ & 	$-104.99 \pm 1.92$ & 	$118.40 \pm 1.87$ \\
4 & Q1.000\_0.00\_0.80 &	$141.16 \pm 0.52$ & 	$-39.33 \pm 4.57$ & 	$146.53 \pm 1.32$ \\
5 & Q1.000\_0.20\_0.80 &	$-76.92 \pm 2.28$ & 	$60.74 \pm 0.52$ & 	$98.01 \pm 1.82$ \\
6 & Q1.000\_0.40\_-0.40 &	$82.56 \pm 2.39$ & 	$164.52 \pm 1.12$ & 	$184.07 \pm 1.47$ \\
7 & Q1.000\_0.40\_0.80 &	$4.99 \pm 0.84$ & 	$-53.98 \pm 0.69$ & 	$54.22 \pm 0.69$ \\
8 & Q1.000\_-0.60\_0.60 &	$-209.74 \pm 2.12$ & 	$177.56 \pm 2.07$ & 	$274.80 \pm 2.10$ \\
9 & Q1.000\_-0.80\_0.80 &	$-167.03 \pm 3.96$ & 	$327.50 \pm 1.27$ & 	$367.63 \pm 2.13$ \\
10 & Q0.750\_0.00\_-0.25 &	$105.53 \pm 0.49$ & 	$-86.34 \pm 0.67$ & 	$136.35 \pm 0.57$ \\
11 & Q0.750\_-0.80\_0.45 &	$208.49 \pm 0.25$ & 	$60.82 \pm 6.60$ & 	$217.18 \pm 1.86$ \\
12 & Q0.750\_0.80\_-0.45 &	$166.34 \pm 1.33$ & 	$-271.75 \pm 0.25$ & 	$318.61 \pm 0.73$ \\
13 & Q0.750\_-0.80\_-0.60 &	$15.40 \pm 1.42$ & 	$94.31 \pm 0.79$ & 	$95.57 \pm 0.81$ \\
14 & Q0.750\_0.80\_0.60 &	$3.89 \pm 1.57$ & 	$-27.08 \pm 0.27$ & 	$27.36 \pm 0.35$ \\
15 & Q0.750\_0.80\_-0.80 &	$-333.94 \pm 1.70$ & 	$258.31 \pm 4.36$ & 	$422.19 \pm 2.99$ \\
16 & Q0.500\_0.00\_-0.50 &	$155.44 \pm 1.36$ & 	$222.33 \pm 1.06$ & 	$271.28 \pm 1.17$ \\
17 & Q0.500\_0.00\_0.50 &	$-18.50 \pm 1.66$ & 	$-34.08 \pm 5.12$ & 	$38.77 \pm 4.56$ \\
18 & Q0.500\_-0.80\_0.20 &	$62.47 \pm 2.45$ & 	$66.40 \pm 6.26$ & 	$91.16 \pm 4.86$ \\
19 & Q0.500\_0.80\_-0.20 &	$254.30 \pm 0.10$ & 	$-68.79 \pm 2.32$ & 	$263.43 \pm 0.61$ \\
20 & Q0.500\_-0.80\_-0.40 &	$-120.44 \pm 2.83$ & 	$-128.91 \pm 0.01$ & 	$176.42 \pm 1.93$ \\
21 & Q0.500\_0.80\_0.40 &	$-80.19 \pm 0.19$ & 	$-8.70 \pm 1.76$ & 	$80.67 \pm 0.26$ \\
22 & Q0.500\_-0.80\_-0.80 &	$126.69 \pm 1.08$ & 	$-235.55 \pm 1.65$ & 	$267.46 \pm 1.54$ \\
23 & Q0.500\_-0.80\_0.80 &	$59.84 \pm 6.28$ & 	$142.31 \pm 11.19$ & 	$154.38 \pm 10.60$ \\
24 & Q0.500\_0.80\_-0.80 &	$231.96 \pm 0.78$ & 	$-350.71 \pm 2.17$ & 	$420.48 \pm 1.86$ \\
25 & Q0.500\_0.80\_0.80 &	$2.12 \pm 4.77$ & 	$-0.39 \pm 2.11$ & 	$2.15 \pm 4.71$ \\
26 & Q0.333\_0.00\_-0.67 &	$-127.75 \pm 2.03$ & 	$-257.15 \pm 0.57$ & 	$287.15 \pm 1.04$ \\
27 & Q0.333\_0.00\_0.67 &	$23.02 \pm 2.89$ & 	$-9.61 \pm 1.69$ & 	$24.94 \pm 2.74$ \\
28 & Q0.333\_-0.80\_0.80 &	$20.09 \pm 4.71$ & 	$69.37 \pm 8.74$ & 	$72.22 \pm 8.50$ \\
29 & Q0.333\_0.80\_-0.80 &	$346.95 \pm 1.00$ & 	$21.11 \pm 1.57$ & 	$347.60 \pm 1.01$ \\
30 & Q0.250\_0.00\_-0.75 &	$-200.80 \pm 0.04$ & 	$143.20 \pm 1.07$ & 	$246.63 \pm 0.62$ \\
31 & Q0.250\_0.00\_0.75 &	$3.53 \pm 2.91$ & 	$11.09 \pm 5.46$ & 	$11.64 \pm 5.27$ \\
32 & Q0.250\_0.80\_-0.80 &	$254.96 \pm 0.06$ & 	$-95.67 \pm 1.83$ & 	$272.32 \pm 0.65$ \\
33 & Q0.200\_0.00\_-0.80 &	$199.51 \pm 0.62$ & 	$46.95 \pm 1.27$ & 	$204.97 \pm 0.67$ \\
34 & Q0.200\_0.00\_0.80 &	$-1.18 \pm 5.14$ & 	$-3.20 \pm 1.49$ & 	$3.41 \pm 2.27$ \\
35 & Q0.167\_0.00\_-0.83 &	$171.27 \pm 0.72$ & 	$9.26 \pm 1.27$ & 	$171.52 \pm 0.73$ \\
36 & Q0.167\_0.00\_0.83 &	$3.33 \pm 3.27$ & 	$0.51 \pm 0.29$ & 	$3.37 \pm 3.23$ \\
\end{tabular}
\end{ruledtabular}
\end{table}

\begin{table}
\caption{The final remnant mass and spin as measured using the IH formalism
and as measured from the radiation of energy and angular momentum.}\label{tab:spinerad}
\begin{ruledtabular}
\begin{tabular}{lccccc}
Run & Config. & $\delta \mathcal{M}^{IH}$ & $\delta \mathcal{M}^{rad}$ & $\alpha_{\mathrm{rem}}^{IH}$ & $\alpha_{\mathrm{rem}}^{rad}$\\
\hline
1 & Q1.000\_0.00\_0.00 &	$0.048379 \pm 0.000001$ & 	$0.047937 \pm 0.000177$ & 	$0.686419 \pm 0.000007$ & 	$0.685034 \pm 0.004747$ \\
2 & Q1.000\_0.00\_0.40 &	$0.054557 \pm 0.000002$ & 	$0.053888 \pm 0.000190$ & 	$0.745985 \pm 0.000058$ & 	$0.745180 \pm 0.004325$ \\
3 & Q1.000\_0.00\_0.60 &	$0.058316 \pm 0.000004$ & 	$0.057438 \pm 0.000250$ & 	$0.774671 \pm 0.000195$ & 	$0.774270 \pm 0.004519$ \\
4 & Q1.000\_0.00\_0.80 &	$0.062821 \pm 0.000007$ & 	$0.061610 \pm 0.000362$ & 	$0.802619 \pm 0.000086$ & 	$0.802453 \pm 0.004977$ \\
5 & Q1.000\_0.20\_0.80 &	$0.067692 \pm 0.000001$ & 	$0.066171 \pm 0.000443$ & 	$0.830671 \pm 0.000015$ & 	$0.830647 \pm 0.005519$ \\
6 & Q1.000\_0.40\_-0.40 &	$0.048437 \pm 0.000000$ & 	$0.047998 \pm 0.000163$ & 	$0.685844 \pm 0.000001$ & 	$0.683873 \pm 0.005278$ \\
7 & Q1.000\_0.40\_0.80 &	$0.073515 \pm 0.000003$ & 	$0.071532 \pm 0.000572$ & 	$0.857465 \pm 0.000046$ & 	$0.857999 \pm 0.006244$ \\
8 & Q1.000\_-0.60\_0.60 &	$0.048780 \pm 0.000000$ & 	$0.048268 \pm 0.000202$ & 	$0.685258 \pm 0.000000$ & 	$0.683850 \pm 0.004611$ \\
9 & Q1.000\_-0.80\_0.80 &	$0.049353 \pm 0.000000$ & 	$0.048593 \pm 0.000304$ & 	$0.684235 \pm 0.000020$ & 	$0.682995 \pm 0.005103$ \\
10 & Q0.750\_0.00\_-0.25 &	$0.042681 \pm 0.000009$ & 	$0.042368 \pm 0.000179$ & 	$0.621171 \pm 0.000025$ & 	$0.618973 \pm 0.006118$ \\
11 & Q0.750\_-0.80\_0.45 &	$0.046525 \pm 0.000001$ & 	$0.045794 \pm 0.000280$ & 	$0.685173 \pm 0.000021$ & 	$0.684202 \pm 0.005942$ \\
12 & Q0.750\_0.80\_-0.45 &	$0.047283 \pm 0.000001$ & 	$0.046683 \pm 0.000270$ & 	$0.662124 \pm 0.000027$ & 	$0.660424 \pm 0.006072$ \\
13 & Q0.750\_-0.80\_-0.60 &	$0.033808 \pm 0.000000$ & 	$0.033435 \pm 0.000204$ & 	$0.451036 \pm 0.000001$ & 	$0.449575 \pm 0.005390$ \\
14 & Q0.750\_0.80\_0.60 &	$0.075774 \pm 0.000000$ & 	$0.073304 \pm 0.000761$ & 	$0.871698 \pm 0.000003$ & 	$0.872990 \pm 0.007658$ \\
15 & Q0.750\_0.80\_-0.80 &	$0.042577 \pm 0.000000$ & 	$0.042037 \pm 0.000329$ & 	$0.586122 \pm 0.000000$ & 	$0.584030 \pm 0.007777$ \\
16 & Q0.500\_0.00\_-0.50 &	$0.031757 \pm 0.000001$ & 	$0.031607 \pm 0.000107$ & 	$0.460169 \pm 0.000001$ & 	$0.458375 \pm 0.004236$ \\
17 & Q0.500\_0.00\_0.50 &	$0.050577 \pm 0.000012$ & 	$0.049510 \pm 0.000268$ & 	$0.778577 \pm 0.000018$ & 	$0.778307 \pm 0.004860$ \\
18 & Q0.500\_-0.80\_0.20 &	$0.038377 \pm 0.000001$ & 	$0.037656 \pm 0.000195$ & 	$0.638918 \pm 0.000018$ & 	$0.638885 \pm 0.004488$ \\
19 & Q0.500\_0.80\_-0.20 &	$0.039610 \pm 0.000013$ & 	$0.039003 \pm 0.000172$ & 	$0.606313 \pm 0.000022$ & 	$0.605347 \pm 0.004643$ \\
20 & Q0.500\_-0.80\_-0.40 &	$0.030257 \pm 0.000001$ & 	$0.029824 \pm 0.000142$ & 	$0.441854 \pm 0.000004$ & 	$0.441744 \pm 0.003427$ \\
21 & Q0.500\_0.80\_0.40 &	$0.054693 \pm 0.000002$ & 	$0.053274 \pm 0.000363$ & 	$0.790499 \pm 0.000012$ & 	$0.791070 \pm 0.005010$ \\
22 & Q0.500\_-0.80\_-0.80 &	$0.026965 \pm 0.000001$ & 	$0.026574 \pm 0.000175$ & 	$0.305299 \pm 0.000000$ & 	$0.304648 \pm 0.004239$ \\
23 & Q0.500\_-0.80\_0.80 &	$0.054259 \pm 0.000001$ & 	$0.052322 \pm 0.000535$ & 	$0.823813 \pm 0.000016$ & 	$0.825549 \pm 0.005449$ \\
24 & Q0.500\_0.80\_-0.80 &	$0.031687 \pm 0.000001$ & 	$0.031281 \pm 0.000219$ & 	$0.410368 \pm 0.000001$ & 	$0.408069 \pm 0.005900$ \\
25 & Q0.500\_0.80\_0.80 &	$0.075669 \pm 0.000021$ & 	$0.071617 \pm 0.001144$ & 	$0.902719 \pm 0.000354$ & 	$0.906457 \pm 0.008499$ \\
26 & Q0.333\_0.00\_-0.67 &	$0.021506 \pm 0.000006$ & 	$0.021379 \pm 0.000114$ & 	$0.240088 \pm 0.000006$ & 	$0.239368 \pm 0.003257$ \\
27 & Q0.333\_0.00\_0.67 &	$0.045862 \pm 0.000010$ & 	$0.044039 \pm 0.000455$ & 	$0.823471 \pm 0.000025$ & 	$0.825458 \pm 0.005127$ \\
28 & Q0.333\_-0.80\_0.80 &	$0.047937 \pm 0.000014$ & 	$0.046294 \pm 0.000398$ & 	$0.855825 \pm 0.000188$ & 	$0.857221 \pm 0.004083$ \\
29 & Q0.333\_0.80\_-0.80 &	$0.022026 \pm 0.000004$ & 	$0.021768 \pm 0.000115$ & 	$0.206316 \pm 0.000002$ & 	$0.205397 \pm 0.003613$ \\
30 & Q0.250\_0.00\_-0.75 &	$0.016007 \pm 0.000001$ & 	$0.015923 \pm 0.000101$ & 	$0.067207 \pm 0.000000$ & 	$0.066201 \pm 0.002989$ \\
31 & Q0.250\_0.00\_0.75 &	$0.041023 \pm 0.000005$ & 	$0.038639 \pm 0.000449$ & 	$0.852368 \pm 0.000083$ & 	$0.855422 \pm 0.004317$ \\
32 & Q0.250\_0.80\_-0.80 &	$0.016470 \pm 0.000000$ & 	$0.016191 \pm 0.000100$ & 	$0.057516 \pm 0.000001$ & 	$0.057117 \pm 0.002698$ \\
33 & Q0.200\_0.00\_-0.80 &	$0.012631 \pm 0.000007$ & 	$0.012556 \pm 0.000091$ & 	$-0.067330 \pm 0.000001$ & 	$-0.067585 \pm 0.002660$ \\
34 & Q0.200\_0.00\_0.80 &	$0.036968 \pm 0.000044$ & 	$0.034071 \pm 0.000500$ & 	$0.872432 \pm 0.000480$ & 	$0.876983 \pm 0.004144$ \\
35 & Q0.167\_0.00\_-0.83 &	$0.010495 \pm 0.000004$ & 	$0.010396 \pm 0.000067$ & 	$-0.172301 \pm 0.000003$ & 	$-0.172286 \pm 0.001438$ \\
36 & Q0.167\_0.00\_0.83 &	$0.033350 \pm 0.000003$ & 	$0.030324 \pm 0.000367$ & 	$0.888377 \pm 0.000166$ & 	$0.893634 \pm 0.003402$ \\
37 & Q0.100\_0.00\_0.00 &        $0.0044   \pm 0.0001  $ &                               &	$0.261 \pm 0.002$       & \\
\end{tabular}
\end{ruledtabular}
\end{table}

\end{widetext}

In Table~\ref{tab:spinerad}, we give the horizon 
mass and spin magnitude of the remnant BH for each 
configuration studied here. We measure these using both 
the isolated
horizon formalism and based on the measured radiated mass and
angular momentum. However, the isolated horizon measurements
are expected to be more accurate, and the differences between the
isolated horizon and radiation quantities is largely due to truncation
errors in the radiation zone that do not affect the accuracy near the
horizons themselves (see Ref.~\cite{Lousto:2013wta}).

\section{New Models of remnant mass, spin and recoil}
\label{sec:fits}

In Refs. \cite{Lousto:2012gt} and \cite{Lousto:2013wta} we
developed a series expansion
for the  mass, spin, and recoil velocity of the remnant BH
produced by the merger of a progenitor BHB with arbitrary BH spin
magnitudes and orientations and arbitrary mass ratio in terms of
the variables $\vec \Delta$, $\vec S$, and $\delta m$.  
For the runs presented here, only terms proportional to
$S_\|$, $\Delta_\|$, and $\delta m$ contribute. 
In addition, only certain combinations are allowed by
symmetry considerations. For more
details see Table IV of Ref.~\cite{Lousto:2012gt}  and Table VI of
Ref.~\cite{Lousto:2013wta}. Here we include all allowed terms up through fourth
order. Here we include powers of $\delta m$ when counting orders.
This differs from our previous conventions~\cite{Lousto:2013wta,
Lousto:2012gt}, where we
only counted powers in the spin variables and allowed the coefficients
of those terms to be arbitrary functions of $\delta m$ (consistent
with the symmetries).

The formula for the mass of the remnant $M_{\rm rem}$ is then
given by,
\begin{eqnarray}\label{eq:p4mass}
\frac{M_{\rm rem}}{m} = \Big\{M_0 + K_1 \Spar+ K_{2a}\,\Dpar\dmt +
                     K_{2b}\,\Spar^2+ \nonumber\\
                     K_{2c}\,\Dpar^2+
                     K_{2d}\,\dmt^2 +
                     K_{3a}\,\Dpar \Spar \dmt+ \nonumber\\
                     K_{3b}\,\Spar \Dpar^2+
                     K_{3c}\,\Spar^3+ \nonumber\\
                     K_{3d}\,\Spar\dmt^2+
                     K_{4a}\,\Dpar\Spar^2\dmt + \nonumber\\
                     K_{4b}\,\Dpar^3\dmt +
                     K_{4c}\,\Dpar^4+
                     K_{4d}\,\Spar^4+\nonumber\\
                     K_{4e}\,\Dpar^2 \Spar^2+
                     K_{4f}\,\dmt^4+
                     K_{4g}\,\Dpar\dmt^3+\nonumber\\
                     K_{4h}\,\Dpar^2\dmt^2 +
                     K_{4i}\,\Spar^2\dmt^2\Big\}+\nonumber\\
                     +{\cal O}(\epsilon^5),
\end{eqnarray} where
${\cal O}(\epsilon^5)$ denotes terms of fifth and higher order in the
expansion variables and where variables with tildes are dimensionless,
that is $\Spar = S_\| / m^2$ and $\Dpar = \Delta_\|/m^2$.
As written, Eq.~(\ref{eq:p4mass}) does not reproduce exactly the particle limit
since $\delta m\to\pm1$ as $\eta\to0$. However, we can add terms of
order ${\cal O}(\epsilon^6)$ and higher to obtain the correct
particle limit behavior while simultaneously producing an expansion
equivalent to Eq.~(\ref{eq:p4mass}). First, we note that in the particle limit, $M_{\rm
rem}$ is given by $M_{\rm rem}/m = 1 + \eta (\tilde E_{\rm isco} -1)+ {\cal
O}(\eta)^2$ (where $ m\eta\tilde E_{\rm isco}$ is the energy of a particle
at the ISCO). To enforce the particle limit for zero spin we add
two terms $K_6\,\delta{m}^6+K_8\,\delta{m}^8$, and then fix the value
of these constants by demanding that a
reexpansion in terms of $\eta$ gives $1 + \eta (\tilde E^{\rm sch}_{\rm isco}
-1) + {\cal
O}(\eta)^2$ ($\tilde E^{\rm sch}_{\rm isco}$ is the Schwarzschild ISCO
energy). We follow a similar procedure for the spin dependent terms.
 For most terms in
Eq.~(\ref{eq:p4mass}), 
the net effect is to simply multiply the given term
by $(4\eta)^2$.
The resulting formula for $M_{\rm rem}$ is
given by,
\begin{eqnarray}\label{eq:4mass}
\frac{M_{\rm rem}}{m} = (4\eta)^2\,\Big\{M_0 + K_1 \Spar+ K_{2a}\,\Dpar\dmt +
                     K_{2b}\,\Spar^2+ \nonumber\\
                     K_{2c}\,\Dpar^2+
                     K_{2d}\,\dmt^2 +
                     K_{3a}\,\Dpar\Spar\dmt+ \nonumber\\
                     K_{3b}\,\Spar\Dpar^2+
                     K_{3c}\,\Spar^3+ \nonumber\\
                     K_{3d}\,\Spar\dmt^2+
                     K_{4a}\,\Dpar\Spar^2\dmt + \nonumber\\
                     K_{4b}\,\Dpar^3\dmt +
                     K_{4c}\,\Dpar^4+
                     K_{4d}\,\Spar^4+\nonumber\\
                     K_{4e}\,\Dpar^2 \Spar^2+
                     K_{4f}\,\dmt^4+
                     K_{4g}\,\Dpar\dmt^3+\nonumber\\
                     K_{4h}\,\Dpar^2\dmt^2 +
                     K_{4i}\,\Spar^2\dmt^2\Big\}+\nonumber\\
                     \left[1+\eta(\tilde{E}_{\rm ISCO}+11)\right]\dmt^6,\quad\,
\end{eqnarray}
Here we take $\tilde{E}_{\rm ISCO}$ from Eq (2.7) of
Ref.~\cite{Ori00} (we replace the variable $a$ in Ref.~\cite{Ori00}
with $\alpha_{\rm rem}$, but similar results are obtained when using 
$S_\|/m^2$ instead).

We then verified that the correct leading power of $4 \eta$ in
Eq.~(\ref{eq:4mass}) is indeed 2 by replacing $(4 \eta)^2$ with $(4
\eta)^p$
and fitting all coefficients and found $p=2.0006$ gives the best fit.
In Table \ref{tab:fitpars}
below, the power $p$ is set to 2 exactly when performing
the fits.

To obtain a phenomenological formula for the remnant spin, we follow
a similar procedure. Prior to enforcing the particle limit we  
have,
\begin{eqnarray}\label{eq:p4spin}
\alpha_{\rm rem} = \frac{S_{\rm rem}}{M^2_{\rm rem}} =
                     \Big\{L_0 + L_{1}\,\Spar+\nonumber\\ 
                     L_{2a}\,\Dpar\dmt+
                     L_{2b}\,\Spar^2+
                     L_{2c}\,\Dpar^2+
                     L_{2d}\,\dmt^2+\nonumber\\
                     L_{3a}\,\Dpar\Spar\dmt+
                     L_{3b}\,\Spar\Dpar^2+
                     L_{3c}\,\Spar^3+\nonumber\\
                     L_{3d}\,\Spar\dmt^2+
                     L_{4a}\,\Dpar\Spar^2\dmt+
                     L_{4b}\,\Dpar^3\dmt+\nonumber\\
                     L_{4c}\,\Dpar^4+
                     L_{4d}\,\Spar^4+
                     L_{4e}\,\Dpar^2\Spar^2+\nonumber\\
                     L_{4f}\,\dmt^4+
                     L_{4g}\,\Dpar\dmt^3+\nonumber\\
                     L_{4h}\,\Dpar^2\dmt^2+
                     L_{4i}\,\Spar^2\dmt^2\Big\}+\nonumber\\
                     {\cal O} (\epsilon^5).
\end{eqnarray}
Once again, we add higher order terms in order to enforce the correct
particle limit behavior. Here 
the new terms are generated by multiplying the existing terms in
Eq.~(\ref{eq:p4spin})
by the next even powers of $\delta m$ that correspond to ${\cal O} (\epsilon^5)$
or higher. For instance for the spin independent terms
we add $L_6\,\dmt^6+L_8\,\dmt^8$ and for the linear in the spin terms
$L_5\,S_\|\delta{m}^4+L_7\,S_\|\delta{m}^6$. We then impose the particle limit 
which is given by
 $\alpha_{\rm rem}  =  \Spar+\eta\tilde{J}_{\rm
 ISCO} + {\cal O}(\eta^2)$.
Again, we use Eq (2.8) of Ref.~\cite{Ori00} to calculate the ISCO
angular momentum, replacing the variable $a$ there with $\alpha_{\rm rem}$.

After enforcing the particle limit we get,
\begin{eqnarray}\label{eq:4spin}
\alpha_{\rm rem} = \frac{S_{\rm rem}}{M^2_{\rm rem}} =
                     (4\eta)^2\Big\{L_0 + L_{1}\,\Spar+\nonumber\\ 
                     L_{2a}\,\Dpar\dmt+
                     L_{2b}\,\Spar^2+
                     L_{2c}\,\Dpar^2+
                     L_{2d}\,\dmt^2+\nonumber\\
                     L_{3a}\,\Dpar\Spar\dmt+
                     L_{3b}\,\Spar\Dpar^2+
                     L_{3c}\,\Spar^3+\nonumber\\
                     L_{3d}\,\Spar\dmt^2+
                     L_{4a}\,\Dpar\Spar^2\dmt+
                     L_{4b}\,\Dpar^3\dmt+\nonumber\\
                     L_{4c}\,\Dpar^4+
                     L_{4d}\,\Spar^4+
                     L_{4e}\,\Dpar^2\Spar^2+\nonumber\\
                     L_{4f}\,\dmt^4+
                     L_{4g}\,\Dpar\dmt^3+\nonumber\\
                     L_{4h}\,\Dpar^2\dmt^2+
                     L_{4i}\,\Spar^2\dmt^2\Big\}+\nonumber\\
                     \Spar(1+8\eta)\dmt^4+\eta\tilde{J}_{\rm ISCO}\dmt^6.
\end{eqnarray}
In order to verify our hypothesis, 
we first replaced $(4 \eta)^2$ with $(4\eta)^p$ and fit for 
all coefficients in Eq.~(\ref{eq:4spin}). We find $p=2.015$, which is 
reasonably close to the expected
power of 2. We then fit again using $p=2$ exactly, and report these
fitting parameters in Table \ref{tab:fitpars} below.

By using $a = \alpha_{\rm rem}$ to evaluate the ISCO quantities, the fitting 
formula for the spin becomes implicit, and the formula for the mass
depends directly on the formula for the spin.  
Therefore, to evaluate the fitting formulas for any given initial binary, 
we use a rapidly converging iterative process where the initial
$a$ is set to $S_\|/m^2$.

Finally, we fit the recoil to the formula,
\begin{eqnarray}\label{eq:4recoil}
v_\perp = H\eta^2\left(\Dpar+ H_{2a} \Spar\dmt    
                     + H_{2b} \Dpar \Spar
                     + H_{3a} \Dpar^2\dmt\right.\nonumber\\
                     \left.+ H_{3b} \Spar^2\dmt
                     + H_{3c} \Dpar\Spar^2
                     + H_{3d} \Dpar^3
                     + H_{3e} \Dpar \dmt^2\right.\nonumber\\
                     + H_{4a} \Spar\Dpar^2\dmt
                     \left.+ H_{4b} \Spar^3 \dmt
                     + H_{4c} \Spar \dmt^3\right.\nonumber\\
                     \left.+ H_{4d} \Dpar \Spar \dmt^2
                     + H_{4e} \Dpar \Spar^3
                     + H_{4f} \Spar \Dpar^3\right)
\end{eqnarray}
\begin{equation}\label{eq:xi}
\tilde\xi=a+b\, \Spar +c\, \dmt\Dpar,
\end{equation}
where we have added a leading power of $\eta^2$ to the expansion.
The issue of the leading power of $\eta$ for the recoil
was discussed in the context of the off-plane recoils
in Ref.~\cite{Baker:2008md, Lousto:2008dn} where the 
possibility of a leading $\eta^3$ versus
$\eta^2$ was studied with full numerical simulations. Further
study of recoils in the small mass 
ratio perturbative regime \cite{Nakano:2010kv}
led to the conclusion that the terms of the recoil
linear in the spin should scale as $\eta^2$ and post-Newtonian
expansions including quadratic terms in the spin also show
a leading $\eta^2$ behavior \cite{Racine:2008kj} (again for low
eccentricity, in-plane orbits). 

Finally, we tested the leading $\eta^2$ dependence in
Eq.~(\ref{eq:4recoil}) by allowing the power of $\eta$ to be free.
 Interestingly, we do
not find $p=2$, but rather we find that the minimum in the fit is quite
shallow with similar results for interval $1.5\leq p \leq 2.5$ (the
minimum is at $p\sim2.29$). Since
$p=2$ gives the correct particle limit behavior for quasicircular
orbits
(at least at moderate mass ratios, see though
Refs.~\cite{Hirata:2010xn, vandeMeent:2014raa} for a
discussion on resonance recoil which scale as $\eta^{1.5}$),
we enforce $p=2$ for the fits presented in Table \ref{tab:fitpars} below.

We note that a factor of $\eta^p$ in the coefficients is
not independent from the expansion proposed in Refs. \cite{Lousto:2012gt} 
and \cite{Lousto:2013wta} since $4\eta=1-\dmt^2$ and this allows
us to recast all powers of $\eta$ into the original form of the expansion.

The 17 constants in Eqs.~(\ref{eq:4recoil})~and~(\ref{eq:xi}), 19 constants in
Eq.~(\ref{eq:4mass}),
and 19 constants in Eq.~(\ref{eq:4spin}) were obtained by a least-squares
fit to the results of our 36 simulations and, in the case of 
the fits of the final mass and spin, additional 
38 SXS runs \cite{SXS:catalog} and a $q=1:10$ simulation.
Note that, as explained above, we allow $v_\perp$ to be positive and
negative and thereby allow $\tilde\xi$ to be continuous
(See Fig.~\ref{fig:xi}).

\begin{figure}
  \includegraphics[angle=270,width=\columnwidth]{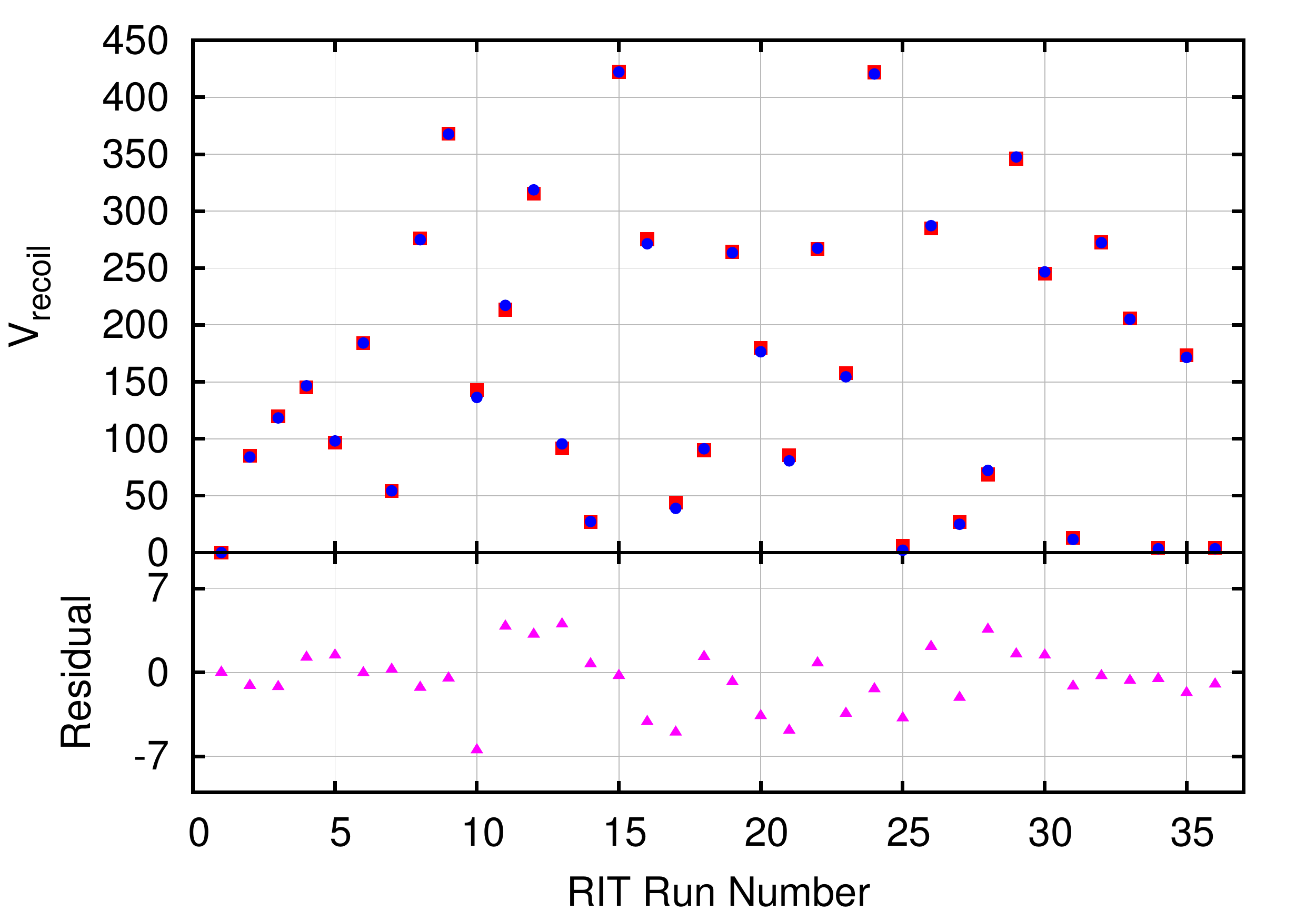}
  \caption{Fit and residuals to 36 RIT data 
(See Table~\ref{tab:kicks}) with a root mean square deviation,
RMS=$2.5\,\KMS$}
  \label{fig:residuals_mp}
\end{figure}

The results of the recoil velocity fit and residuals to the entire set of
36 runs is shown in Fig.~\ref{fig:residuals_mp}. We observe that
the residuals are below $7\,\KMS$.

\begin{figure}
  \includegraphics[angle=270,width=\columnwidth]{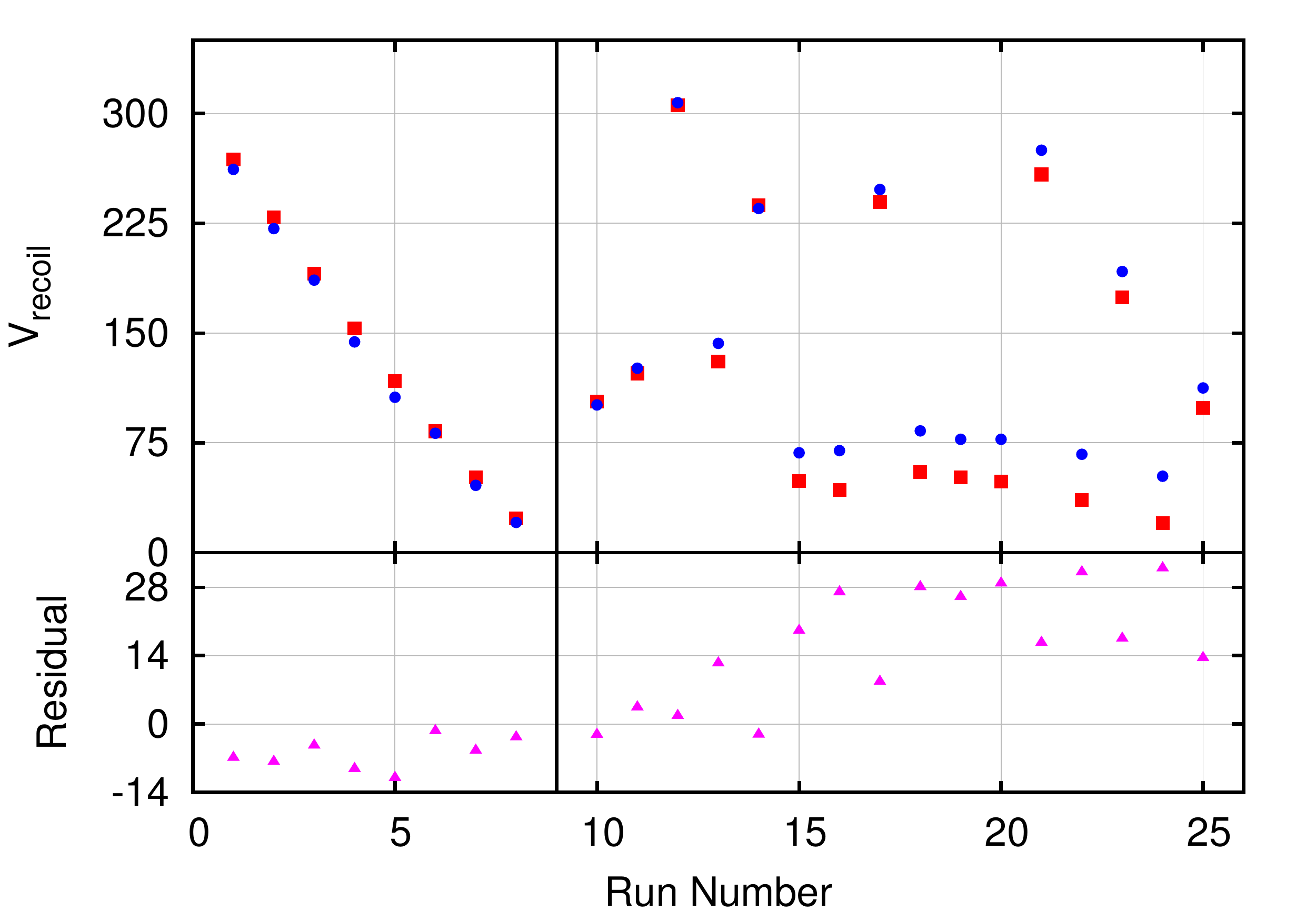}
  \caption{Recoil Fit and residuals to 8 AEI (RMS=$6.8\,\KMS$) simulations and 
16 SXS (RMS=$20\,\KMS$) simulations.}
  \label{fig:SXSfit}
\end{figure}

\begin{figure}
  \includegraphics[angle=270,width=\columnwidth]{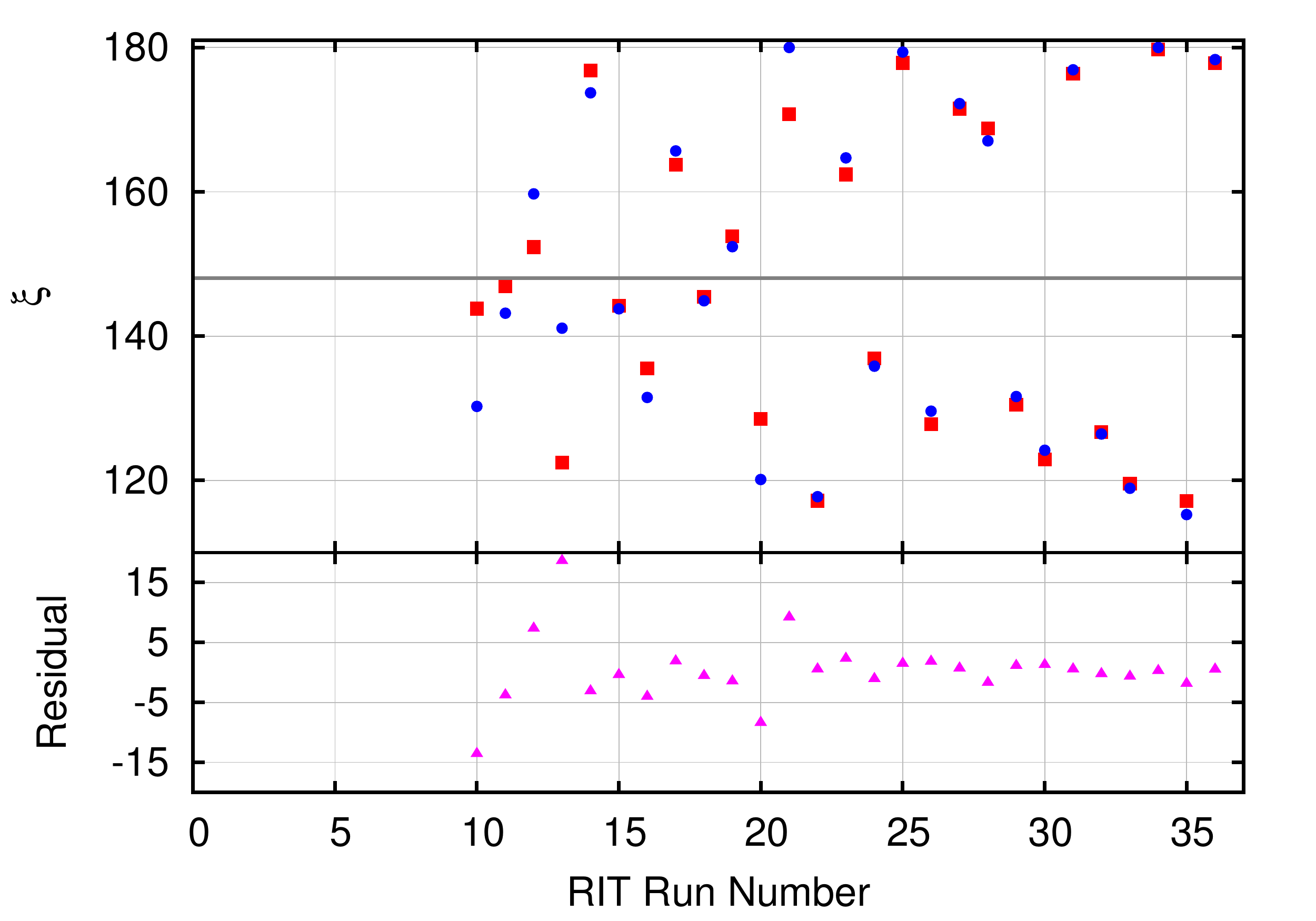}
  \caption{The angle $\tilde\xi$ as defined in Eq.~(\ref{eq:xi}), 
computed from the measured magnitude of the recoil (circles), 
compared to the fitted formula (squares). 
The angle is not defined for the (first) 9 equal mass runs in Table~\ref{tab:kicks}.
The average of $\tilde\xi\sim148^\circ$ (gray line) is near the previously 
measured value of $\tilde\xi\sim145^\circ$, but the new spin-dependent
formula for $\tilde\xi$ significantly reduces the residuals.}
  \label{fig:xi}
\end{figure}

In order to assess the accuracy of our formula,
we compare its predictions for 8 independent runs from the group at
AEI reported in Ref.~\cite{Pollney:2007ss} and 16 from the SXS collaboration 
in Ref.~\cite{SXS:catalog}.
The results are shown in Fig.~\ref{fig:SXSfit}.
We observe that while the residuals of our runs and
those of AEI are similar and relatively small (i.e., within
$10\,\KMS$), the residuals with respect to the SXS runs are roughly
3 times larger. We note that while in this paper (and
Ref.~\cite{Pollney:2007ss})
recoils are computed using the radiated linear momentum, 
the SXS catalog reports coordinate velocities.

Note that while we use the recoils of Ref.~\cite{Pollney:2007ss}
as an independent test of our fitting formula, the recoil formula 
proposed in \cite{Pollney:2007ss} [Eq. (42) there] does not
respect the symmetry of exchange of black hole 
labels $1\leftrightarrow2$, hence we would expect it to be less
accurate
outside of the region of parameter space used to generate that fit.

Interestingly, by allowing $v_\perp$ in Eq.~(\ref{eq:vplane}) to
take on positive and negative values, the angle $\tilde\xi$ can be
restricted to the interval $90^\circ\leq\tilde\xi\leq 180^\circ$ and
its average value,
as shown in Fig.~\ref{fig:xi}, is $\tilde\xi\sim148^\circ$, which is
very close to the estimate $\tilde\xi\sim145^\circ$ 
in Ref.~\cite{Lousto:2007db}. Note that the dispersion
is quite large though.
As part of our fitting of the recoil, we must simultaneously fit
 $\tilde\xi$ to Eq.~(\ref{eq:xi}). Interestingly,
the choice of coefficients in Eq.~(\ref{eq:xi}) that optimizes
the fit is close to $c/b=3/7$. This ratio is significant
because the  effective spin defined
in Ref.~\cite{Damour:2001tu} is given by
$\vec{S}_\text{effective}=\vec{S}+\frac{3}{7}\vec{\Delta}\delta m$. Thus it
appears that the functional form of  $\tilde\xi$ that minimizes the
residuals is essentially a linear function in the effective spin.

The fitted values of all coefficients in
Eqs.~(\ref{eq:4recoil})~and~(\ref{eq:xi}), as well as the
uncertainties in these values are
given in Table~\ref{tab:fitpars}.
We estimate the errors in the fitting parameters by adding
Gaussian distributed random noise to the fitting function.  
Each data point is given a different random number, and the
width of the Gaussian is determined by the estimated error
in that data.  By performing the fit 50,000 times, each
with a different set of noise, a distribution in the fitting
parameters is found.  The standard deviation of these 
distributions is recorded as the error in the fitting
parameter in Table~\ref{tab:fitpars}.

\begin{widetext}

\begin{table}
\caption{Table of fitting parameters for recoil, mass, and spin formulas.}\label{tab:fitpars}
\begin{ruledtabular}
\begin{tabular}{lc|lc|lc}
H& $7367.250029 \pm 66.122336$ & M0  & $0.951507 \pm 0.000030$  & L0  & $0.686710 \pm 0.000039$\\
H2a & $-1.626094 \pm 0.053888$ & K1  & $-0.051379 \pm 0.000193$ & L1  & $0.613247 \pm 0.000168$\\
H2b & $-0.578177 \pm 0.055790$ & K2a & $-0.004804 \pm 0.000514$ & L2a & $-0.145427 \pm 0.000473$\\
H3a & $-0.717370 \pm 0.077605$ & K2b & $-0.054522 \pm 0.000690$ & L2b & $-0.115689 \pm 0.000761$\\
H3b & $-2.244229 \pm 0.137982$ & K2c & $-0.000022 \pm 0.000010$ & L2c & $-0.005254 \pm 0.000332$\\
H3c & $-1.221517 \pm 0.176699$ & K2d & $1.995246 \pm 0.000497$  & L2d & $0.801838 \pm 0.000514$\\
H3d & $-0.002325 \pm 0.021612$ & K3a & $0.007064 \pm 0.002680$  & L3a & $-0.073839 \pm 0.002986$\\
H3e & $-1.064708 \pm 0.133021$ & K3b & $-0.017599 \pm 0.001678$ & L3b & $0.004759 \pm 0.001374$\\
H4a & $-0.579599 \pm 0.297351$ & K3c & $-0.119175 \pm 0.001054$ & L3c & $-0.078377 \pm 0.000911$\\
H4b & $-0.455986 \pm 0.302432$ & K3d & $0.025000 \pm 0.001951$  & L3d & $1.585809 \pm 0.001777$\\
H4c & $0.010963 \pm 0.174289$  & K4a & $-0.068981 \pm 0.004251$ & L4a & $-0.003050 \pm 0.001910$\\
H4d & $1.542924 \pm 0.274459$  & K4b & $-0.011383 \pm 0.001709$ & L4b & $-0.002968 \pm 0.001431$\\
H4e & $-4.735367 \pm 0.430869$ & K4c & $-0.002284 \pm 0.000192$ & L4c & $0.004364 \pm 0.000532$\\
H4f & $-0.284062 \pm 0.174087$ & K4d & $-0.165658 \pm 0.003100$ & L4d & $-0.047204 \pm 0.003250$\\
a  & $2.611988 \pm 0.028327$  & K4e & $0.019403 \pm 0.003220$  & L4e & $-0.053099 \pm 0.003682$\\
b  & $1.383778 \pm 0.092915$  & K4f & $2.980990 \pm 0.001197$  & L4f & $0.953458 \pm 0.001210$\\
c  & $0.549758 \pm 0.113300$  & K4g & $0.020250 \pm 0.002524$  & L4g & $-0.067998 \pm 0.002369$\\
    &                          & K4h & $-0.004091 \pm 0.002057$ & L4h & $0.001629 \pm 0.000980$\\
    &                          & K4i & $0.078441 \pm 0.003263$  & L4i & $-0.066693 \pm 0.003289$\\

\end{tabular}
\end{ruledtabular}
\end{table}

\end{widetext}

\begin{figure}
\includegraphics[angle=270,width=\columnwidth]{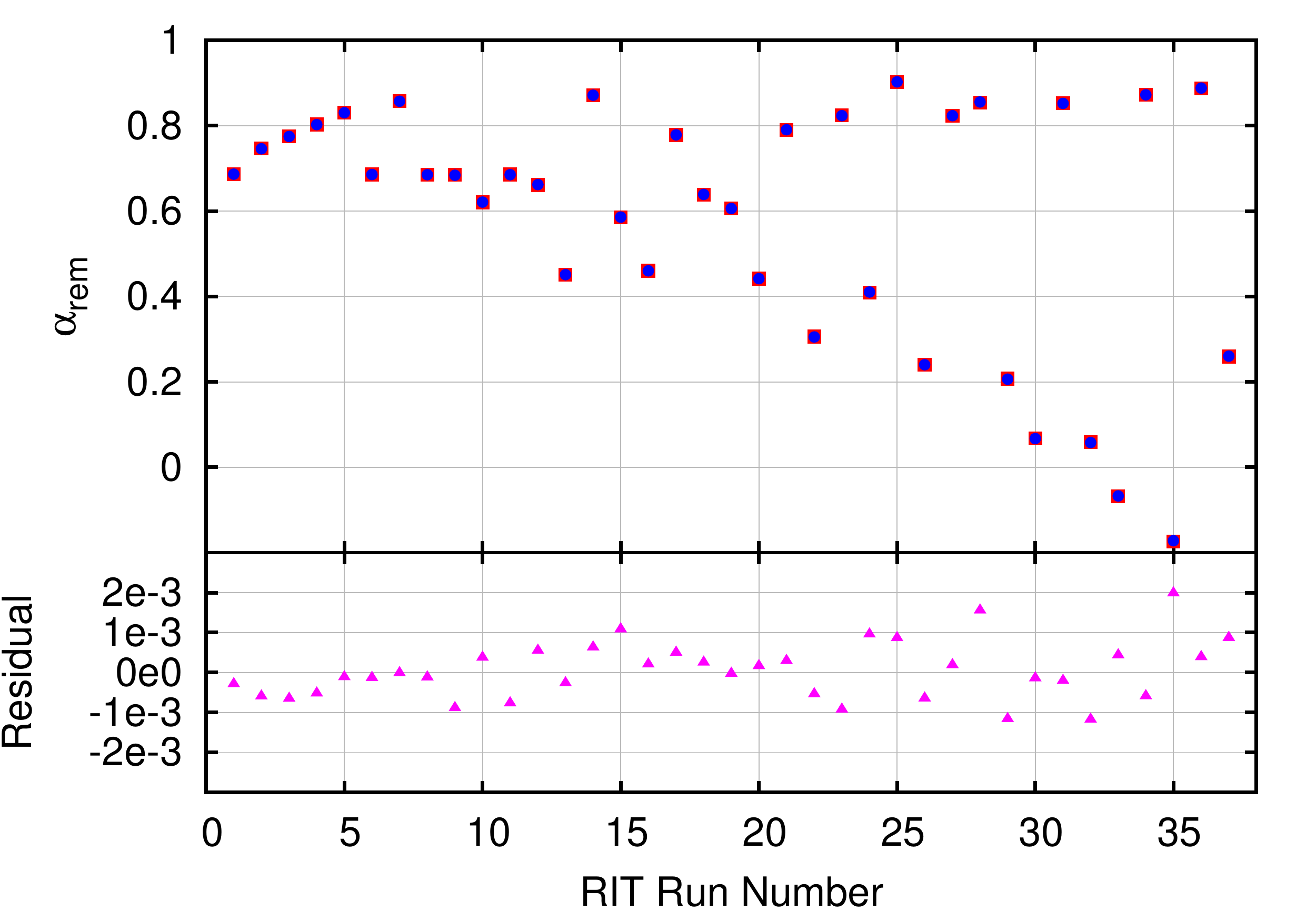}
  \caption{Fit to the remnant spin using RIT+SXS data and residuals of RIT runs as
labeled by run number (see Table~\ref{tab:spinerad}). RMS=$7.16\times 10^{-4}$.}
  \label{fig:spin_fit_RIT}
\end{figure}
\begin{figure}
  \includegraphics[angle=270,width=\columnwidth]{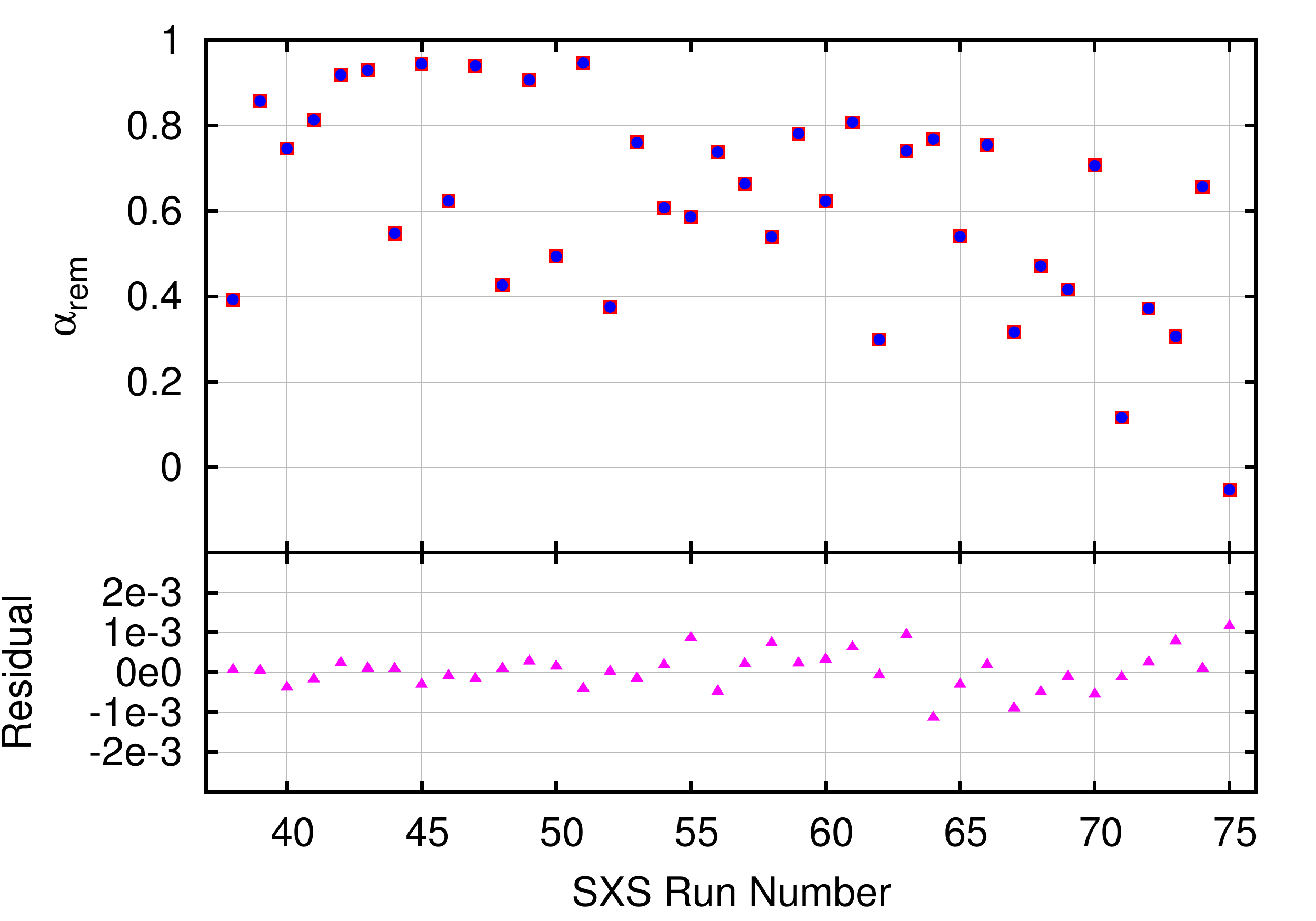}
  \caption{Fit to the remnant spin using RIT+SXS data residuals for
the SXS data. RMS=$4.73\times10^{-4}$.}
  \label{fig:spin_fit_SXS}
\end{figure}

We use a similar procedure to fit the final remnant mass and spin
to Eqs.~(\ref{eq:4mass})~and~(\ref{eq:4spin}). Here, however, we
add the data from the SXS catalog~\cite{SXS:catalog} (which
include results from highly spinning BHBs) into our fits,
as well as results from a non-spinning binary with mass
ratio $q=1/10$ from Refs.~\cite{Lousto:2010qx, Nakano:2011pb}.
The resulting fitting parameters
are given in Table~\ref{tab:fitpars}.
The data, fit, and residuals for the remnant spins are shown in 
Figs.~\ref{fig:spin_fit_RIT} and
\ref{fig:spin_fit_SXS}. The residuals are below $6\times10^{-4}$.

\begin{figure}
  \includegraphics[angle=270,width=0.49\columnwidth]{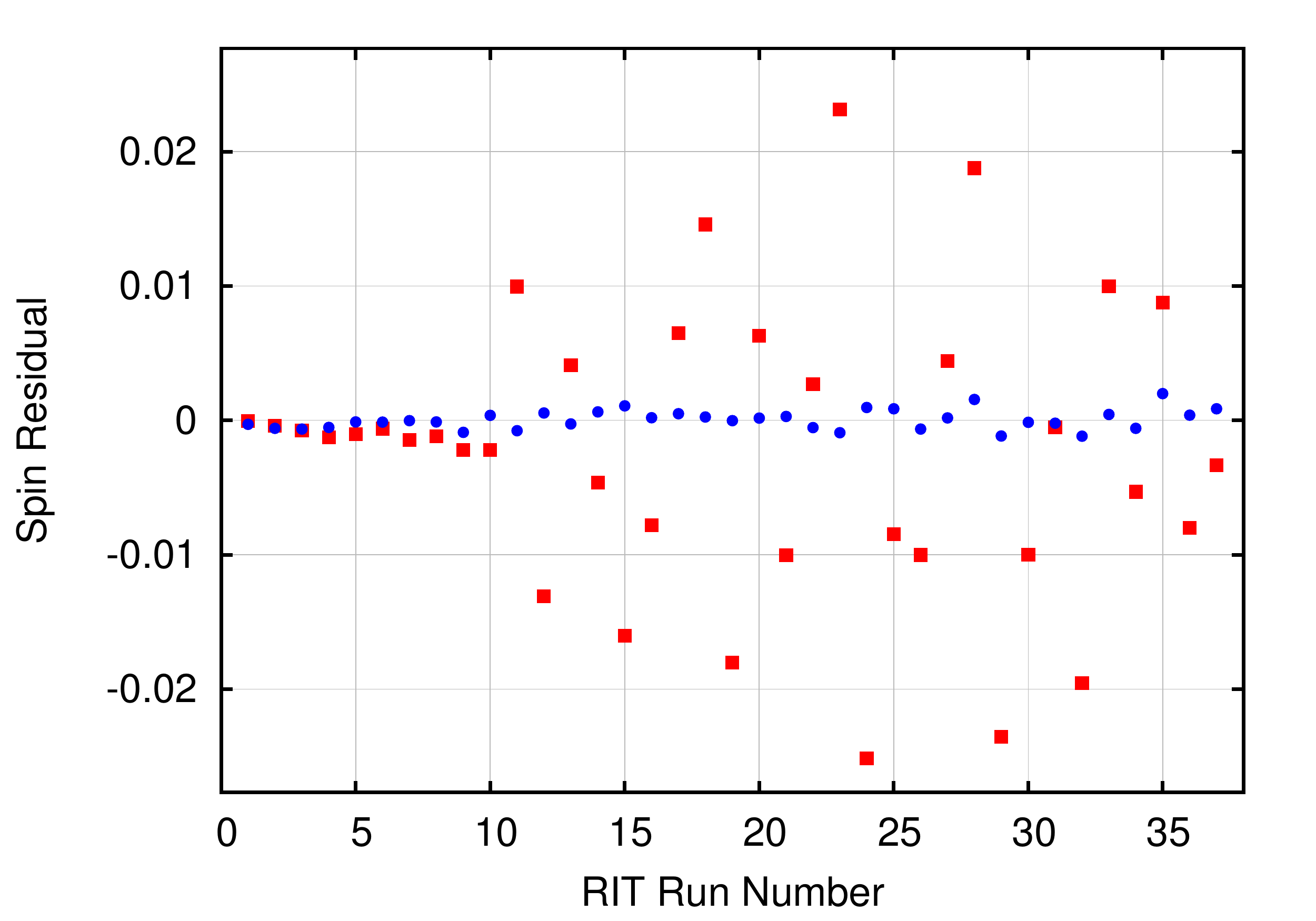}
  \includegraphics[angle=270,width=0.49\columnwidth]{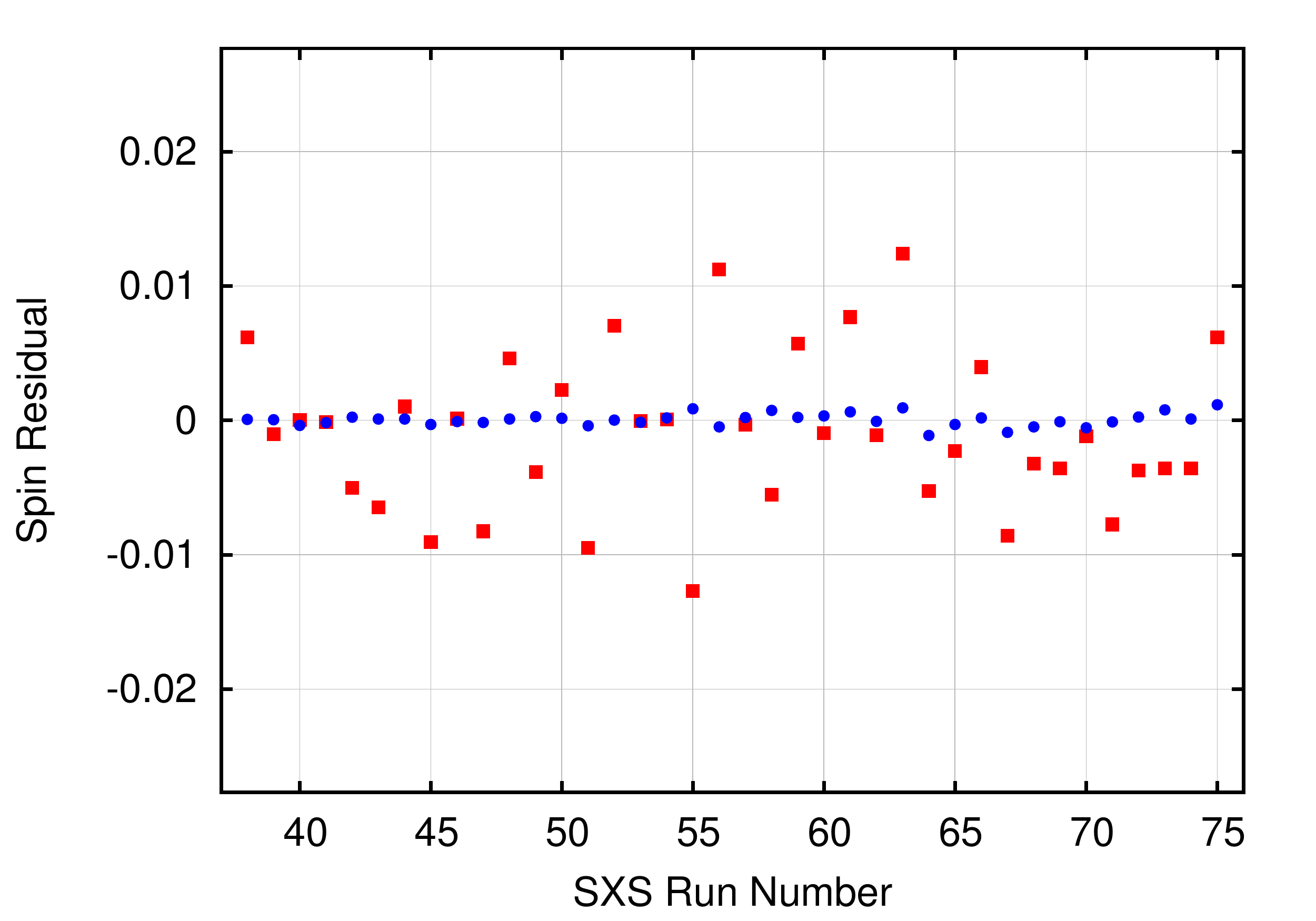}
  \caption{A comparison of the residuals from our spin fit and the AEI fit
for the  RIT data (left) and
SXS data (right).
Residuals from
our formula are denoted by (blue) circles and residuals from AEI
formula are denoted by (red) squares.}
  \label{fig:RezzVsOurs_RIT_2009}
\end{figure}

We can compare the residuals of our fit with other fitting formulas
in the literature, for instance, for the final spin of the merged
black hole given in Ref.~\cite{Barausse:2009uz} (we denote this fit
by AEI).
The results are shown in Fig.~\ref{fig:RezzVsOurs_RIT_2009}
for the current data (RIT) and 
the SXS data. We observe a clear improvement of our fitting
formula (\ref{eq:4spin}), with over an order of magnitude reduction
in the residuals.

We see that our fitting formula for the final remnant spin is
remarkably accurate over a wide range of mass ratios.
In addition, we constructed the formula such that it gives the correct
small-mass limit behavior. We thus expect that our formula will be
reasonably accurate for all mass ratios, at least for moderate spins
($\alpha \leq 0.9$).

\begin{figure}
  \includegraphics[angle=270,width=\columnwidth]{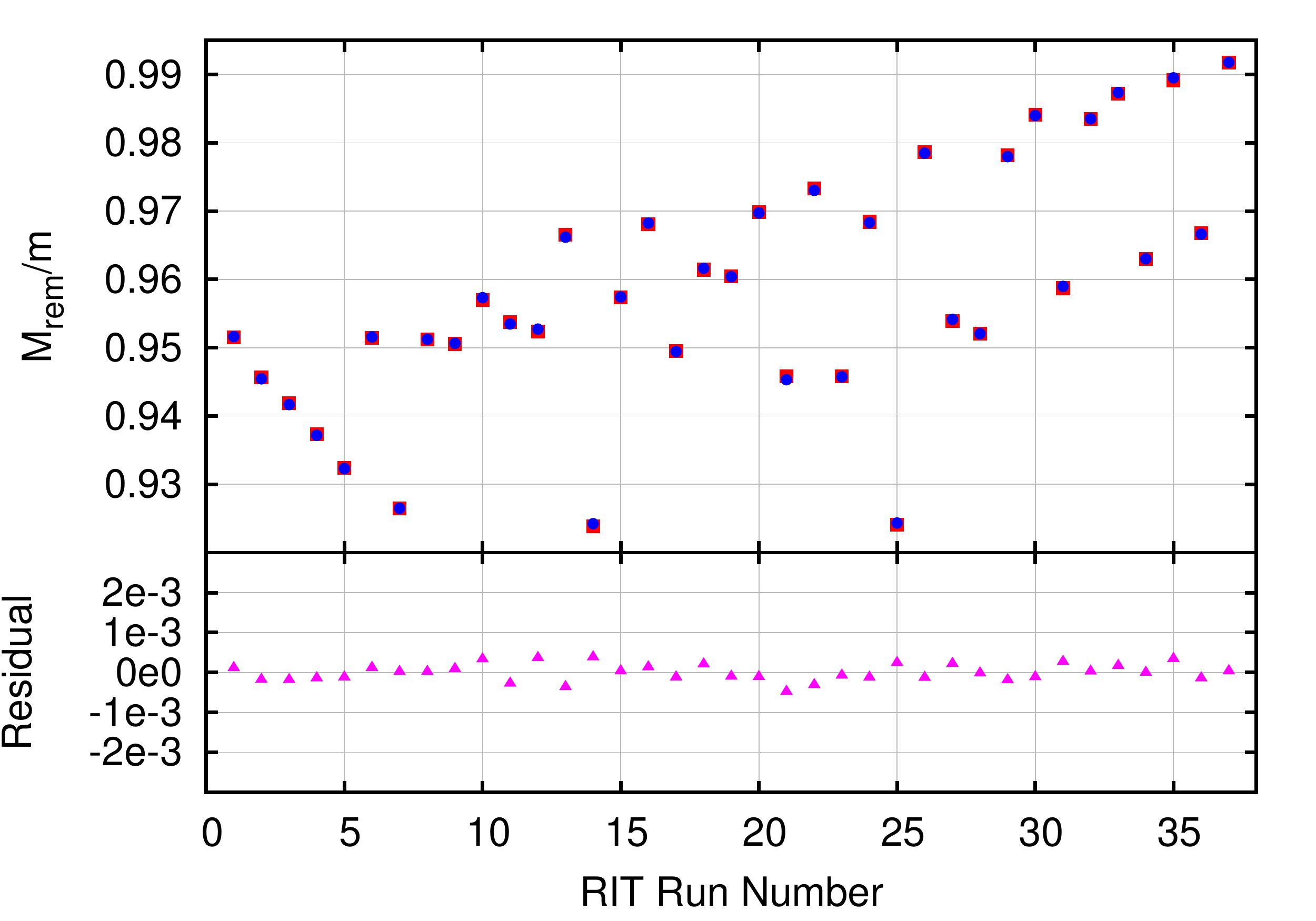}
  \caption{Predicted and measured remnant mass for
RIT data versus run number. RMS=$2.07\times10^{-4}$.}
  \label{fig:mass_fit_RIT}
\end{figure}
\begin{figure}
  \includegraphics[angle=270,width=\columnwidth]{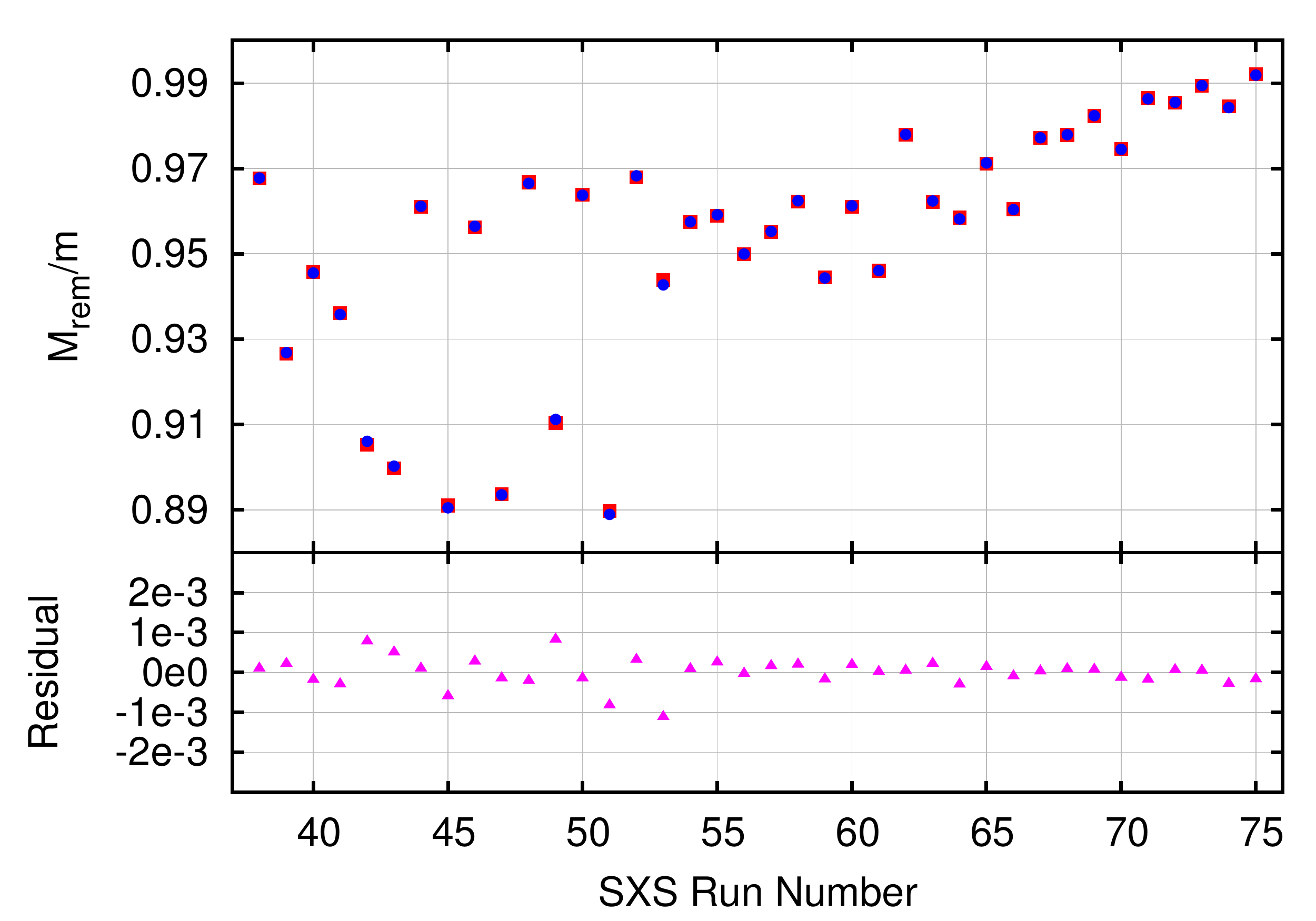}
  \caption{Predicted and measured remnant mass for the
 SXS data. RMS=$3.56\times10^{-4}$.}
  \label{fig:mass_fit_SXS}
\end{figure}
\begin{figure}
  \includegraphics[angle=270,width=0.49\columnwidth]{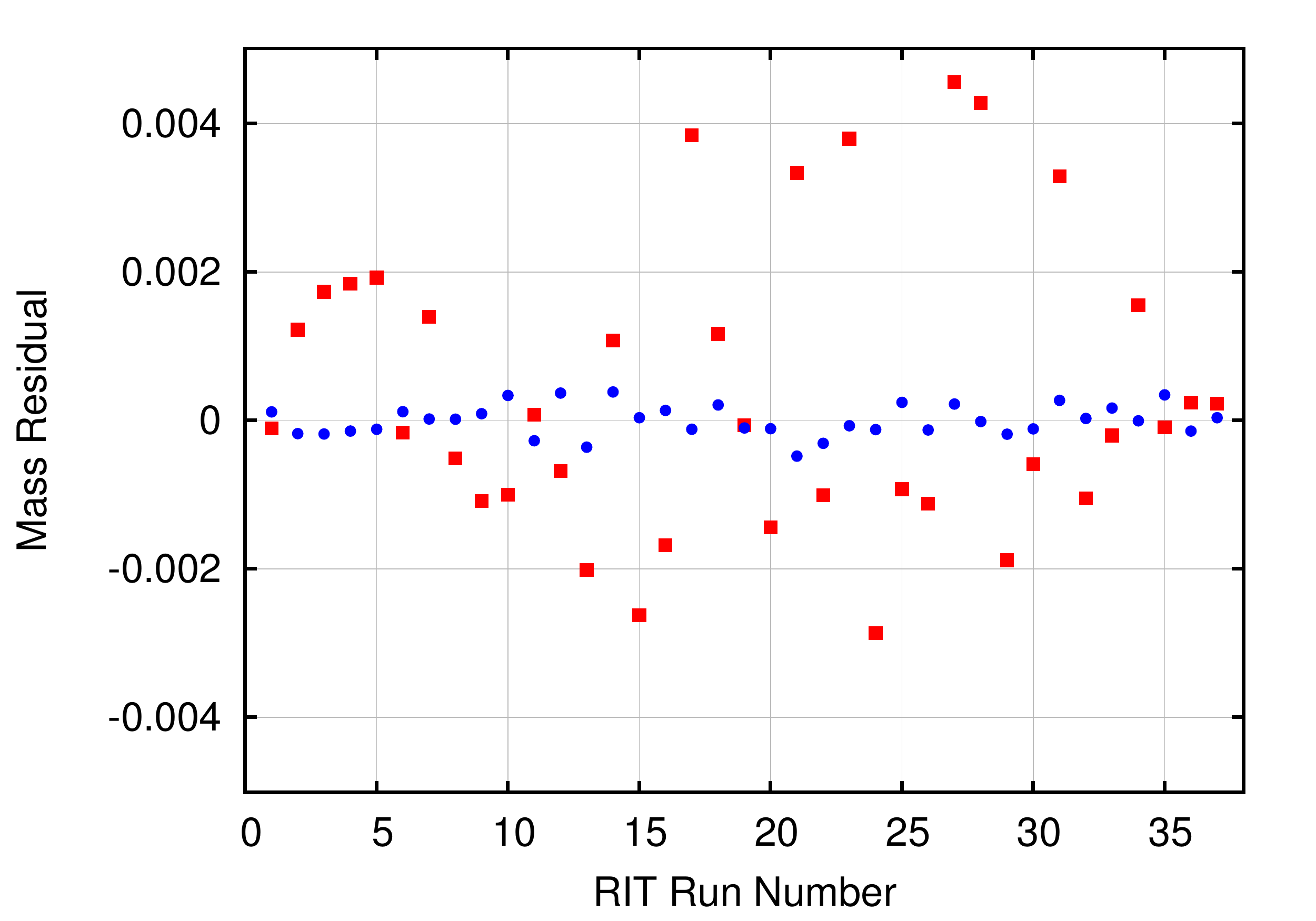}
  \includegraphics[angle=270,width=0.49\columnwidth]{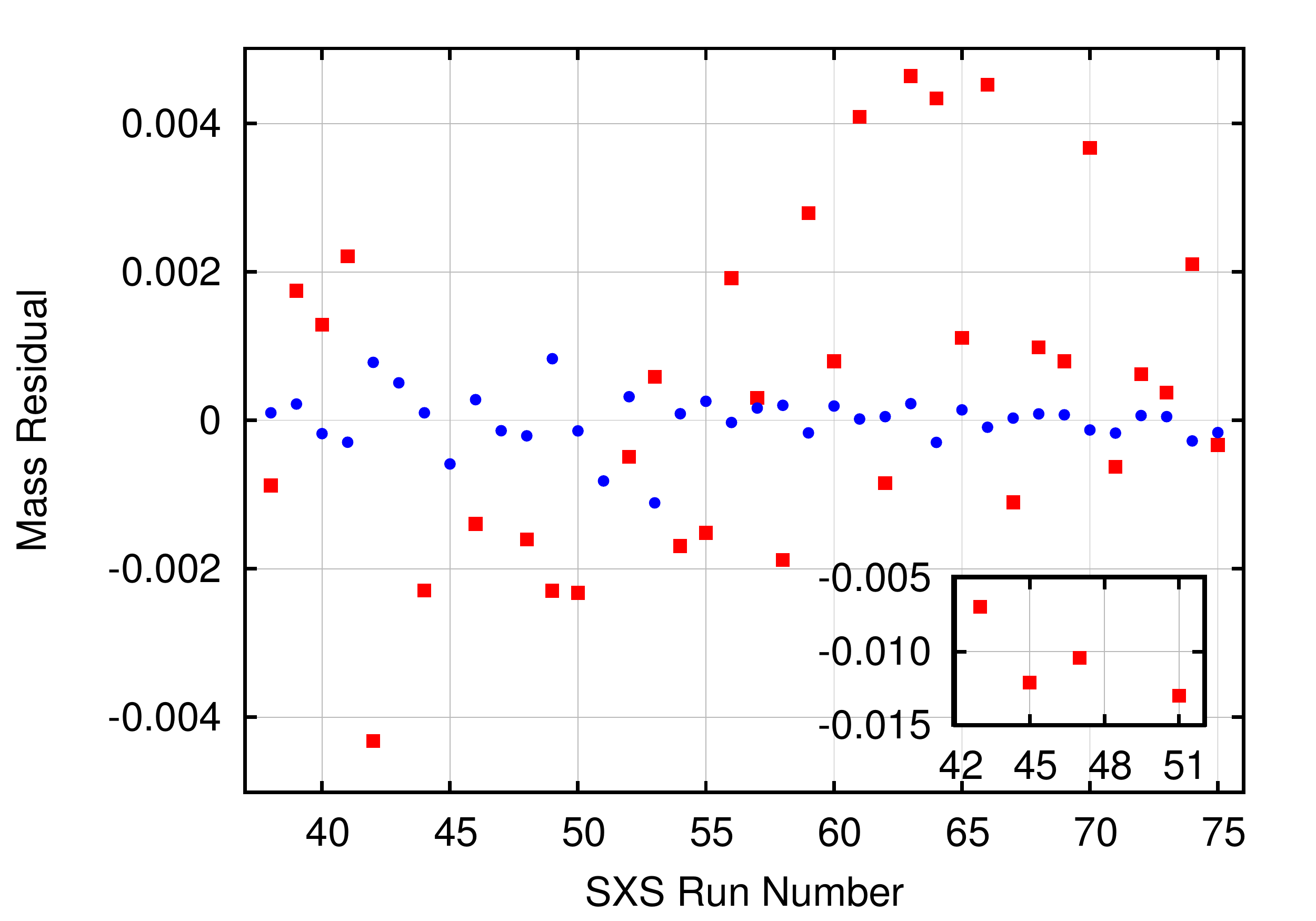}
  \caption{Residuals for our mass fit (small blue circles)
to the RIT (left) and SXS data (right) compared to the AEI fitting formula
(large red squares).}
  \label{fig:RezzVsOurs_Mass_RIT}
\end{figure}

A similar analysis for the final remnant mass is shown
in Figs.~\ref{fig:mass_fit_RIT}, \ref{fig:mass_fit_SXS},
and~\ref{fig:RezzVsOurs_Mass_RIT}. Once again we used the SXS data
in generating our fits.
 Here we
see residuals of order $3\times10^{-4}$ for our new formula and
residuals several times larger for the AEI formula
\cite{Barausse:2012qz}.

\begin{figure}
  \includegraphics[angle=270,width=\columnwidth]{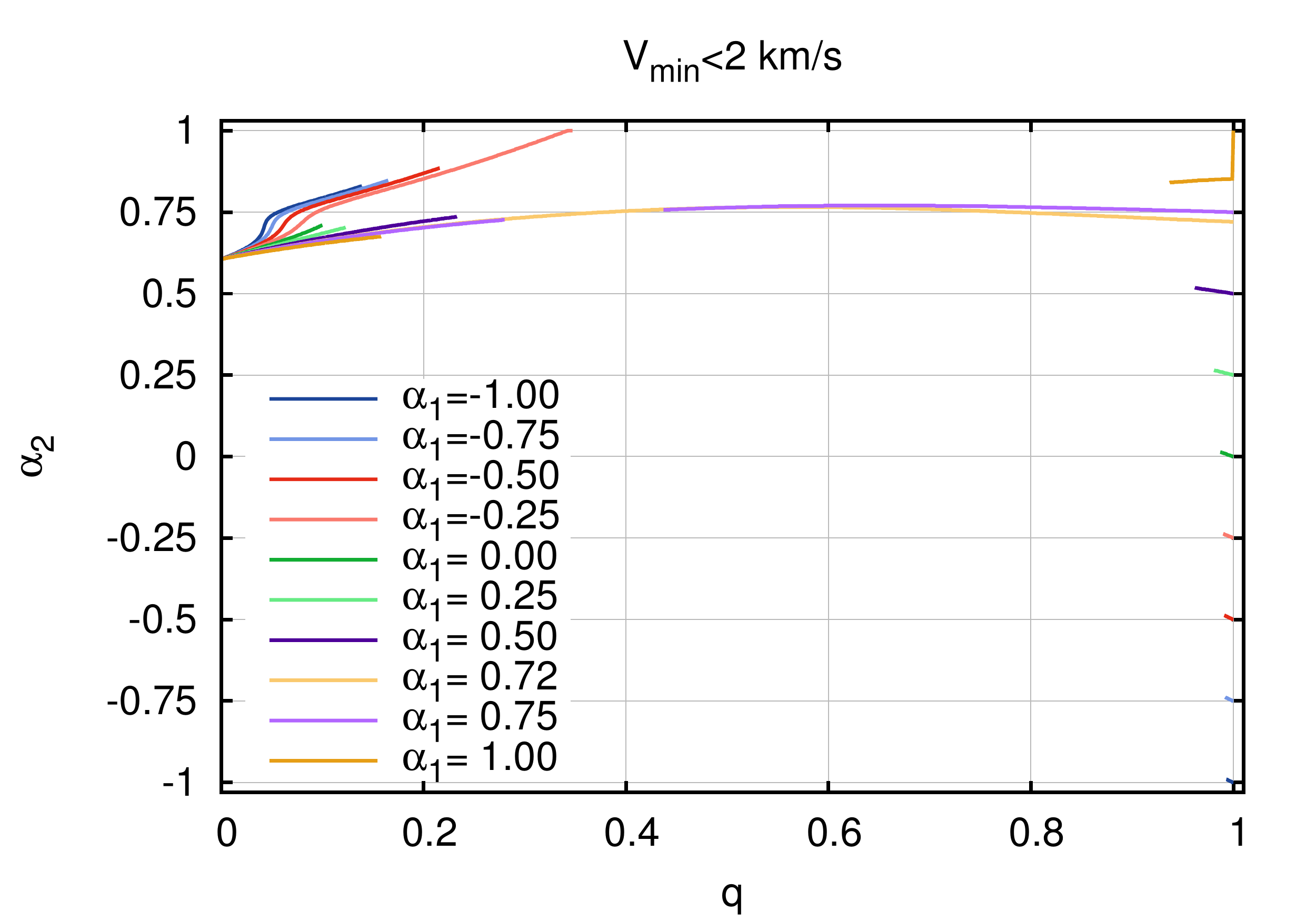}
  \caption{The BHB configurations that lead to a final remnant
black hole with zero recoil. Additions zero recoils exist for $q=0$ and
for $q=1$ with $\alpha_1=\alpha_2$.}\label{fig:V0}
\end{figure}

An interesting consequence of the form of the recoil velocity
(\ref{eq:4recoil}) is that, for certain combinations of the spins
and the mass ratio the total magnitude can be very small. In
Fig.~\ref{fig:V0}, we plot the values of $\alpha_2$ and $q$ that lead
to small recoils for a given $\alpha_1$. Apart from the zero recoil
imposed by symmetry, i.e. $q\to0$ and $q=1,\,\alpha_1=\alpha_2$,
there appear to be two branches that lead to vanishing recoils. One
branch spans all mass ratios with $0.6\lesssim \alpha_2 \lesssim 0.75$ and the 
other branch only spans the smaller mass ratio regime
$q\lesssim 0.4$ and larger spins $0.74\lesssim\alpha_2 \leq 1$.

\begin{figure}
  \includegraphics[angle=270,width=\columnwidth]{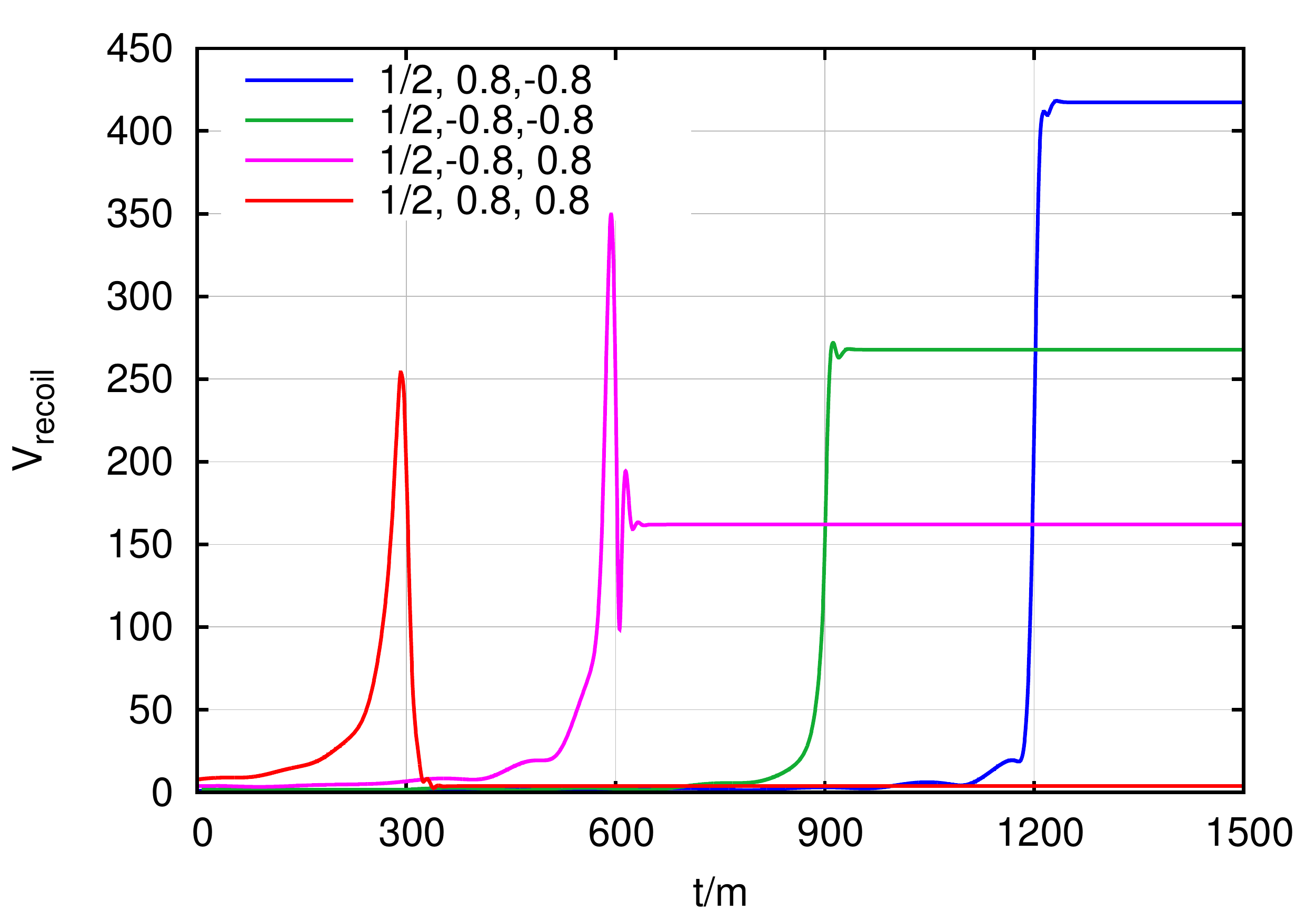}
  \caption{The radiated linear momentum for the  Q0.500\_0.80\_0.80
configuration (red).
 We observe the wobbling from
the inspiral, a sudden raise to above $250\,\KMS$ 
due to the merger, and the final antikick from the
ringdown phase which reduces the final velocity of the remnant
to $2\,\KMS$. For comparison we also show the Q$0.500\_\pm0.80\_\pm0.80$
configurations to see the different effects of the superposition of spin
(signs) and unequal mass components of the recoil.
}\label{fig:V0t}
\end{figure}

Interestingly, the vanishing of the recoil velocity does not arise from any 
symmetry, but rather from a cancellation of processes that involve
a wobbling of the center of mass as the BHB slowly
inspirals, the recoil generated during the rapid plunge, and the
post-merger anti-kick \cite{Rezzolla:2010df}, which is generated
 during the ring-down phase. 
All of these three stages combine to produce a 
non moving final BH, but in
the process the BH is displaced from the original center
of mass of the binary.
In Fig.~\ref{fig:V0t} we provide an explicit example for
run \#25 Q0.500\_0.80\_0.80 of a near zero final recoil. 

It is interesting to recall here the zero-recoil superkicks
seen in~\cite{Campanelli:2007cga, Lousto:2011kp}. In that case the
bobbing of the BHs up and down can be tuned by choosing the 
azimuthal orientation of the spin such that the merger occurs
when the bobbing velocity is instantaneously zero, which leads to a
vanishing recoil.

In Fig.~\ref{fig:V0t} we show the velocity of the center
of mass versus time for the four possible combinations of signs of the
spins for a $q=1/2$ binary with spin magnitudes $\alpha_i=0.8$
(Q$0.500\_\pm0.80\_\pm0.80$).
The UD configuration Q0.500\_0.80\_-0.80 recoils at
$420\ \KMS$ while the DD configuration recoils at a more
modest  $267\ \KMS$. Reversing the spin directions for both BH leads
to a DU configuration that recoils at  $154\ \KMS$. Finally, a UU
configuration recoils at a very small $2\ \KMS$.
Notably, the recoil of the DU configuration agrees with
the purely unequal mass recoil $v_m(q=1/2)$, and
the difference between the  recoil velocities of the UD and DU
configurations, as well as the differences in the recoil
between the DD and UU configurations 
are both around $266\ \KMS$.

\begin{figure}
  \includegraphics[angle=270,width=\columnwidth]{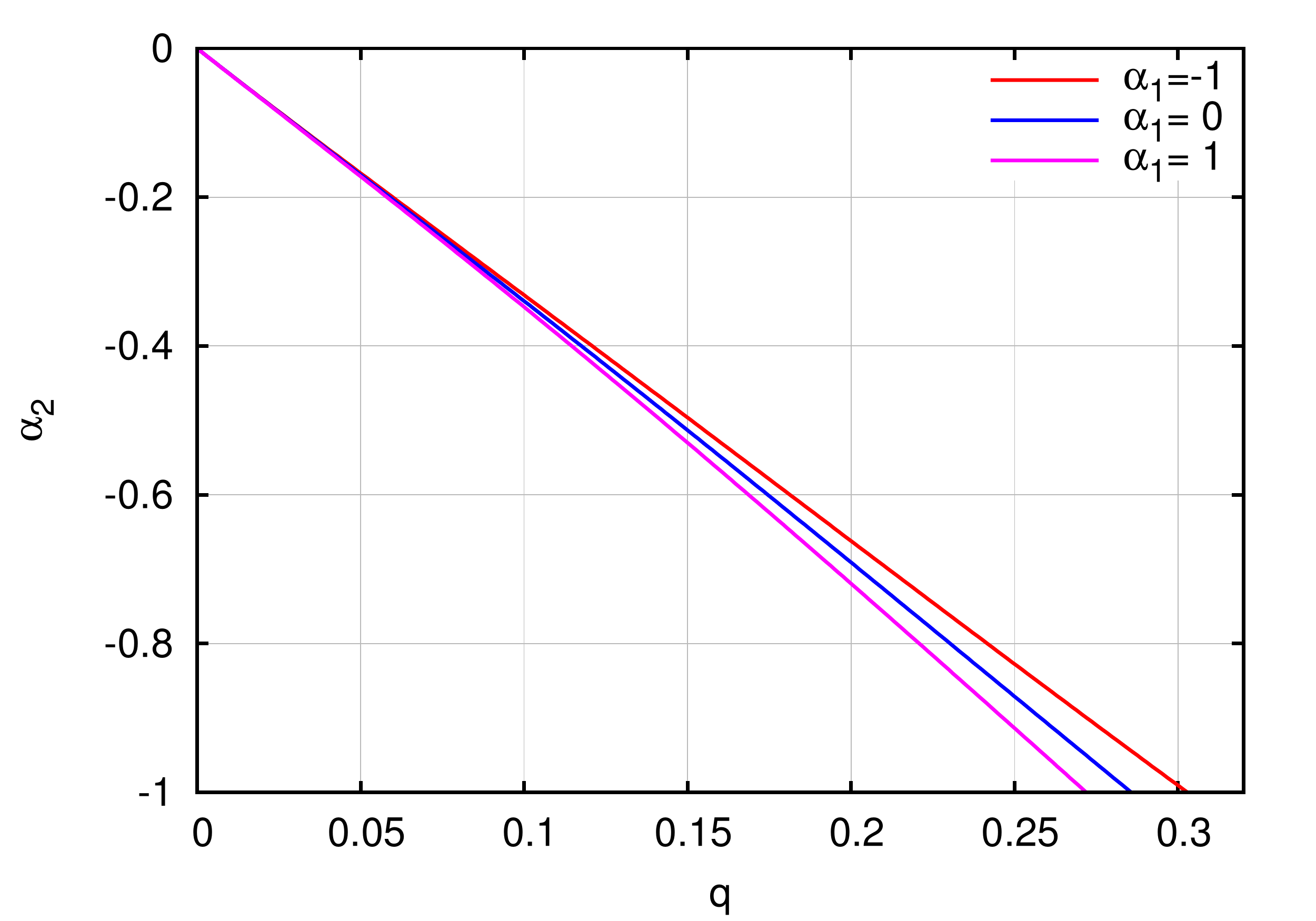}
  \caption{The BHB configurations that lead to a final Schwarzschild
black hole. We note that the spin of the small hole has little
influence on the values of the mass ratio and spin of the large
black hole that lead to a vanishing final spin.}\label{fig:S0}
\end{figure}

It is also interesting to see which configurations lead to a remnant
with vanishing spin (as was done in Ref.~\cite{Rezzolla:2007rd}, Fig
4a). Our results are shown in Fig.~\ref{fig:S0}.
They show that in order to have a final Schwarzschild black hole
the larger hole must be counteraligned with the orbital angular
momentum and the smaller hole must bear a mass ratio less than $0.3$.
This small mass ratio also explains the relatively weak dependence
on the spin of the secondary black hole.

In the equal-mass regime, Eq.~ (\ref{eq:4mass}) predicts a maximum
amount of radiated energy of $M_{\rm rem}^{\rm max}/m = 0.88693 \pm
0.00027$, i.e., a maximum radiated energy of $11.3\%$, and Eq.~ (\ref{eq:4spin}) predicts a maximum
remnant spin of  $\alpha_{\rm rem}^{\rm max} = 0.95166 \pm 0.00027$,
both of which closely agree with the predictions of
Ref.~\cite{Hemberger:2013hsa}.

\section{Statistical Distributions}\label{sec:stats}

\begin{figure}
  \includegraphics[width=\columnwidth]{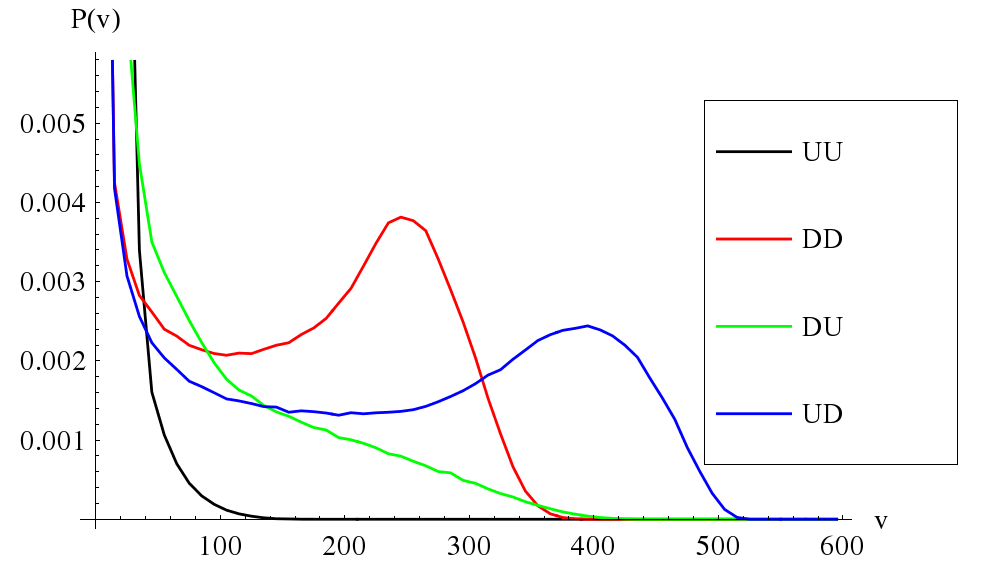}
  \includegraphics[width=\columnwidth]{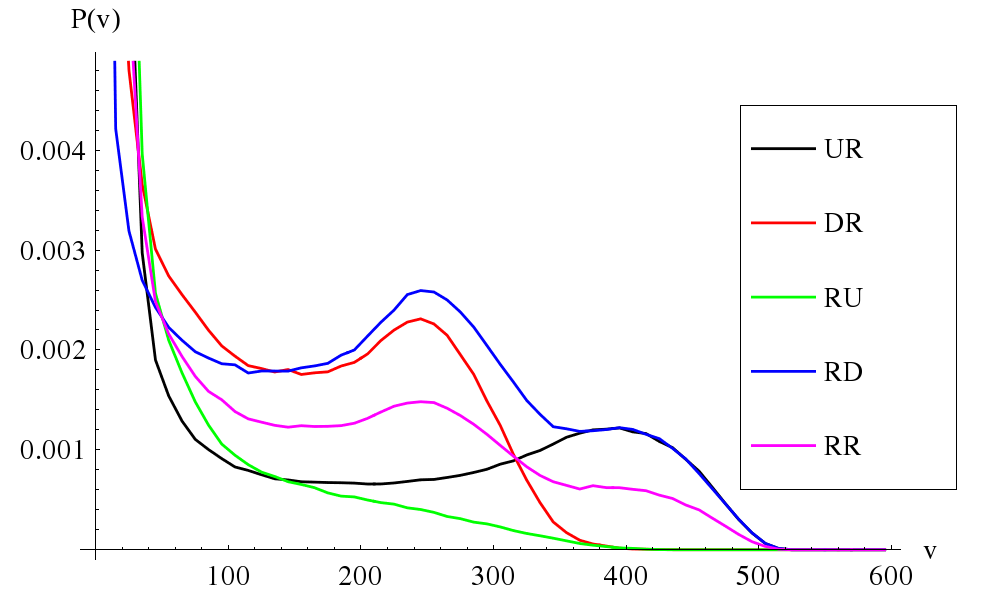}
  \caption{The probability $P(v)$ of the remnant BH
recoiling with speed $v$ assuming
a distribution of progenitor binaries 
 with spin-magnitude given by the cold accretion model
of~\cite{Lousto:2012su} and mass ratio distribution given
by~\cite{Yu:2011vp}, and assuming
the first (smaller) BH or second (larger) BH is always
aligned (U),
always counteraligned (D), or randomly (R) distributed with equal
probability of being aligned or counteraligned.}\label{fig:p_v}
\end{figure}

\begin{figure}
  \includegraphics[width=\columnwidth]{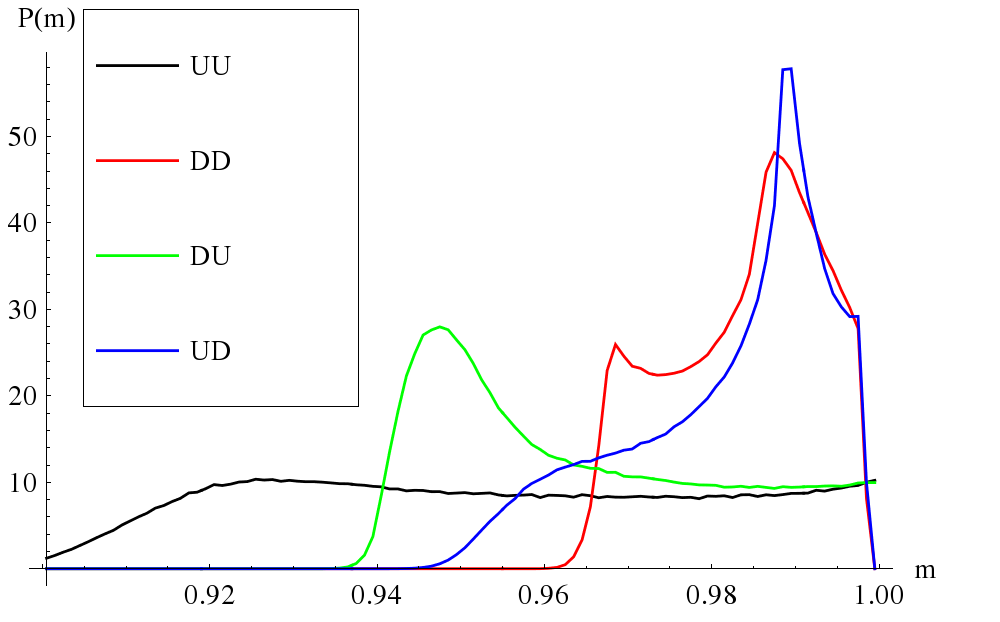}
  \includegraphics[width=\columnwidth]{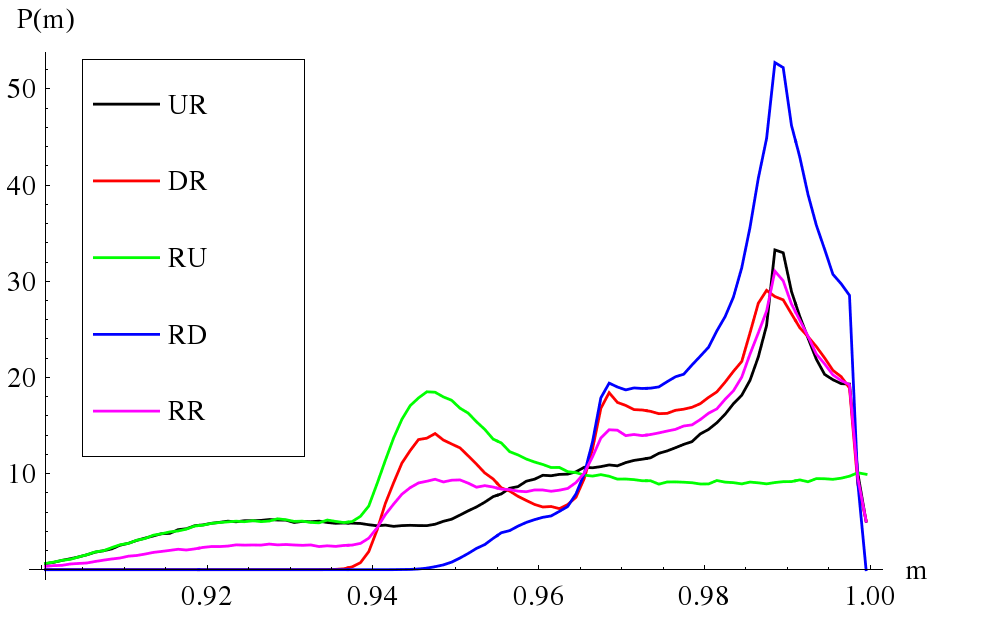}
  \caption{The probability $P(m)$ of the remnant BH having mass
$m$ (in units of the initial mass $M_1+M_2$) assuming a distribution
of progenitor binaries
with spin-magnitude given by the cold accretion model
of~\cite{Lousto:2012su} and mass ratio distribution given
by~\cite{Yu:2011vp} and assuming
the first (smaller) BH or second (larger) BH is always
aligned (U),
always counteraligned (D), or randomly (R) distributed with equal
probability of being aligned or counteraligned.}\label{fig:p_m}
\end{figure}

\begin{figure}
  \includegraphics[width=\columnwidth]{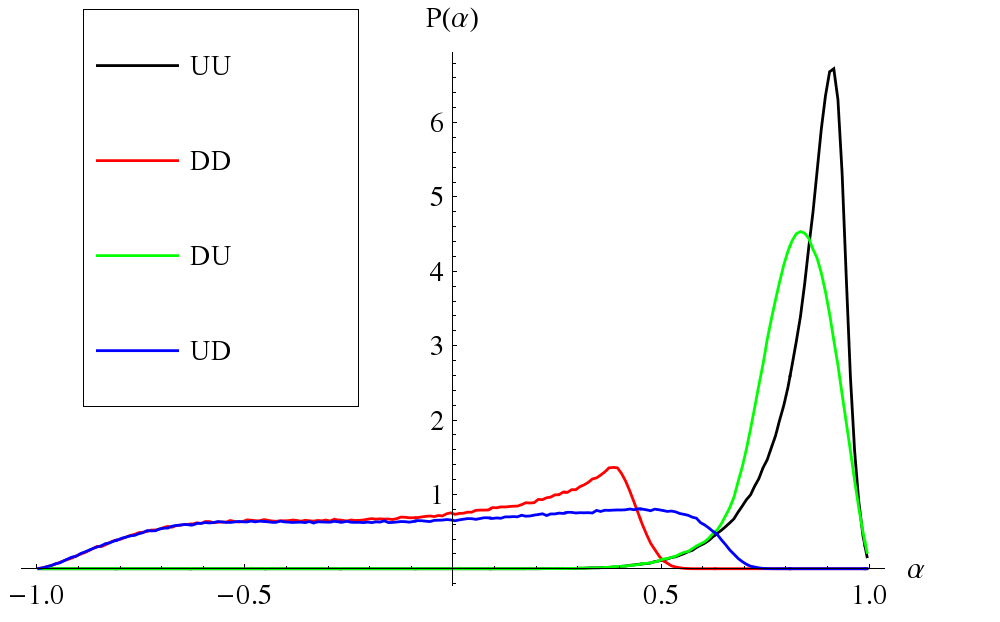}
  \includegraphics[width=\columnwidth]{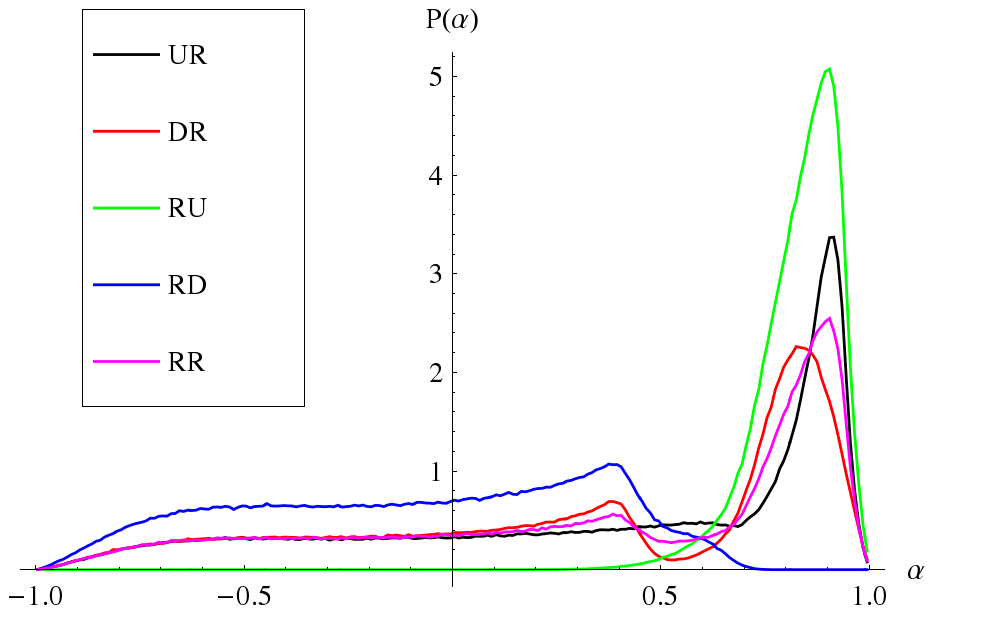}
  \caption{The probability $P(\alpha)$ of the remnant BH having
dimensionless spin $\alpha$
assuming a distribution
of progenitor binaries
with spin-magnitude given by the cold accretion model
of~\cite{Lousto:2012su} and mass ratio distribution given
by~\cite{Yu:2011vp} and assuming
the first (smaller) BH or second (larger) BH is always
aligned (U),
always counteraligned (D), or randomly (R) distributed with equal
probability of being aligned or counteraligned.}\label{fig:p_a}
\end{figure}

In order to visualize the consequences of Eqs.~(\ref{eq:4mass}),
(\ref{eq:4spin}),
(\ref{eq:4recoil}), and
(\ref{eq:xi}), we study the distributions of recoils and remnant masses and
spins from BHBs where the individual BH spins are either aligned or
counteraligned with the orbital angular momentum.  We study 9 families
of distributions of progenitors:
both BH spins aligned with the orbital angular momentum, both
counteraligned, two families where one BH  is aligned and the other
counter aligned, four families where one BH spin direction is chosen
randomly (aligned or counteraligned) and the other direction is 
fixed, and one family
where both BH orientations are chosen randomly. In all cases the
spin-magnitudes are chosen from the cold accretion distribution
in~\cite{Lousto:2012su}, which is 
represented by $P(\alpha) = (1-\alpha)^{b-1}\alpha^{a-1}$,
where $a=5.935$ and $b=1.856$,
and the mass ratio distribution from~\cite{Yu:2011vp,
Stewart:2008ep,
Hopkins:2009yy}, which is given by $P(q) \propto q^{-0.3}
(1-q)$.
In Figs.~\ref{fig:p_v},\ref{fig:p_m}, and \ref{fig:p_a} we show the
probabilities for a remnant recoiling with speed $v$, having mass
$M_{\rm rem}$, and spin $\alpha_{\rm rem}$. 

There are several interesting things to note from Fig.~\ref{fig:p_v}.
First, the probability for large recoils is much larger for the UD
family of configurations than for any of the other (nonrandom) configurations.
The UR and RD (here R denotes that the spin orientation is chosen
randomly) families both show the same probabilities at high
velocities. The reason for this is that high velocities can only come
from a UD type configuration. Both the UR and RD families have a 50\%
probability for a given configuration to be UD.

From Fig.~\ref{fig:p_m}, we see that the UU families show
significant probabilities for smaller remnant mass. Small remnant
masses occur for near-equal-mass UU systems, as this maximizes the
radiated energy. The UR, RU, and RR families show
a similar tail at smaller remnant masses. The UR and RU configurations have a 50\% probability
of being UU and the RR configurations have a 25\% probability of being
UU. Hence the $P(m)$ for the RR configurations is half the value of
$P(m)$ for the UR and RU families.

Finally, in Fig.~\ref{fig:p_a}, we see that the probability for a
final remnant spin counteraligned with the binary's orbital angular
momentum is nearly equal for the DD and UD families. The reason
is this can only happen in the small mass ratio regime with a
counteralign larger BH. Similarly the UR, DR, RR
configurations show similar tails near $\alpha=-1$. Again, these
configurations have a 50\% probability of being an XD configuration
(here X just means that the orientation of the smaller BH is
unimportant). 

\begin{figure}
  \includegraphics[width=\columnwidth]{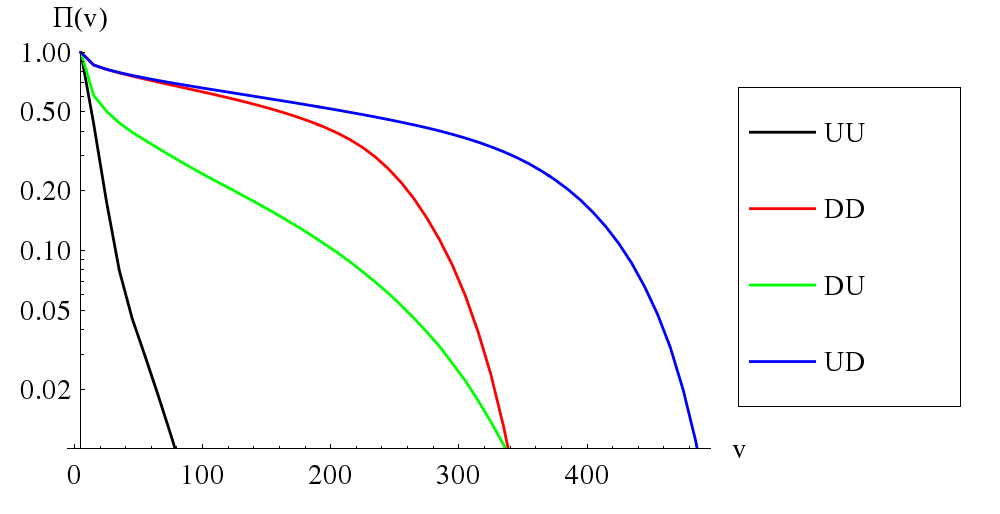}
  \includegraphics[width=\columnwidth]{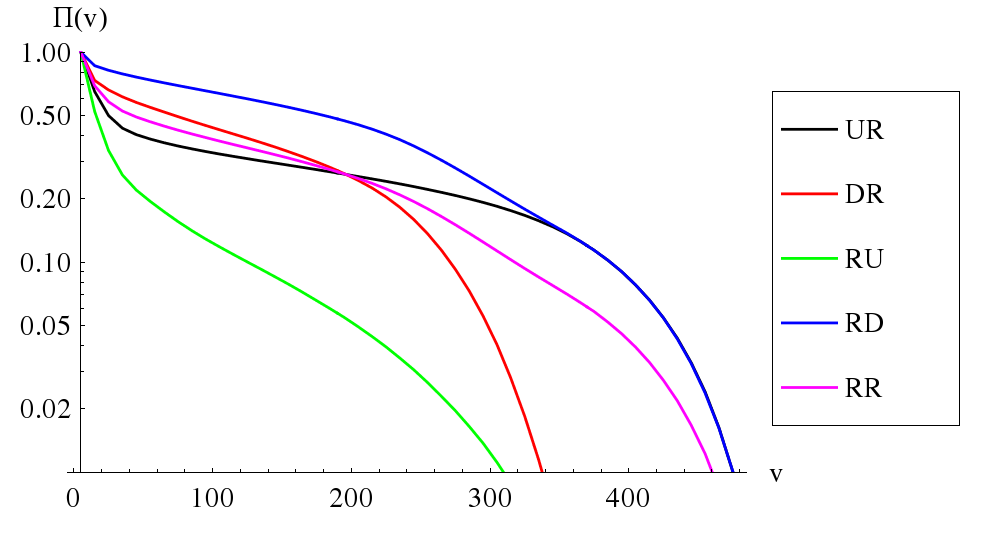}
  \caption{The integrated probability $\Pi(v)$ of the remnant BH
recoiling with speed $v$ or larger assuming
a distribution of progenitor binaries 
 with spin-magnitude given by the cold accretion model
of~\cite{Lousto:2012su} and mass ratio distribution given
by~\cite{Yu:2011vp} and assuming
the first (smaller) BH or second (larger) BH is always
aligned (U),
always counteraligned (D), or randomly (R) distributed with equal
probability of being aligned or counteraligned.}\label{fig:p_Iv}
\end{figure}

In Fig.~\ref{fig:p_Iv}, we show the integrated probability $\Pi(v)$ for a
recoil $v$ or larger, where 
$$
\Pi(v) = \int_v^\infty P(\nu)d\nu,
$$
and $P(\nu)$ is the probability for a recoil with speed $\nu < v < \nu
+ d\nu$.

We can thus consider a scenario where coherent accretion
aligned the smaller BH spin but left the spin of larger BH either
aligned or counteraligned (with equal probability). 
For such a UR configuration, we find the probability 
for $V>250\,\KMS$ is nearly $23\%$, while the probability for
$V>400\,\KMS$ is $8.4\%$.
If we assume that both BHs are equally likely to be aligned or
counteraligned,
the 
probabilities reduce to
$19\%$ and $4.2\%$, respectively. While these recoil velocities
are enough to expel the merged BHs from galaxies similar to
the milky way, they are not enough for the BHs to escape from 
much larger galaxies.
Nevertheless, these recoils  can still produce observational effects such
as displacement of the central BH from the galactic core
or a disturbance in the velocity field
of nearby stars \cite{Campanelli:2007ew}.
We also note that these recoil velocities probabilities represent
a lower bound for large recoils
 since we assumed exact alignment (or counteralignment)
of the spins with the orbital angular momentum and components
of the spin on the orbital plane can lead to very large recoils
\cite{Lousto:2011kp,Lousto:2012gt} even for relatively 
small misalignment angles, i.e., a few degrees.

\section{Conclusions and Discussion}\label{sec:discussion}

We studied the merger remnant of nonprecessing BHBs as a function
of the individual BH spins and mass ratio.  As accretion
\cite{Miller:2013gya, Sorathia:2013pca}
and resonances \cite{Schnittman:2004vq, Gerosa:2013laa}
align spins with the orbital angular momentum, this represents an important
subcase of the more general, 7 dimensional parameter space of binaries,
that includes arbitrary orientation of the spins.
The study performed here allowed us to use the unified
phenomenological description of a BHB merger developed in
\cite{Lousto:2012gt, Lousto:2013wta} to model the recoil
(\ref{eq:4recoil}), remnant mass (\ref{eq:4mass}), and spin
(\ref{eq:4spin}), with expected accuracies to within
 $3\%$, $1\%$, and $1\%$
relative errors, respectively.

We found that the spin contribution to the recoil can add to, or
subtract from, the component of the recoil due to unequal masses, with
(partial)
cancellation occurring when the larger BH spin is aligned with the
orbital angular momentum. On the other hand, when the larger BH spin
is counteraligned, the two components of the recoil add, leading to
larger recoils at intermediate mass ratios. We find that the maximum
recoil occurs for $q\sim0.62$.

\begin{figure}
  \includegraphics[angle=270,width=\columnwidth]{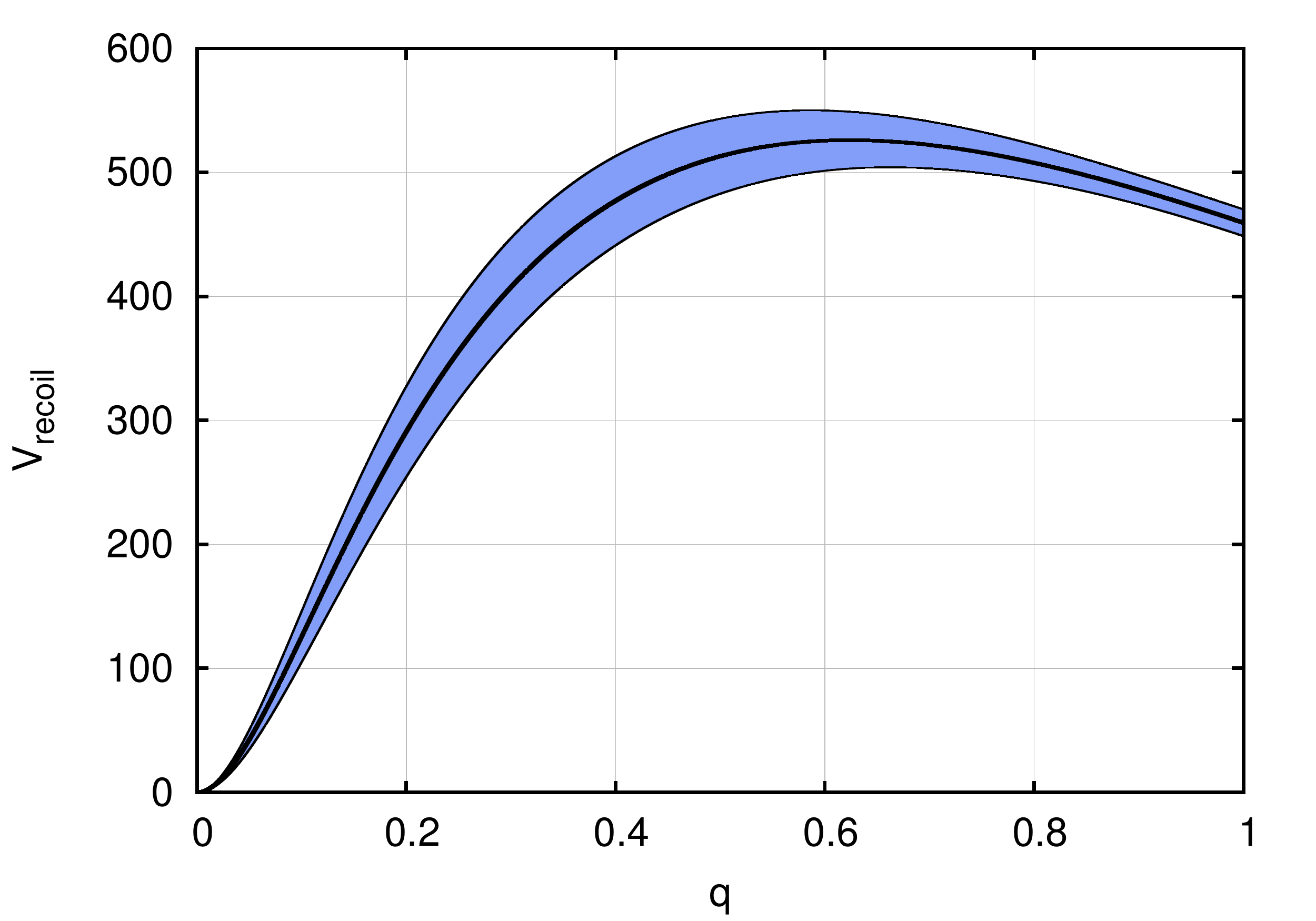}
  \caption{The recoil velocity for the UD configuration with the
small black hole spin aligned with the orbital angular
momentum and the large hole spin counteraligned as a function
of the mass ratio as predicted by Eq.~(\ref{eq:4recoil}. 
The maximum recoil of 
$526\pm 23\,\KMS$  is reached at $q = 0.6235 \pm 0.038$, 
with maximally spinning holes, $\alpha_1=1.0,\ \alpha_2=-1.0.$
The shaded area represents the estimated errors.}
\label{fig:kmax_vs_q}
\end{figure}

Also note that the new maximum of the recoil 
(See Fig.~\ref{fig:kmax_vs_q}) represents a modest
increase in the maximum value itself (nearly $17\%$). However,
just like for the case of the hangup kicks \cite{Lousto:2011kp}, 
the most important effect is that the volume of parameter space
leading to large recoils is much larger, i.e., the UD configurations
have $(V_{recoil}>200\,\KMS)$ with a $52\%$ probability.

Another similarity with the hangup kick effect is the need to
incorporate terms beyond linear in the spins (and mass ratio)
to accurately model the final recoil (and mass and spin). This underlines the
inherently nonlinear nature of general relativity, in particular
when modeling the highly dynamical regime of BHB mergers.

This provides the opportunity for an important test of general relativity
in its strong field realm. Searches for observational
effects from recoiling black holes are well underway.
This includes searches for large differential red/blue shifts
from AGN (see \cite{Komossa:2012cy} for a review),
and distortions in the dynamics of the core of galaxies
(see \cite{Gerosa:2014gja} for the latest
observation that the lack of black holes in bright cluster galaxies
might be the result of large kicks).

We finally note that 
the use of gravitational waveforms from aligned and antialigned
spins proves to be of great help for detection 
algorithms~\cite{Privitera:2013xza} used by laser interferometer 
observatories. Our models for the final mass and spin
from the merger of two black holes can be used to produce more accurate
semianalytic models of such waveform templates, which may also be used for
parameter estimation.

\acknowledgments 

The authors gratefully acknowledge the NSF for financial support from Grants
PHY-1305730, PHY-1212426, PHY-1229173,
AST-1028087, PHY-0969855, OCI-0832606, and
DRL-1136221. Computational resources were provided by XSEDE allocation
TG-PHY060027N, and by NewHorizons and BlueSky Clusters 
at Rochester Institute of Technology, which were supported
by NSF grant No. PHY-0722703, DMS-0820923, AST-1028087, and PHY-1229173.


\appendix
\section{Analysis of the sources of errors and robustness of remnant
properties in the BHB simulations}
\label{app:appendix}

In order to assess the robustness of our results with respect to the
different sources of errors and the various approximations that we use
in our simulations, we study in detail an equal-mass BHB in a UD
configuration (with spins $\alpha=0.8$)
starting from two different initial separations. We vary the
resolutions, grid structure, waveform extraction radii, and the number
of $\ell$ modes used in the construction of the radiated linear
momentum.  The initial data parameters for the two configurations,
denoted here by A and B, are given in Table~\ref{tab:appID}.

Case A represents a prototypical configuration of the runs in this paper 
while the  case B was first studied in Ref.~\cite{Lousto:2013wta},
where we also performed a convergence study of that configuration.
In this work, we use a grid structure with between 9 and 11 levels of
refinement, depending on mass ratio and spin. For all new simulations,
the outer boundary was placed at 400M with a resolution of $4M$ on the
coarsest level and a resolution of $1M$ in the wavezone.
The finest level around each BH was as wide as twice the
diameter of the relaxed horizon (the number of points across each
horizon was between 28 and 60). In addition, for the
highly-spinning horizons, we added an additional level inside the
horizon of width roughly half of the horizon diameter.
We also performed similar runs but with resolutions in
the wavezone of $M/0.88$ and $M/1.2$.

Since in the current work,  we use a different refinement level
 grid structure than in 
Ref.~\cite{Lousto:2013wta},
we also perform a new set of convergence simulations for case B using
the newer grid structures.

\begin{figure}
  \includegraphics[angle=270,width=\columnwidth]{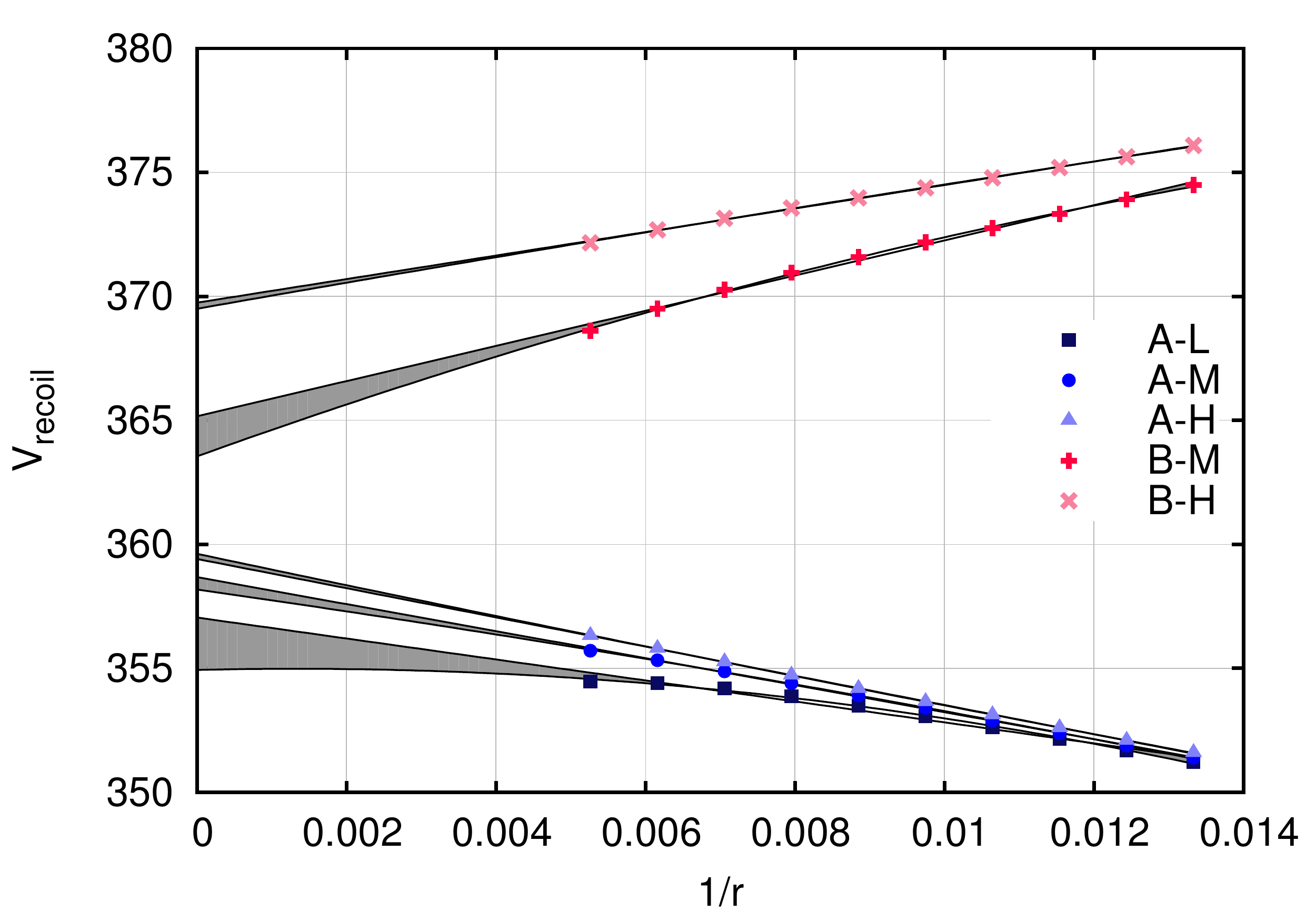}
  \caption{The recoil velocity as computed at a given extraction
radius: $75M - 190M$ and extrapolations to infinity. The different
curves correspond to the two initial separations labeled as A and B
and as a function of resolution (Low - Medium - High)
refined by a global factor 1.2\,. The shaded regions are
those points contained between a linear and quadratic extrapolation of
the data (least squares fit).}
\label{fig:kick_rinv}
\end{figure}

Aside from truncation errors due to finite resolution, the simulation
results will depend on the extraction radii.
Hence we also consider different extraction radii and extrapolations
to null infinity.
While the location of the observers in the set of
runs in Ref.~\cite{Lousto:2013wta} was restricted to the $R_{obs}/m=60 - 100$
range in the runs of this paper we extended this to 
$R_{obs}/m=190$ and, in addition, locate the extraction radii
equidistant in $1/R$, with $R_{obs}/m=75,80.4,86.7,94.0,102.6,113.0,125.7,141.7,162.3,190.0.$ 
The results of such studies is displayed in Fig.~\ref{fig:kick_rinv}.

\begin{widetext}

\begin{table}
\caption{A and B case studies with BHs at two different initial separations
with two sets of parameters from estimated quasicircular orbits.}\label{tab:appID}
\begin{ruledtabular}
\begin{tabular}{lcccccccccccc}
Config.   & $x_1/m$ & $x_2/m$  & $P/m$    & $m^p_1/m$ & $m^p_2/m$ & $S_1/m^2$ & $S_2/m^2$ & $m^H_1/m$ & $m^H_2/m$ & $M_{\rm ADM}/m$ & $a_1/m_1^H$ & $a_2/m_2^H$\\
\hline
A\_DU0.8  & -4.9832 & 4.5267 & 0.09905 & 0.30178 & 0.30168 & -0.2     & 0.2     & 0.5    & 0.5    & 0.98951 & -0.8 & 0.8 \\
B\_DU0.8  & -4.5465 & 4.4303 & 0.10557 & 0.30377 & 0.30366 & -0.20465 & 0.20465 & 0.5053 & 0.5053 & 1       & -0.8 & 0.8 \\
\end{tabular}
\end{ruledtabular}
\end{table}

\end{widetext}

Interestingly, we see that while the measured recoils from the
A and B simulations differ by $\sim 30\,\KMS$ at lower resolution and
smaller radii, they approach each other as both the resolution
and extraction radii increases. The differences in the extrapolated
recoil for the highest resolution A and B configuration is smaller
than $10\KMS$.

\begin{figure}
  \includegraphics[angle=270,width=\columnwidth]{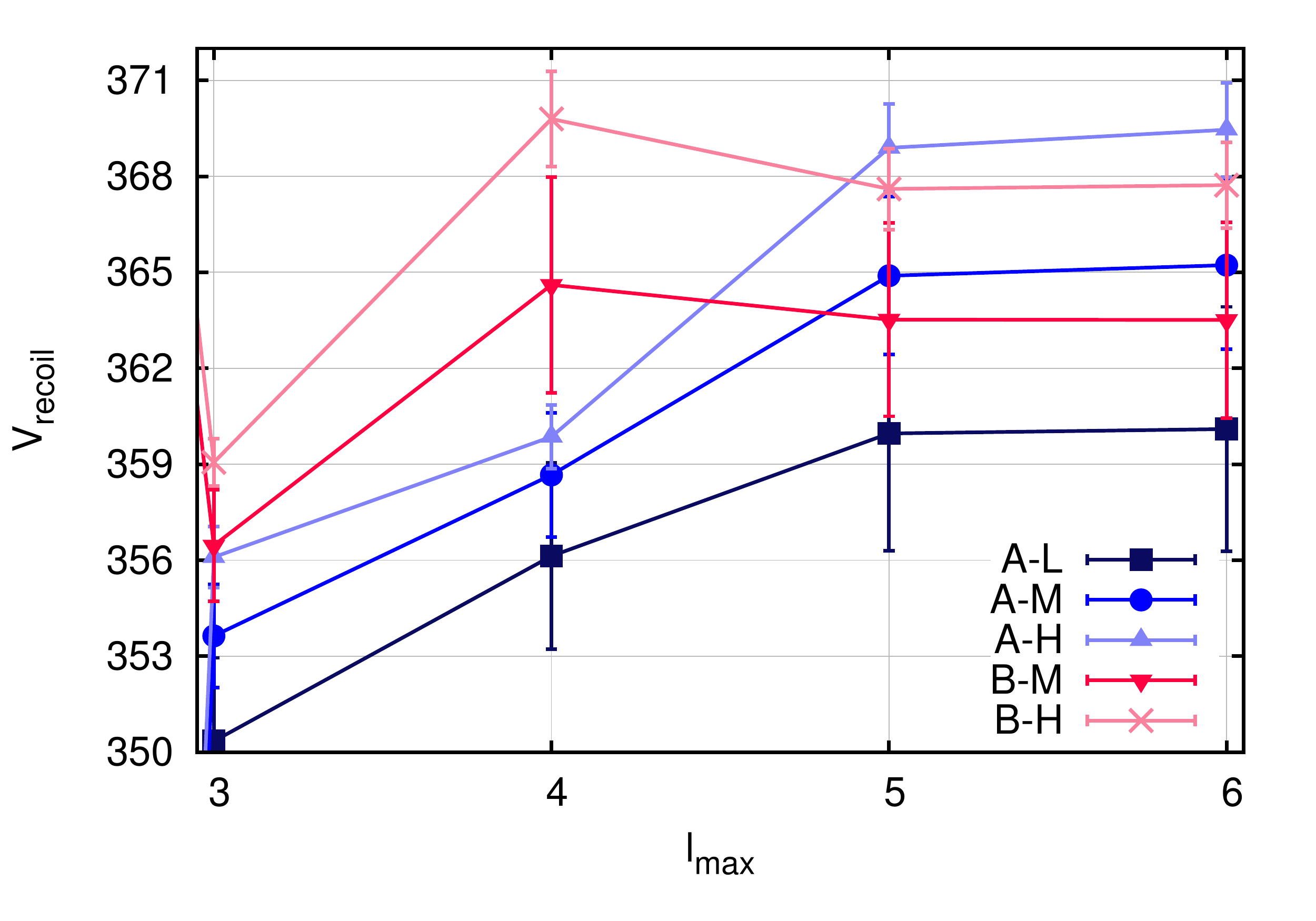}
  \caption{The dependence of the computed recoil velocity on  the
number of $\ell$ modes used to construct the radiated linear
momentum. Here all modes with $\ell \leq \ell_{\rm max}$ were used
and we show the recoil for the A and B configurations for the Low,
Medium, and High resolution runs.
}
  \label{fig:kickVslmax}
\end{figure}

In Fig.~\ref{fig:kickVslmax} we show the recoil extrapolated to
infinity versus the number of $\ell$ modes. Interestingly, for the B
configuration, using all modes up through $\ell=4$ appears to be
sufficient, while for the A configuration, there is a noticeable change
in the recoil when adding the $\ell=5$ modes.

Based on the results in Figs.~\ref{fig:kick_rinv} 
and \ref{fig:kickVslmax}, we used the medium
resolution grid structure (i.e., A-M in the figures) and summed all
modes up through $\ell=6$ when calculating the recoils given in
the tables and figures of this work. We also note that, because the
recoils from the A and B configurations did approach each other,
effects due to finite starting separation are reasonably mitigated for
initial separations of between $9M$ and $10M$ and above.

In conclusion we see that in order to have a robust measure
of the recoil
we need to consider BHBs with sufficiently large initial separations,
medium resolution, and we need to sum over modes up through $\ell=6$.

\begin{figure}
  \includegraphics[angle=270,width=\columnwidth]{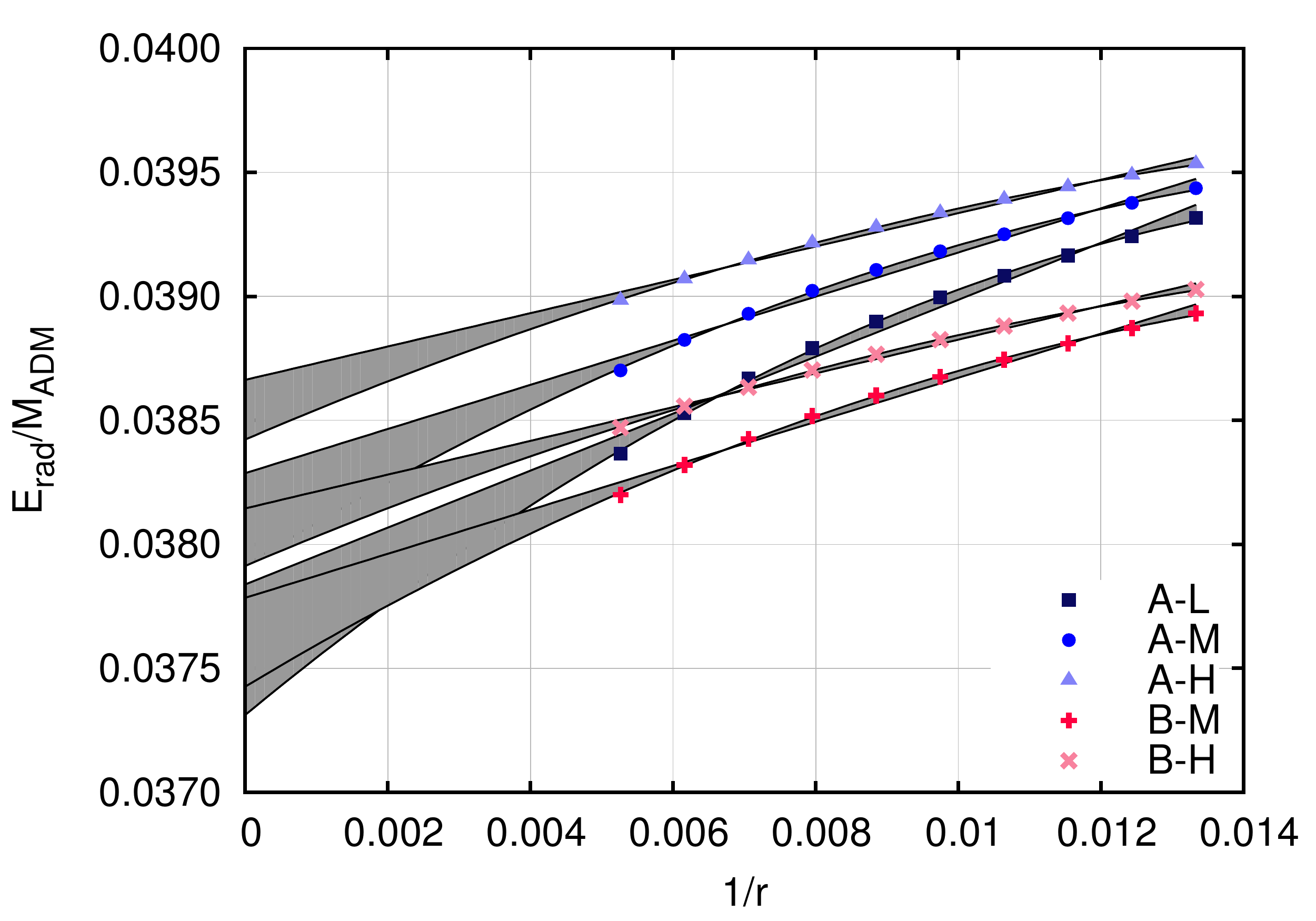}\\
  \includegraphics[angle=270,width=\columnwidth]{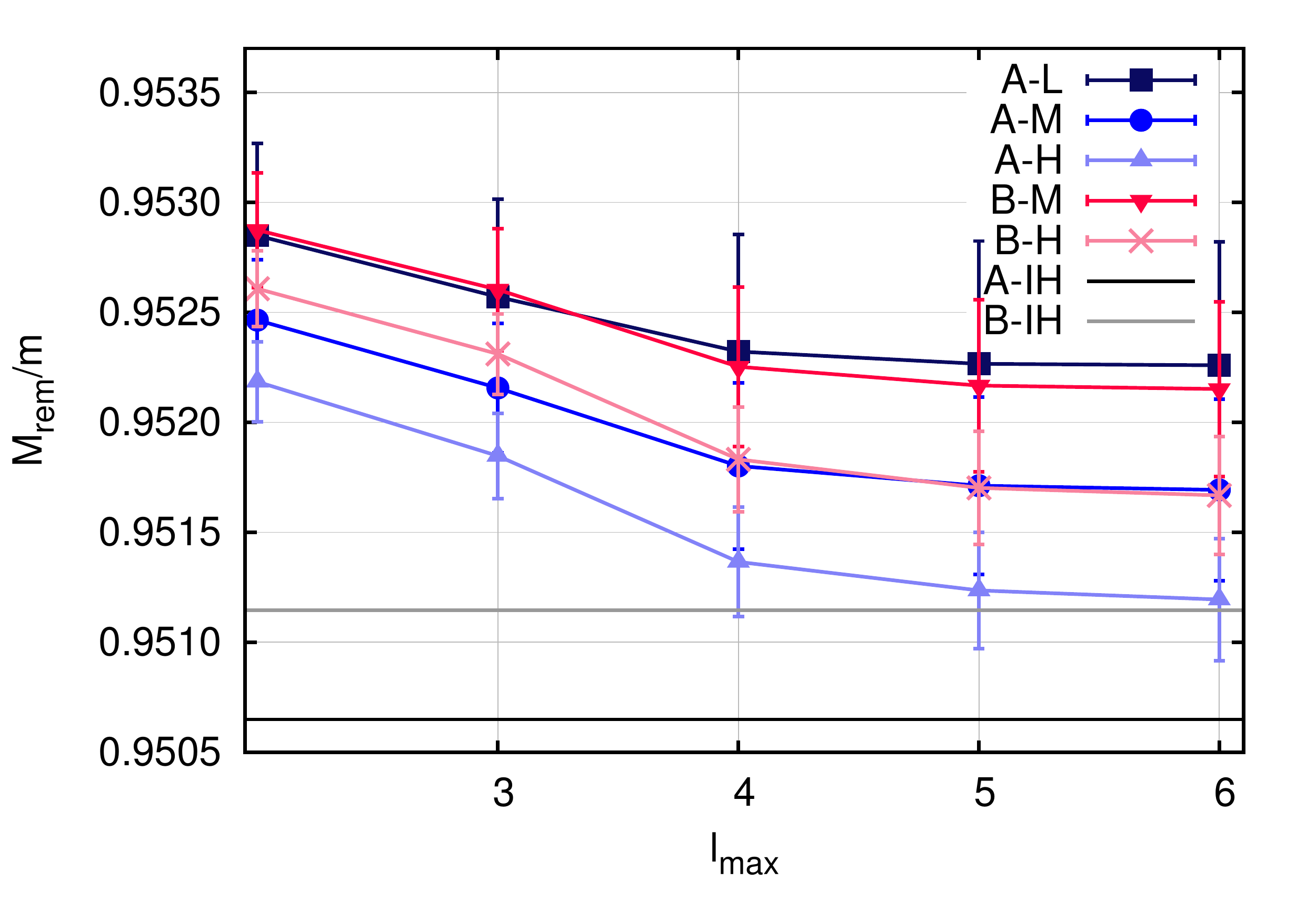}
  \caption{Above: The radiated energy as computed at a given extraction
radius: $75M - 190M$ and extrapolations to infinity. The different
curves correspond to the two initial separations labeled as A and B
and as a function of resolution (Low - Medium - High)
refined by a global factor 1.2\,. The shaded regions are
those points contained between a linear and quadratic extrapolation of
the data (least squares fit).
Below: The dependence of the computed radiated energy on  the
number of $\ell$ modes used to construct it.
Here all modes with $\ell \leq \ell_{\rm max}$ were used.  The black
and gray lines labeled with ``IH" are the associated final mass calculated
from the BH horizon.  On this scale, all resolutions are on top of
one another, so only one line is shown.
}
  \label{fig:erad}
\end{figure}

\begin{figure}
  \includegraphics[angle=270,width=\columnwidth]{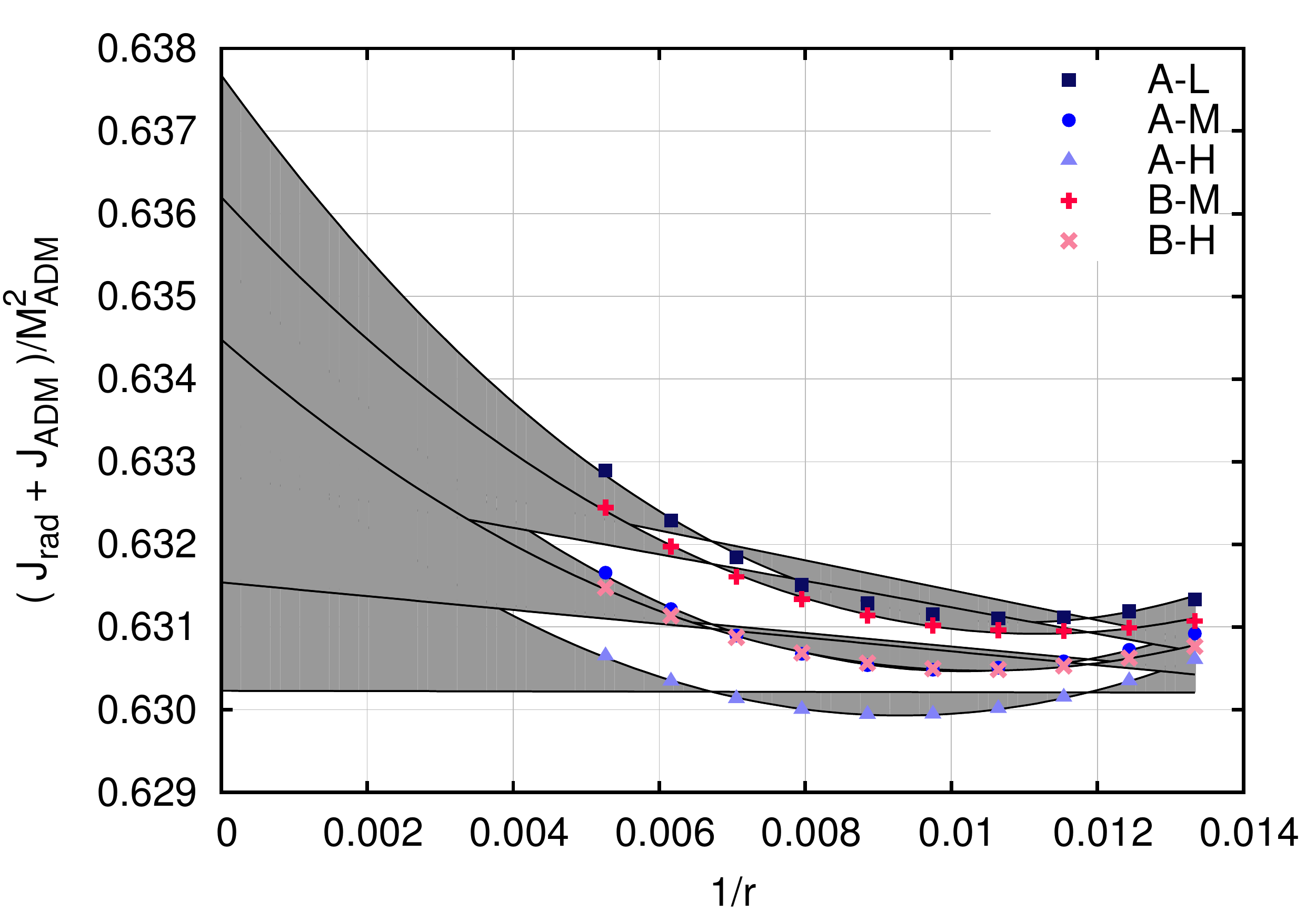}\\
  \includegraphics[angle=270,width=\columnwidth]{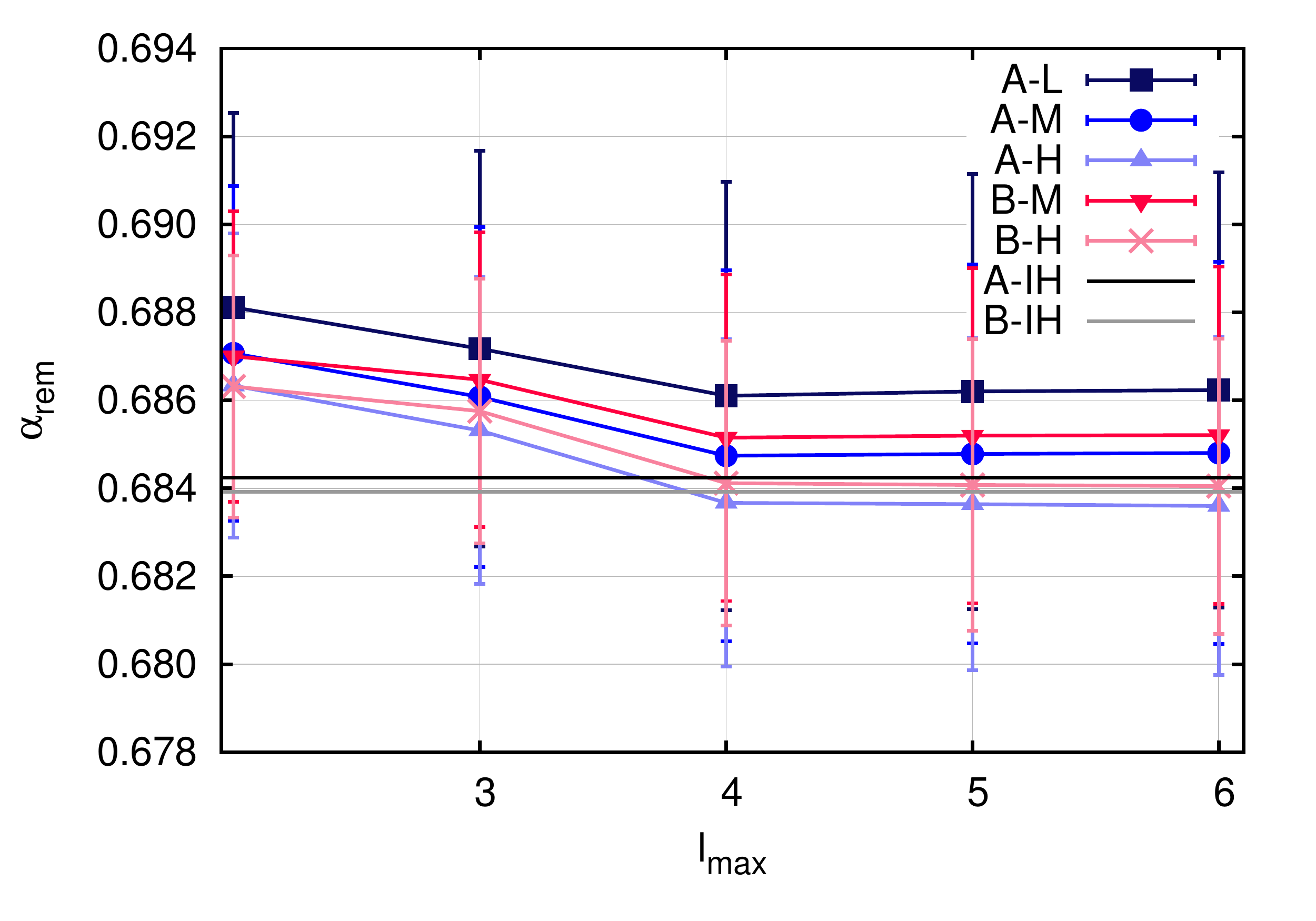}
  \caption{Above: The radiated angular momentum
as computed at a given extraction
radius: $75M - 190M$ and extrapolations to infinity. The different
curves correspond to the two initial separations labeled as A and B
and as a function of resolution (Low - Medium - High)
refined by a global factor 1.2\,. The shaded regions are
those points contained between a linear and quadratic extrapolation of
the data (least squares fit).
Below: The dependence of the computed radiated angular momentum on  the
number of $\ell$ modes used to construct it.
Here all modes with $\ell \leq \ell_{\rm max}$ were used.  The black
and gray lines labeled with ``IH" are the associated final spin calculated
from the BH horizon.  On this scale, all resolutions are on top of
one another, so only one line is shown.
}
  \label{fig:jrad}
\end{figure}

In Fig.~\ref{fig:erad}, we plot the radiated energy and the ratio of the final
to initial mass derived from the radiated energy as a function of
$1/r$ and $\ell_{\rm max}$. We see that A and B configuration approach
each other with an increase in resolution and larger observer radii.
We also note that there are only small errors introduced by using
$\ell_{\rm max}$ as small as $\ell=4$. For the radiated angular
momentum (see Fig.~\ref{fig:jrad}), the extrapolation error
dominates, and while the A and B configurations results seems
to converge to each other, the error bars are quite large.
Again, we see that summing up through $\ell=4$ is sufficient to obtain
the final remnant spin.

One final note, for the runs presented in the main body of the paper,
we use extraction radii up to $r=102.6m$, rather than $r=190m$. We did
this because we observed, at our working resolutions,
a dissipation effect at larger radii in
which the amplitude of $r\psi_4$ steadily decreases with radius in
the outer zone.

We apply all
these criteria to the rest of the new simulations we perform in this
paper to ensure similar error bars as the ones presented in this Appendix.
Note also that a similar, but independent, waveform-error analysis
was carried out in 
Ref.~\cite{Hinder:2013oqa}.

\bibliographystyle{apsrev}
\bibliography{../../../Bibtex/references}

\end{document}